\documentstyle[11pt,epsf]{article}

\newcommand{\newsection}{ \setcounter{equation}{0} \section}

\newcommand{\beq}{\begin{equation}} \newcommand{\eeq}{\end{equation}}
\newcommand{\bea}{\begin{eqnarray}} \newcommand{\eea}{\end{eqnarray}}

  \newcommand
{\Romannumeral}[1]{\uppercase\expandafter{\romannumeral#1}}

\newcommand{\be}{\begin{enumerate}} \newcommand{\ee}{\end{enumerate}}
\newcommand{\bi}{\begin{itemize}} \newcommand{\ei}{\end{itemize}}
\newcommand{\ba}{\begin{array}} \newcommand{\ea}{\end{array}}
\newcommand{\bc}{\begin{center}} \newcommand{\ec}{\end{center}}
\newcommand{\bt}{\begin{tabular}} \newcommand{\et}{\end{tabular}}

\def\lsim{\mathrel{\rlap{\lower4pt\hbox{\hskip1pt$\sim$}}
    \raise1pt\hbox{$<$}}}           % less than or approx. symbol
\def\gsim{\mathrel{\rlap{\lower4pt\hbox{\hskip1pt$\sim$}}
    \raise1pt\hbox{$>$}}}           % greater than or approx. symbol

\newcommand{\Tr}{\mathop{\rm Tr}}           % Trace
\newcommand{\half}{\textstyle {1\over2} \displaystyle}    % One half
\newcommand{\third}{\textstyle {1\over3} \displaystyle}   % One third
\newcommand{\Dslash}{{\hbox{D}\kern-0.6em\raise0.15ex\hbox{/}}} % D slash

\renewcommand{\et}{\eta}

\newcommand{\k}{\kappa}

\begin{document}

\setlength{\oddsidemargin}{0cm} \setlength{\baselineskip}{7mm}

\input epsf

\begin{normalsize}\begin{flushright}

CERN-PH-TH/2006-145 \\
% DAMTP-2006-xx \\
% UCI-2004-xx \\
July 2006 \\

\end{flushright}\end{normalsize}

\begin{center}
  
\vspace{5pt}

{\Large \bf Renormalization Group Running of Newton's G :}

{\Large \bf The Static Isotropic Case}

\vspace{40pt}

{\sl H. W. Hamber}
$^{}$\footnote{On leave from the
Department of Physics, University of California, Irvine Ca 92717, USA.}
and 
{\sl R. M. Williams}
$^{}$\footnote{Permanent address:
Department of Applied Mathematics and Theoretical Physics,
Wilberforce Road, Cambridge CB3 0WA, United Kingdom.} \\

\vspace{30pt}

Theory Division \\
CERN \\
CH-1211 Geneva 23, Switzerland \\

\vspace{20pt}

\end{center}

\begin{center} {\bf ABSTRACT } \end{center}

\noindent

Corrections are computed to the classical static isotropic solution of general
relativity, arising from non-perturbative quantum gravity effects.
A slow rise of the effective gravitational coupling with distance is shown to involve
a genuinely non-perturbative scale, closely connected with the gravitational vacuum
condensate, and thereby, it is argued, related to the observed effective
cosmological constant.
Several analogies between the proposed vacuum condensate picture of quantum gravitation,
and non-perturbative aspects of vacuum condensation in strongly coupled
non-abelian gauge theories are developed.
In contrast to phenomenological approaches, the
underlying functional integral formulation of the theory severely
constrains possible scenarios for the renormalization group evolution of couplings.
The expected running of Newton's constant $G$ is compared 
to known vacuum polarization induced effects in QED and QCD.
The general analysis is then extended to a set of covariant non-local
effective field equations, intended to incorporate the full scale dependence of $G$,
and examined in the case of the static isotropic metric. 
The existence of vacuum solutions to the effective field equations in
general severely restricts the possible values of the scaling exponent $\nu$.

% \vspace{15pt}

% \begin{center} {\it (Submitted to the Physical Review D)} \end{center}

% \noindent PACS Numbers: 04.60.-m, 04.60.Gw, 04.60.Nc, 98.80.Qc

% Above are: Quant Grav, Cov Sum over Hist Quant, Latt Grav, Quant Cosm 

\vfill

%% no page number on 1-st page
\pagestyle{empty}

\newpage

\pagestyle{plain}

\vskip 10pt
\newsection{Introduction}
\hspace*{\parindent}

Over the last few years evidence has mounted to suggest that quantum gravitation,
even though plagued by meaningless infinities in standard weak coupling perturbation theory,
might actually make sense, and lead to a consistent theory at the non-perturbative level.
As is often the case in physics, the best evidence does not come from
often incomplete and partial results in a single model,
but more appropriately from the level of consistency that various, often quite unrelated, 
field theoretic approaches provide.
While it would certainly seem desirable to obtain a closed form analytical solution
for the euclidean path integral of quantum gravity, experience with other field theories
suggests that this goal might remain unrealistic in the foreseeable future, and 
that one might have to rely in the interim on partial results and reasoned analogies
to obtain a partially consistent picture of what the
true nature of the ground state of non-perturbative gravity might be.

One aspect of quantum gravitation that has stood out for some time is the
rather strident contrast between the naive picture one gains from perturbation
theory, namely the possibility of an infinite set of counterterms, uncontrollable 
divergences in the vacuum energy of just about any field including the
graviton itself, and typical curvature scales comparable to the Planck mass [1-3],
and, on the other hand, the new insights gained from non-perturbative
approaches, which avoid reliance on an expansion in a small parameter (which
does not exist in the case of gravity) and which would suggest instead
a surprisingly rich phase structure, non-trivial ultraviolet fixed points [4-8]
and genuinely non-perturbative effects such as the appearance of 
a gravitational condensate.
The existence of non-perturbative vacuum condensates does not necessarily
invalidate the wide range of semi-classical results [9-11] obtained in gravity so
far, but re-interprets the gravitational background fields as suitable
quantum averages, and further adds to the effective gravitational
Lagrangian the effects of the (finite) scale dependence of the gravitational
coupling, in a spirit similar to the Euler-Heisenberg corrections to
electromagnetism. 

Perhaps the goals that are sometimes set for quantum gravity and related extensions,
that is, to explain and derive, from first principles, the values of Newton's
constant and the cosmological constant, are placed unrealistically high.
After all, in other well understood quantum field theories like QED and QCD the renormalized
parameters ($\alpha$, $\alpha_S$, ...) are fixed by experiments, and 
no really compelling reason exists yet as to why they should
take on the actual values observed in laboratory experiments.
More specifically in the case of gravity, Feynman has given
elaborate arguments as to why quantities such as Newton's constant 
(and therefore the Planck length) might have cosmological origin,
and therefore unrelated to any known particle physics phenomenon \cite{feylec}.

In this paper we will examine a number of issues connected with the
renormalization group running of gravitational couplings.
We will refrain from considering more general frameworks (higher derivative couplings,
matter fields etc.), and will focus instead on basic aspects of the
pure gravity theory by itself.
Our presentation is heavily influenced by the numerical and analytical
results from the lattice theory of quantum gravity (LQG), which have,
in our opinion, helped elucidate numerous details of the non-perturbative phase structure
of quantum gravity, and allowed a first determination of the scaling
dimensions directly in $d=4$.
The lattice provides a well defined ultraviolet regulator, reduces the
continuum functional integral to a finite set of convergent integrals,
and allows statistical field theory methods, including numerical ones,
to be used to explore the nature of ground state averages and correlations. 

The scope of this paper is therefore to explore the overall consistency of the picture 
obtained from the lattice, by considering a number of core issues, one of
which is the analogy with a much better understood class of theories, non-abelian
gauge theories and QCD (Sec. 2).
We will argue that, once one takes for granted a set of basic lattice results,
it is possible to discuss a number of general features without having to explicitly
resort to specific aspects of the lattice cutoff or the lattice action. 
For example, it is often sufficient to assume that a cutoff $\Lambda$
is operative at very short distances, without having to involve in
the discussion specific aspects of its implementation.
In fact the use of continuum language, in spite of its
occasional ambiguities when it comes to the proper, regulated definition
of quantum entities, provides a more transparent language for presenting 
and discussing basic results.

The second aspect we wish to investigate in this paper is the nature of the
rather specific predictions about the running of Newton's constant $G$.
A natural starting point is the solution of the non-relativistic
Poisson equation (Sec. 3), whose solutions for a point source
can be investigated for various values of the exponent $\nu$.
We will then show that a scale dependence of $G$ can be consistently embedded
in a relativistic covariant framework, whose consequences can then be worked
out in detail for specific choices of metrics (Sec. 4).
For the static isotropic metric, we then derive the leading quantum correction
and show that, unexpectedly, it seems to restrict the possible
values for the exponent $\nu$, in the sense that in some instances
no consistent solution to the effective non-local
field equations can be found unless $\nu^{-1}$ is an integer.

To check the overall consistency of the results, a slightly different
approach to the solution of the static isotropic
metric is discussed in Sec. 5, in terms of an effective vacuum density
and pressure.
Again it appears that unless the exponent $\nu$ is close to $1/3$,
a consistent solution cannot be obtained.
At the end of the paper we add some general comments on two subjects 
we discussed previously.
We first make the rather simple observation that a running of Newton's
constant will slightly distort the gravitational wave spectrum
at very long wavelengths (Sec. 6).
We then return to the problem (Sec. 7) of finding solutions of the effective
non-local field equations in a cosmological context \cite{effective}, wherein
quantum corrections to the Robertson-Walker metric and the basic
Friedman equations are worked out, and discuss some of the simplest
and more plausible scenarios for the growth (or lack thereof)
of the coupling at very large distances, past the deSitter horizon.
Sec. 8 contains our conclusions.

% \newpage

\vskip 30pt
\newsection{Vacuum Condensate Picture of Quantum Gravitation}
\hspace*{\parindent}

The lattice theory of quantum gravity provides a well defined 
and regularized framework in which non-perturbative quantum
aspects can be systematically investigated.

Let us recall here some of the main results of the lattice quantum
gravity (LQG) approach, and their relationship to related approaches.

(i) The theory is formulated via a discretized Feynman functional integral [13-26].
Convergence of the euclidean lattice path integral {\it requires}
in dimensions $d>2$ a positive bare cosmological constant $\lambda_0 >0$ \cite{hw84}. 
The need for a bare cosmological constant is in line with renormalization
group results in the continuum, which also imply that radiative
corrections will inevitably generate a non-vanishing $\lambda$ term.

(ii) The lattice theory in four dimensions is characterized by {\it two phases},
one of which appears for $G$ less than some critical value $G_c$, and
can be shown to be physically unacceptable as it describes a collapsed manifold with
dimension $d \simeq 2$.  
The quantum gravity phase for which $G > G_c$ can be shown instead to describe 
smooth four-dimensional manifolds at large distances, and remains therefore physically viable.
The continuum limit is taken in the standard way,
by having the bare coupling $G$ approach $G_c$.
The two phase structure persists in three dimensions \cite{hw3d}, and even
at $d=\infty$ \cite{larged}, whereas in two dimensions one has only one phase \cite{hw2d}. 

(iii) The presence of two phases in the lattice theory is consistent with
the continuum $2+\epsilon$ expansion result, which also predicts the existence
of two phases above dimensions $d=2$. 
The presence of a nontrivial ultraviolet fixed point in the continuum above $d=2$, with
nontrivial scaling dimensions, relates to the existence of a phase transition
in the lattice theory. 
The lattice results further suggest that the weakly coupled phase is in fact non-perturbatively
{\it unstable}, with the manifold collapsing into a two-dimensional degenerate geometry.
The latter phase, if it had existed, would have described gravitational screening. 

(iv) One key quantity, the critical exponent $\nu$, characterizing the non-analyticity in the 
vacuum condensates at $G_c$, is naturally related to the derivative of
the beta function at $G_c$ in the $2+\epsilon$ expansion. 
The value $\nu \simeq 1/3$ in four dimensions, found by numerical evaluation of the lattice path
integral, is close but somewhat smaller than the lowest order $\epsilon$ expansion result
$\nu=1/(d-2)$.
An analysis of the strongly coupled phase of the lattice theory further gives $\nu=0$ at
$d=\infty$ \cite{larged}.

(v) The genuinely non-perturbative scale $\xi$, specific to the strongly coupled
phase of gravity for which $G>G_c$, can be shown to be related to the vacuum expectation
value of the curvature via $\langle {\cal R} \rangle \sim 1 / \xi^2$, 
and is therefore presumably macroscopic \cite{lines}.
It is naturally identified with the physical (scaled) cosmological constant $\lambda$;
$\xi$ therefore appears to play a role analogous to the non-perturbative 
scaling violation parameter $\Lambda_{\overline{MS}}$ of QCD.

(vi) The existence of a non-trivial ultraviolet fixed point (a phase transition
in statistical mechanics language) implies a scale dependence for
Newton's constant in the physical, strongly coupled phase $G>G_c$.
To leading order in the vicinity of the fixed point the scale dependence
is determined by the exponent $\nu$, and the overall size of the corrections is set
by the condensate scale $\xi$.
Thus in the strongly coupled phase, gravitational vacuum polarization effects
should cause the physical Newton's constant to grow slowly with distances.

\subsection{Non-Trivial Fixed Point and Scale Dependence of $G(\mu^2)$ }

This section will establish basic notation and provide some key
results and formulas, some of which will be discussed further in
the following sections.
For more details the reader is referred to the more
recent papers \cite{critical,larged,effective}.

For the running gravitational coupling one has in the vicinity of the
ultraviolet fixed point
\beq
G(k^2) \; = \; G_c \left [ \; 1 \, 
+ \, a_0 \left ( { m^2 \over k^2 } \right )^{1 \over 2 \nu} \, 
+ \, O ( \, ( m^2 / k^2 )^{1 \over \nu} ) \; \right ]
\label{eq:grun_latt}
\eeq
with $m=1/\xi$, $a_0 > 0$ and $\nu \simeq 1/3$ \cite{critical}.
We have argued that the quantity $G_c$ in the above expression should in
fact be identified with the laboratory scale value, 
$ \sqrt{G_c} \sim \sqrt{G_{phys}} \sim 1.6 \times 10^{-33} cm$,
the reason being that the scale $\xi$ can be very large.
Indeed in \cite{lines,ttmodes,effective} it was argued that $\xi$ should be
of the same order as the scaled cosmological constant $\lambda$.
Quantum corrections on the r.h.s are therefore quite small as
long as $ k^2 \gg m^2 $, which in real space corresponds to the ``short distance''
regime $ r \ll \xi$.

The above expression diverges as $k^2 \rightarrow 0$, and the infrared
divergence needs to be regulated.
A natural infrared regulator exists in the form of $m=1/\xi$, and
therefore a properly infrared regulated version of the above expression is
\beq
G(k^2) \; \simeq \; G_c \left [ \; 1 \, 
+ \, a_0 \left ( { m^2 \over k^2 \, + \, m^2 } \right )^{1 \over 2 \nu} \, 
+ \, \dots \; \right ]
\label{eq:grun_m}
\eeq
with $m=1/\xi$ the (tiny) infrared cutoff.
Then in the limit of large $k^2$ (small distances)
the correction to $G(k^2)$ reduces to the expression 
in Eq.~(\ref{eq:grun_latt}), namely 
\beq
G(k^2) \; \mathrel{\mathop\sim_{ k^2 / m^2 \, \rightarrow \, \infty }} \; 
G_c \, \left [ \, 1 \, + 
\, a_0 \left ( { m^2 \over k^2 } \right )^{1 \over 2 \nu} 
\left ( 1 \, - \, { 1 \over 2 \, \nu } \, { m^2 \over k^2 } \, 
+ \, \dots \right) \, + \, \dots \, \right ]
\eeq
whereas its limiting behavior for small $k^2$ (large distances) is now given by
\beq
G(k^2) \; \mathrel{\mathop\sim_{ k^2 / m^2 \, \rightarrow \, 0 }} \; 
G_\infty \left [ \, 1 \, - \, 
\left ( { a_0 \over 2 \, \nu \, (1 \, + \, a_0) } \, + \, \dots \right )
\, { k^2 \over m^2 } \, + \, O ( k^4 / m^4 ) \; \right ]
\eeq
implying that the gravitational coupling approaches the finite value 
$G_\infty = ( 1 + a_0 + \dots ) \, G_c $, independent of $m=1/ \xi$,
at very large distances $r \gg \xi$.
At the other end, for large $k^2$ (small distances) one has, from 
either Eqs.~(\ref{eq:grun_m}) or (\ref{eq:grun_latt}), 
\beq
G(k^2) \; \mathrel{\mathop\sim_{ k^2 / m^2 \, \rightarrow \, \infty }} \; G_c 
\eeq
meaning that the gravitational coupling approaches  
the ultraviolet (UV) fixed point value $G_c$ at ``short distances'' $r \ll \xi$.
Since the theory is formulated with an explicit ultraviolet cutoff, the latter must
appear somewhere, and indeed $G_c = \Lambda^{-2} \, \tilde{G_c}$, with 
the UV cutoff of the order of the Planck length 
$\Lambda^{-1} \sim 1.6 \times 10^{-33} cm$,
and $\tilde{G_c}$ a dimensionless number of order one.
Note though that in Eqs.~(\ref{eq:grun_latt}) or (\ref{eq:grun_m}) the cutoff
does not appear explicitly, it is ``absorbed'' into the definition of $G_c$.

The non-relativistic, static Newtonian potential is defined as
\beq
\phi (r) \; = \; ( - M ) \int { d^3 {\bf k} \over (2 \pi)^3 } \, 
e^{i \, {\bf k} \cdot {\bf x} } \, G( {\bf k^2} ) \, { 4 \pi \over {\bf k^2} } 
\label{eq:pot_def}
\eeq
and therefore proportional to the $3-d$ Fourier transform of 
\beq
{ 4 \pi \over {\bf k^2} } \; \rightarrow \;
{ 4 \pi \over {\bf k^2} } \left [ \; 1 \, 
+ \, a_0 \left ( { m^2 \over {\bf k^2} } \right )^{1 \over 2 \nu} \, 
+ \, \dots  \; \right ]
\label{eq:pot}
\eeq
But, as we mentioned before, proper care has to be exercised in providing
a properly infrared regulated version of the above expression,
which, from Eq.~(\ref{eq:grun_m}), reads 
\beq
{ 4 \pi \over ( {\bf k^2} \, + \, \mu^2 ) } \; \rightarrow \; \;
{ 4 \pi \over ( {\bf k^2} \, + \, \mu^2 ) } \left [ \; 1 \, 
+ \, a_0 \left ( { m^2 \over {\bf k^2} \, + \, m^2 } \right )^{1 \over 2 \nu} \, 
+ \, \dots  \; \right ]
\label{eq:pot_m}
\eeq
where the limit $\mu \rightarrow 0$ should be taken at the end of the calculation.
We wish to emphasize here that the regulators $\mu \rightarrow 0$ and $m$ are quite
distinct.
The distinction  originates
in the condition that $m$ arises due to strong infrared effects and 
renormalization group properties in the quantum
regime, while $\mu$ has nothing to do with quantum effects: it is required
to make the Fourier transform of the classical, Newtonian $4 \pi / {\bf k^2}$
well defined.
This is an important issue to keep in mind, and to which we will return later.

\subsection{Renormalization group properties of $G(\mu^2)$ }

This section will discuss the relationship between the running of the coupling $G$,
the renormalization group beta function $\beta(G)$, the lattice coupling $G(\Lambda)$
(the bare coupling, or equivalently, the running coupling at the scale
of the cutoff $\Lambda$), and the parameter $m=1/\xi$.
Differentiation of Eq.~(\ref{eq:grun_latt}) with respect to $k \rightarrow \Lambda$
gives
\beq
\Lambda \, { \partial \, G ( \Lambda ) \over \partial \, \Lambda } \; \equiv \; 
\beta ( G ) \; = \;
- \, { 1 \over \nu } \, ( G \, - \, G_c ) \, + \, \dots 
\label{eq:beta_latt}
\eeq
Here and in the following, unless stated otherwise, $G$ will refer to the dimensionless
gravitational coupling, i.e. $G_{phys} = \Lambda^{2-d} \, G (\Lambda) $.
In four dimensions, on laboratory scales, $\sqrt{G_{phys}} \simeq 10^{-33} cm$.
Therefore the exponent $ { 1 \over \nu } \equiv - \, \beta' (G_c) $ is related
to the derivative of the beta function for $G$ evaluated at the fixed point.
Here the dots account for higher order terms not included
in either Eq.~(\ref{eq:grun_latt}) or Eq.~(\ref{eq:grun_m}).

The above scaling form for $G(k^2)$ and the non-perturbative
exponent $\nu$ are determined as follows \cite{lines,critical}. 
Scaling around the fixed point originates in the divergence of the correlation
length $\xi=1/m$, related to the bare (lattice) couplings by
\beq
m \; \mathrel{\mathop\sim_{ G(\Lambda) \, \rightarrow \, G_c}} \; \Lambda \;
\left [ { G(\Lambda) \, - \, G_c \over a_0 \, G_c } \right ]^{\nu}
\label{eq:m_latt}
\eeq
where $\Lambda$ is the ultraviolet cutoff (the inverse lattice spacing).
The continuum limit is approached in the standard way by having 
$G \rightarrow G_c$ and $\Lambda$ very large, with $m$ kept fixed.
This last equation is recognized as being just Eq.~(\ref{eq:grun_latt})
here with the scale $k^2 \rightarrow \Lambda^2$, and
solved for the renormalization group invariant $m$ (the inverse of the physical
correlation length $\xi$) in terms of the bare (lattice) coupling $G(\Lambda)$,
at the ultraviolet cutoff scale $\Lambda$.
Scaling arguments in the vicinity of the non-trivial ultraviolet 
fixed point then allow one to determine the scaling
behavior of correlation functions from the critical exponents characterizing
the singular behavior of local averages.
Since the physical quantity $m=1/\xi$ is kept fixed and
is not supposed to depend on the ultraviolet cutoff $\Lambda$, which is
sent to infinity, one requires that $G(\Lambda)$ change in accordance to 
\beq
\Lambda \, {d \over d \, \Lambda} \, m ( \Lambda, G (\Lambda) ) \; = \; 0 
\eeq
It is useful to introduce the dimensionless function $F(G)$ via
\beq
m \equiv \xi^{-1} = \Lambda \, F(G)
\label{eq:functionF}
\eeq
By differentiation of the renormalization-group invariant quantity $m$,
one then obtains an expression for the Callan-Symanzik beta function 
$\beta (G)$ in terms of $F$. 
From its definition 
\beq
\Lambda \, { \partial \over \partial \, \Lambda } \, G ( \Lambda ) \; = \; 
\beta ( G ( \Lambda ) ) \;\; ,
\label{eq:betaf}
\eeq
one has
\beq
\beta (G) \; = \; - \, { F(G) \over 
\partial F(G) / \partial G } \;\; .
\eeq
One concludes that the knowledge of the dependence of $m$ on $G$, encoded in
the function $F(G)$, implies a specific form for the $\beta$ function.
In terms of the function $\beta (G) $ the result of
Eq.~(\ref{eq:m_latt}) is then equivalent to
\beq
\beta (G) \, \mathrel{\mathop\sim_{G \, \rightarrow \, G_c }} \, 
- \, {1 \over \nu} \, (G - G_c) \, + \, O( (G \, - \, G_c)^2 )
\eeq
with $ \beta ' (G_c) \, = \, - 1/ \nu  $.
In general one therefore expects the scaling behavior in the vicinity
of the fixed point 
\beq
m \, \sim \, \Lambda \, \exp \left ( { - \int^G \, {d G' \over \beta (G') } }
\right )
\, \mathrel{\mathop\sim_{G \, \rightarrow \, G_c }} \,
\Lambda \, | \, G - G_c |^{ - 1 / \beta ' (G_c) } \;\;\; .
\label{eq:m_cont}
\eeq
where here $\sim$ indicates up to a constant of proportionality.
The main conclusion is that the function $F(G)$ determines
the beta function $\beta (G)$, which in turn determines
the scale evolution of the coupling (obtained from Eq.~(\ref{eq:betaf}),
for any $\mu$,
\beq
\mu \, { \partial \over \partial \, \mu } \, G ( \mu ) \; = \; 
\beta ( G ( \mu ) ) \;\; ,
\label{eq:beta_mu}
\eeq
The latter can in principle be integrated in the vicinity of the fixed point, and
leads to a definite relationship between the relevant coupling 
$G$, the renormalization-group invariant (cutoff independent)
quantity $m=1/\xi$, and the arbitrary sliding scale $\mu^2 = k^2$,
as outlined in the preceding section.

Let us add a comment on the so-called corrections to scaling in the vicinity
of the fixed point at $G_c$.
We note that whereas Eq.~(\ref{eq:grun_m}) follows from the 
result Eq.~(\ref{eq:m_latt}) and Eq.~(\ref{eq:beta_latt}), 
the infrared regulated running coupling of $G$ in Eq.~(\ref{eq:grun_m})
is equivalent to assuming the following correction to scaling to Eq.~(\ref{eq:m_latt}),
\beq
m \; \mathrel{\mathop\sim_{ G(\Lambda) \, \rightarrow \, G_c } } \; 
{ \Lambda \; \left ( { G(\Lambda) \, - \, G_c \over a_0 \, G_c } \right )^{ \nu }
\over 
\sqrt{ 1 \, - \, \left ( { G(\Lambda) \, - \, G_c \over a_0 \, G_c } \right ) ^{2 \nu } } }
\; \simeq \;
\Lambda \; \left ( { G(\Lambda) \, - \, G_c \over a_0 \, G_c } \right ) ^{ \nu }
\left [ 1 \, + \, {1 \over 2 } 
\left ( { G(\Lambda) \, - \, G_c \over a_0 \, G_c } \right ) ^{2 \nu } \, 
+ \, \dots \right ]  
\label{eq:m_latt_corr}
\eeq
which gives by differentiation the previously quoted result, 
Eq.~(\ref{eq:grun_latt}), plus a small corrections close to $G_c$
\beq
\beta ( G ) \; = \;
- \, { 1 \over \nu } \, ( G \, - \, G_c ) \, { G \over G_c } \,
\left [ \, 1 \, - \,  \left ( { G \, - \, G_c \over a_0 \, G } \right )^{2 \nu}
\; \right ] \, + \, \dots
\label{eq:beta_latt1}
\eeq
where the dots account for higher order terms not included
in Eq.~(\ref{eq:grun_m}), and, implicitly, in Eq.~(\ref{eq:grun_latt}) as well.

\subsection{Lattice gravity determination of the universal exponent $\nu$ }

This section summarizes the connection between the lattice regularized quantum
gravity path integral $Z$, the singular part of the corresponding
free energy $F$ (which determines the scaling behavior in the vicinity of
the fixed point), and the universal critical exponent $\nu$
determining the scale dependence of the gravitational coupling in Eq.~(\ref{eq:grun_m}).
For more details the reader is referred to \cite{critical}, and references therein.

An important alternative to analytic analyses in the continuum is an
attempt to solve quantum gravity directly via numerical simulations.
The underlying idea is to perform the gravitational functional integral
by discretizing the action on a space-time lattice, and then evaluate the partition
function $Z$ by summing over a suitable finite set of representative
field configurations.
In principle such a method, given enough configurations and a fine enough
lattice, can provide an arbitrarily accurate solution to the
original quantum gravity theory.

In practice there are several important factors to consider, which effectively
limit the accuracy that can be achieved today in a practical calculation. 
Perhaps the most important one is the enormous amounts of computer time
that such calculations can use up.
This is particularly true when correlations of operators at fixed geodesic
distance are evaluated.
Another practical limitation is that one is mostly interested in the behavior
of the theory in the vicinity of the critical point at $G_c$, where the
correlation length $\xi$ can be quite large and significant correlations
develop both between different lattice regions, as well as among
representative field configurations, an effect known as critical slowing down.
Finally, there are processes which are not well suited to a lattice
study, such as problems with several different length (or energy) scales.
In spite of these limitations, the progress in lattice field theory has
been phenomenal in the last few years, driven in part by
enormous advances in computer technology, and in part by the
development of new techniques relevant to the problems of lattice
field theories. 

In practice the exponent $\nu$ in Eqs.~(\ref{eq:m_latt}) or (\ref{eq:m_latt_corr})
(and therefore also in Eqs.~(\ref{eq:grun_latt}) and (\ref{eq:grun_m}), which
follow from these)
is determined from the singularities that arise in the free energy
$F=-{1 \over V}\ln Z$, with the euclidean path integral for pure quantum gravity
$Z$ defined as
\beq
Z \; = \; \int \left [ d \, g_{\mu\nu} \right ] \; 
e^{ - \lambda_0 \, \int d^d x \, \sqrt g \, + \, 
{ 1 \over 16 \pi G } \int d^d x \sqrt g \, R  } \;\; .
\label{eq:zc}
\eeq
in the presence of a divergent correlation length $\xi \rightarrow \infty$
in the vicinity of the fixed point at $G_c$.
This is the scaling hypothesis, the basis of many important results of statistical
field theory \cite{zinn}.

On purely dimensional grounds, for the singular part of the free
energy one has $F_{sing} (G) \sim \xi^{-d}$.
Standard arguments then give, assuming a divergence $\xi \sim (G - G_c)^{-\nu}$
close to $G_c$, for the first derivative of $F$ (here
proportional to the average curvature ${\cal R} (G)$)
\beq
- \, \frac{1}{V} \frac{\partial}{\partial G} \ln Z  \; \sim \; 
{ < \int d x \, \sqrt{ g } \, R(x) > \over < \int d x \, \sqrt{ g } > } 
\; \mathrel{\mathop\sim_{ G \rightarrow G_c}}
A_{\cal R} \, ( G - G_c )^{d \nu -1 } \;\;\;\; .
\label{eq:rsing}
\eeq
An additive constant could appear as well, but the evidence up to now
points to this constant being zero \cite{critical}).
Similarly, for the second derivative of $F$, proportional to the fluctuation in
the scalar curvature $ \chi_{\cal R} (G)$, one has
\beq
- \, \frac{1}{V} \frac{\partial^2}{\partial G^2} \ln Z \; \sim \; 
{ < ( \int d x \sqrt{g} \, R )^2 > - < \int d x \sqrt{g} \, R >^2
\over < \int d x \, \sqrt{g} > } 
\; \mathrel{\mathop\sim_{ G \rightarrow G_c}} \;
A_{\chi_{\cal R}} \; ( G - G_c ) ^{ -( 2 - d \nu ) } \;\;\;\; .
\label{eq:chising}
\eeq
The above curvature fluctuation is related to the connected
scalar curvature correlation at zero momentum,
\beq
\chi_{\cal R} (k) 
\sim { \int d x \int d y \, < \sqrt{g} R (x) \, \sqrt{g} R (y) >_c
\over < \int d x \, \sqrt{g} > }
\eeq
and a divergence in the scalar curvature fluctuation is indicative of long range
correlations, corresponding to scale invariance and the presence of 
massless modes.

From the computation of such averages one can determine by standard methods
the numerical values for $\nu$, $G_c$ and $a_0$ to reasonably good 
accuracy \cite{critical}.
It is often advantageous to express results in the cutoff (lattice)
theory in terms of physical (i.e. cutoff independent) quantities.
By the latter we mean quantities for which the cutoff dependence has been
re-absorbed, or restored, in the relevant definition.
Thus, for example, Eqs.~(\ref{eq:grun_latt}) and (\ref{eq:grun_m}) will
include an overall factor of $\Lambda^{-2}$ if they refer to the
dimensionful, physical Newton's constant;
the cutoff is still present, but is ``hidden'' in the definition of
physical quantities, and cannot be set equal to infinity as the dimensionless
fixed point value $G_c$ is a finite number.

As an example, the result equivalent to Eq.~(\ref{eq:rsing}), 
relating the vacuum expectation value of the local scalar curvature 
(computed therefore for infinitesimal loops)
to the physical correlation length $\xi$ , is
\beq
{ < \int d x \, \sqrt{ g } \, R(x) > \over < \int d x \, \sqrt{ g } > } 
\; \mathrel{\mathop\sim_{ G \rightarrow G_c}} \; {\rm const.} \,
\left ( l_P^2 \right )^{1 - { 1 \over 2 \nu} } \,
\left ( {1 \over \xi^2 } \right )^{ {d \over 2} - {1 \over 2 \nu} }
\;\; ,
\label{eq:curvature}
\eeq
and which is simply obtained from Eqs.~(\ref{eq:m_latt}) and (\ref{eq:rsing}).
Matching of dimensionalities in this last equation has been restored by
supplying appropriate powers of the Planck length $l_P=\sqrt{G_{phys}}$.
For $\nu=1/3$ the result of Eq.~(\ref{eq:curvature})
becomes particularly simple \cite{critical,ttmodes}
\beq
{ < \int d x \, \sqrt{ g } \, R(x) > \over < \int d x \, \sqrt{ g } > } 
\; \mathrel{\mathop\sim_{ G \rightarrow G_c}} \; 
{\rm const.} \; \, {1 \over l_P \, \xi }  
\label{eq:curvature1}
\eeq
The naive estimate, based on a simple dimensional argument, would have
suggested the (incorrect) result $\sim 1 / l_P^2 $.
This shows that $\nu$ can also play the role of an anomalous dimension,
giving the magnitude of the deviation from naive dimensional arguments.
From the divergences of the free energy $F$ one can determine the universal exponent
$\nu$ appearing for example in Eqs.~(\ref{eq:grun_latt}) and (\ref{eq:grun_m}),
but not the amplitude $a_0$. 
The latter requires a direct determination of $m=1/\xi$ in terms of
bare lattice quantities (as in Eq. (\ref{eq:m_latt})), which we discuss next.

There are several correlation functions one can compute to extract $\nu$
and $a_0$ directly, either through the decay
of euclidean invariant correlations at fixed geodesic distance \cite{corr},
or, equivalently, from the correlations of Wilson lines associated with the
propagation of heavy spinless particles \cite{lines}.
In either case one expects the scaling result of Eq.~(\ref{eq:m_latt}) close to the
fixed point, namely
\beq
m \; \simeq \; A_m \, \Lambda \, | \, k \, - \, k_c \, |^{ \nu } \; ,
\;\;\;\; \Lambda \rightarrow \infty \; , 
\;\;\;\; k \rightarrow k_c \; ,
\;\;\;\; m \; {\rm fixed} \; ,
\label{eq:m_latt1}
\eeq
with the bare coupling $k (\Lambda) \equiv 1/( 8 \pi G(\Lambda) )$, and $A_m$
a calculable numerical constant.
Detailed knowledge of $m(k)$ allows one to independently estimate the
exponent $\nu$, but the method is generally extremely time
consuming (due to the appearance of geodesic distances in the correlation
functions), and therefore so far not very accurate.

But, more importantly, from the knowledge of the dimensionless constant
$A_m$ one can estimate from first principles the value of $a_0$ in 
Eqs.~(\ref{eq:grun_latt}) and (\ref{eq:grun_m}).
The first lattice results gave $A_m \simeq 0.56 $ \cite{corr} and
$A_m \simeq 0.87 $ \cite{lines}, with some significant uncertainty in
both cases (perhaps by as much as an order of magnitude, due to
the difficulties inherent in computing correlations at fixed geodesic
distance), which then, 
combined with the more recent estimate $k_c \simeq 0.0636 $ and 
$ \nu \simeq 0.335$ in four dimensions \cite{critical},
gives $a_0 = 1 /( k_c \, A_m^{1/\nu}) \simeq 42$.
The rather surprisingly large value for $a_0$ appears here perhaps as a consequence
of the relatively small value of the lattice $k_c$ in four dimensions.
A new determination of $a_0$ with significantly reduced errors would 
clearly be desirable.

The direct numerical determinations of $k_c=1/8 \pi G_c$ in $d=3$ and $d=4$
space-time dimensions are in fact quite close to the analytical prediction
of the lattice $1/d$ expansion \cite{larged},
\beq
k_c \; = \; { \lambda_0^{d - 2 \over d} \over \, d^3 } \, \left [ 
{2 \over d } \, { d! \, 2^{d/2} \over \sqrt{d+1} } \right ]^{2/d}
\label{eq:kcd}
\eeq
The latter gives for a bare $\lambda_0=1$ the estimate 
$k_c = \sqrt{3} / (16 \cdot 5^{1/4} ) = 0.0724$ in $d=4$,
to be compared to the direct determination of $k_c = 0.0636(11)$ of \cite{critical},
and $k_c= 2^{5/3} / 27 = 0.118$ in $d=3$, to be compared with 
the direct determination $k_c=0.112(5)$ in \cite{hw3d}
(these estimates will be compared later in Fig. 1).

\subsection{How many independent bare couplings for pure gravity?}

In this section it will be argued that the scaling behavior of pure quantum gravity
is determined by {\it one} dimensionless combination of $\lambda_0$ and $G$ only.
We will then argue that the only sensible scenario, from a renormalization
group point of view, is one in which the scaled
cosmological constant $\lambda$ is kept fixed, and only $G$ is allowed to run, as
in Eq.~(\ref{eq:grun_m}).
 
At first it might appear that in pure gravity one has two independent couplings
($\lambda_0$ and $G$), but in reality a simple scaling argument shows that
there can only be one, which can be taken to be a suitable dimensionless ratio \cite{hw84}.
Consider the (euclidean) Einstein-Hilbert action with a cosmological term
in $d$ dimensions
\beq
I_E [g] \; = \; \lambda_0 \, \Lambda^d \int dx \, \sqrt{g} \, - \, 
{ 1 \over 16 \pi \, G_0 } \, \Lambda^{d-2} \int dx \, \sqrt{g} \, R
\label{eq:ac_cont}
\eeq
Here $\lambda_0$ is the bare cosmological constant and $G_0$ the bare Newton's
constant, both measured in units of the cutoff
(we follow customary notation used in cutoff field theories, and
denote by $\Lambda$ the ultraviolet cutoff, not to be confused with 
the scaled cosmological constant)
\footnote{
We also deviate in this paper from the convention used in our previous work.
Due to ubiquitous ultraviolet cutoff $\Lambda$, we reserve here the symbol
$\lambda_0$ for the cosmological constant, and $\lambda$ for the {\it scaled}
cosmological constant $\lambda \equiv 8 \pi G \cdot \lambda_0$
}.
Convergence of the lattice regulated euclidean path integral requires
$\lambda_0 >0$ \cite{hw84}.
The natural expectation is for the bare microscopic, dimensionless
couplings to have magnitudes of order one in units of the cutoff, 
$\lambda_0 \sim G_0 \sim O(1) $.

Next one rescales the metric so as the obtain a unit coefficient
for the cosmological constant term,
\beq
g_{\mu\nu}' \; = \; \lambda_0^{2/d} \, g_{\mu\nu}
\;\;\;\;\;\;\;
{g'}^{\mu\nu} \; = \; \lambda_0^{-2/d} \, g^{\mu\nu}
\eeq
to obtain
\beq
I_E [g] \; = \; \Lambda^d \int dx \, \sqrt{g'} \, - \, 
{ 1 \over 16 \pi \, G_0 \, \lambda_0^{d-2 \over d} } \, \Lambda^{d-2} \int dx \, \sqrt{g'} \, R'
\label{eq:ac_cont1}
\eeq
The (euclidean) Feynman path integral, defined as
\beq
Z \; = \; \int [ d \, g_{\mu\nu} ] \; e^{- I_E [g] }
\label{eq:z-cont}
\eeq
includes as well a functional integration
over all metrics, with functional measure given for example by \cite{dewitt,misner}
\beq
\int [ d \, g_{\mu\nu} ] \; = \; 
 \int \prod_x \; \left [ g(x) \right ]^{ (d-4)(d+1) \over 8 } \;
\prod_{\mu \ge \nu} \, d g_{\mu \nu} (x)
\label{eq:dewitt}
\eeq
Therefore under a rescaling of the metric 
the functional measure only picks up a multiplicative constant.
It does not drop out when computing vacuum expectation values,
such as the one in Eq.~(\ref{eq:curvature}), but cannot give rise
to singularities at finite $G$ such as the ones in Eqs.~(\ref{eq:rsing}) and
(\ref{eq:chising}).

Equivalently, one can view a rescaling of the metric as simply a 
redefinition of the ultraviolet cutoff $\Lambda$,
$\Lambda \rightarrow \lambda_0^{1/d} \Lambda$.
As a consequence, the non-trivial part of the gravitational functional
integral over metrics only depends on $\lambda_0$ and
$G_0$ through the {\it dimensionless} combination \cite{hw84}
\beq
\tilde G \; \equiv \; G_0 \, \lambda_0^{(d-2)/d}
\eeq
The existence of an ultraviolet fixed point is then entirely controlled
by this dimensionless parameter only, both on the lattice ~\cite{hw84,critical,ttmodes}
and in the continuum ~\cite{epsilon} : the non-trivial part of the
functional integral only depends on this specific combination.
One has the Ward identity for the singular part of the generating function,
\beq
\left \{ \, G_0 \, { \partial \, \over \partial \, G_0 } \, - \, { d \over d-2 } \,
\lambda_0 \, { \partial \, \over \partial \, \lambda_0 } \, \right \}
\, F_{sing} \left ( G, \, \lambda_0, \, \dots  \right ) \; = \; 0
\eeq
Thus the individual scaling dimensions of the cosmological
constant and of the gravitational coupling constant do not have
separate physical meaning; only the relative scaling dimension,
as expressed through their dimensionless ratio, is physical.

Physically, the parameter $\lambda_0$ controls the overall scale of the
problem (the volume of space-time), while the $G_0$ term provides the 
necessary derivative or coupling term.
Since the total volume of space-time 
is normally not considered a physical observable, quantum
averages are computed by dividing out by the total
space-time volume.
For example, for the quantum expectation value of the Ricci scalar one 
has the expression of Eq.~(\ref{eq:curvature}).

Without any loss of generality one can therefore fix the overall
scale in terms of the ultraviolet cutoff, and
set the bare cosmological constant $\lambda_0$ equal to one in units
of the ultraviolet cutoff. 
\footnote{
These considerations are not dissimilar from the case of a 
self-interacting scalar field, where one might want to introduce three
couplings for the kinetic term, the mass term and the quartic
coupling term, respectively. A simple rescaling of the field would then
reveal that only two coupling ratios are in fact physically relevant.}

The addition of matter field does not change the conclusions
of the previous discussion, it is just that additional rescalings are
needed.
Thus for a scalar field with action
\beq
I_S [\phi] \; = \; \half \, \int dx \sqrt{g} \, \left ( 
g^{\mu\nu} \, \partial_\mu \, \phi \, \partial_\nu \, \phi 
\, + \, m_0^2 \, \phi^2
\, + \, R \, \phi^2  \right )
\eeq
and functional measure (for a single field)
\beq
\int [ d \, \phi] \; = \; \int \prod_x \,
\left [\, g(x) \, \right ]^{1/4} \, d \phi (x)
\eeq
the metric rescaling is to be followed by a field rescaling 
\beq
\phi ' (x) \; = \; \phi (x) \, \lambda_0^{1/d-1/2}
\eeq
with the only surviving change being a rescaling of the bare mass
$m_0 \rightarrow m_0 / \lambda_0^{1/d} $. 
Again  the scalar functional measure acquires an irrelevant multiplicative factor
which does not affect quantum averages.

The same results are obtained if one considers a 
lattice regularized version of the original (euclidean)
path integral of Eq.~(\ref{eq:ac_cont}), which reads \cite{hw84}
\beq 
Z_L \; = \;  \int [ d \, l^2 ] \, \; e^{- I_L [l^2] }
\label{eq:zlatt} 
\eeq
with lattice Regge action \cite{regge}
\beq 
I_L \; = \;  \lambda_0 \sum_h V_h (l^2) -  2 \, \k_0 \sum_h \delta_h (l^2 ) \, A_h (l^2) 
\label{eq:ilatt} 
\eeq
and regularized lattice functional measure \cite{cms1,hartle,hw84}
\beq
\int [ d \, l^2 ] \; \equiv \;
\int_0^\infty \; \prod_{ ij } \, d l_{ij}^2 \; 
\prod_s \; \left [ V_d (s) \right ]^{\sigma} \; \Theta ( l_{ij}^2 )
\label{eq:mlatt} 
\eeq
with $k_0 = 1 / ( 16 \pi G_0 )$, and $\Theta$ a function incorporating
the effects of the trian
gle inequalities.
As is customary in lattice field theory, the lattice ultraviolet cutoff is
set equal to one (i.e. all lengths and masses are measured in units of the cutoff).
Convergence of the euclidean lattice functional integral
requires a positive bare cosmological constant $\lambda_0 > 0$ \cite{hw84,monte}.
On can show again by a trivial rescaling of the edges that, as in the continuum,
non-trivial part of the lattice regularized path
integral only depends, in the absence of matter, on the single dimensionless 
parameter $ \tilde G \; \equiv \; G_0 \, \lambda_0^{(d-2)/d} $.
Without loss of generality therefore the bare coupling $\lambda_0$ can be
set equal to one
\footnote{
In lattice quantum gravity, the average edge length $l_0 = \langle l^2 \rangle^{1/2}$
is, for non-singular measures, largely a function of $\lambda_0$. 
In the large $d$ limit an explicit formula can be given for the measure
of Eq.~(\ref{eq:mlatt}) with $\sigma=0$,
\beq
l_0^2 = { 1 \over \lambda_0^{2/d} } \left [ 
{2 \over d } \, { d! \, 2^{d/2} \over \sqrt{d+1} } \right ]^{2/d}
\label{eq:l2d}
\eeq
which agrees well with numerical estimates for finite $d$ \cite{larged}.
}.

The question that remains open is then the following: which coupling should be
allowed to run within the renormalization group framework?
Since the path integral in four dimensions only depends on the dimensionless
ratio $ \tilde G^2 = G_0^2 \, \lambda_0 $ (which is expected to be scale dependent), one has
several choices; for example $G$ runs and the cosmological constant $\lambda_0$ is fixed.
Alternatively, $G$ runs and the {\it scaled} cosmological constant 
$ \lambda \equiv G \lambda_0$
is kept fixed; or $G$ is fixed and $\lambda$ runs etc. 

At first thought, it would seem that the coupling $\lambda_0$ should not be
allowed to run, as the overall space-time volume should perhaps be considered fixed,
not to be rescaled under a renormalization group transformation.
After all, in the spirit of Wilson \cite{wilson}, a renormalization group transformation
provides a description of the original physical system in terms
of a new coarse-grained Hamiltonian, whose new operators are
interpreted as describing averages of the original system on the
finer scale, but within the {\it same} physical volume.
The new effective Hamiltonian is then supposed to still
describe the original physical system, but more
economically, in terms of a reduced set of degrees of freedom.
 
These considerations are to some extent implicit in the correct
definition of gravitational averages, for example in Eq.~(\ref{eq:curvature}).
Physical, observable averages such as the one in Eq.~(\ref{eq:curvature})
in general have some rather non-trivial dependence on the bare coupling
$G_0$, more so in the presence of an ultraviolet fixed point.
Renormalization in the vicinity of the ultraviolet
fixed point invariably leads to the introduction
of a new dynamically generated, non-perturbative scale for $G > G_c$. 

It appears though that the correct answer is that the combination
$\lambda \equiv 8 \pi G \cdot \lambda_0$, corresponding to the
{\it scaled} cosmological constant,
should be kept {\it fixed}, while Newton's constant is allowed to run
in accordance to the scale dependence obtained from $\tilde G$. 
The reasons for this choice are three-fold. 
First, in the weak field expansion it is the combination $ \lambda \equiv G \lambda_0$
that appears as a mass-like term (and not $\lambda_0$ or $G$ separately).
A similar conclusion is reached if one just compares the appearance of the field
equations for gravity to say QED (massive via the Higgs mechanism), 
or a self-interacting scalar field,
\bea
R_{\mu\nu} \, - \, \half \, g_{\mu\nu} \, R \, + \, \lambda \, g_{\mu\nu}
\; & = & \; 8 \pi G  \, T_{\mu\nu}
\nonumber \\
\partial^{\mu} F_{\mu\nu} \, + \, \mu^2 \, A_\nu \, & = & \; 4 \pi e \, j_{\nu} 
\nonumber \\
\partial^{\mu} \partial_{\mu} \, \phi \, + \, m^2 \, \phi \; & = & \; {g \over 3!} \, \phi^3
\label{eq:masses}
\eea
Secondly, the scaled cosmological constant represents a 
measure of physical curvature,
as should be clear from how the scaled cosmological constant relates
for example to the expectation values of the scalar curvature at short distances
(i.e. for infinitesimally small loops, whose size is comparable to the cutoff
scale),
\beq
{ < \int d x \, {\textstyle {\sqrt{g(x)}} \displaystyle} \, R(x) >
\over < \int d x \, {\textstyle {\sqrt{g(x)}} \displaystyle} > } 
\; \simeq \; R_{\, class} \; = \; 4 \lambda 
\eeq
in the case of pure gravity.
But perhaps the most convincing argument for the scaled cosmological constant 
$ \lambda \equiv 8 \pi \, G \lambda_0$ to be kept fixed
is given in the following section.

\subsection{The value of $\xi$ dilemma - small or large ?}

In this section we will argue that the scale $\xi$, which determines
the running of $G$ according to Eq.~(\ref{eq:grun_m}),
should be identified with the observed scaled
cosmological constant $\lambda$.

The lattice quantum gravity result of Eq.~(\ref{eq:curvature}) 
(and Eq.~(\ref{eq:curvature1}) for $\nu=1/3$ ) suggests a deep relationship
between the correlation length $\xi=1/m$ determining the size
of scale dependent corrections of Eqs.~(\ref{eq:grun_latt})
and the curvature.
Small averaged curvatures correspond to very large length scales $\xi$.
In gravity, curvature is detected by parallel transporting vectors around
closed loops.
This requires the calculation of a path dependent product of
(Lorentz) rotations, ${\bf R}^\alpha_{\;\; \beta}$, elements of $SO(4)$
in the euclidean case.
On the lattice, the above rotation is directly related to the 
path-ordered (${\cal P}$) exponential of the integral of the lattice
affine connection $ \Gamma^{\lambda}_{\mu \nu}$ via
\beq
{\bf R}^\alpha_{\;\; \beta} \; = \; \Bigl [ \; {\cal P} \, e^{\int_
{{\bf path \atop between \; simplices}}
\Gamma^\lambda d x_\lambda} \; \Bigr ]^\alpha_{\;\; \beta}  \;\; .
\label{eq:rot1}
\eeq
as discussed clearly for example in \cite{caselle,froh}, and more recently in \cite{bianchi}.
Now, in the strongly coupled gravity regime ($G>G_c$) large fluctuations in the
gravitational field at short distances will be reflected in large fluctuations
of the ${\bf R}$ matrices, which deep in the strong coupling regime
should be reasonably well described by a uniform (Haar) measure \cite{larged,fadpop}.

Borrowing still from the analogy with Yang-Mills theories,
one might therefore worry that the effects of large strong coupling
fluctuations in the ${\bf R}$ matrices might lead to a phenomenon similar
to confinement in non-Abelian lattice gauge theories \cite{wilson-lgt} .
That this is {\it not} the case can be seen from the fact
that the gravitational analog of the Wilson loop $W(\Gamma)$, defined here as a path-ordered
exponential of the affine connection $\Gamma_{\mu\nu}^{\lambda}$ around a closed
planar loop,
\beq
W(\Gamma) \, \sim \, \Tr \, {\cal P} \, \exp \, \left [ \;
\int_C \, \Gamma^{\lambda}_{\;\; \cdot \, \cdot} \, dx_\lambda  \; \right ]
\label{eq:wloop}
\eeq
does {\it not} give the static gravitational potential.
The static gravitational potential is determined instead from the correlation of
(exponentials of) geodesic line segments, as in
\beq
\exp \left [ \, - M_0 \int d \tau \sqrt{ \textstyle g_{\mu\nu} (x)
{d x^{\mu} \over d \tau} {d x^{\nu} \over d \tau} \displaystyle } \;\,
\right ] 
\label{eq:lines}
\eeq
where $M_0$ is the mass of the heavy source,
as discussed already in some detail in \cite{moda,lines}.
Indeed a direct lattice calculation of the potential between heavy sources via the 
correlation of geodesic lines showed no sign of confinement \cite{lines}.
Borrowing from the well-established results in non-abelian lattice gauge theories
with compact groups (and to which no exceptions are known), 
it is easy to show that the expected decay of near-planar Wilson loops with area
$A$ is then given by  
\beq
W(\Gamma) \, \sim \, \exp \, \left [ \;
\int_{S(C)}\, R^{\, \cdot}_{\;\; \cdot \, \mu\nu} \, A^{\mu\nu}_{C} \; \right ]
\; \sim \; \exp ( - A / \xi^2 )
\label{eq:wloop_curv}
\eeq
\cite{peskin}, where $A$ is the minimal physical area spanned by
the near-planar loop.
The rapid decay of the Wilson loop as a function of the area is then seen 
simply as a general and direct consequence of the disorder in the fluctuations
of the ${\bf R}$ matrices at strong coupling.
One concludes therefore that the Wilson loop in gravity provides 
a measure of the magnitude of
the large-scale, averaged curvature, operationally determined by
the process of parallel-transporting test vectors around very large loops,
and which therefore, from the above expression, is computed to be of the 
order $R \sim 1 / \xi^2 $.  

One important assumption in the above result is the identification 
of the correlation length $\xi$ in Eq.~(\ref{eq:wloop_curv})
with the correlation length $\xi$ in Eqs.~(\ref{eq:m_latt}) and (\ref{eq:curvature}).
This is the scaling hypothesis, at the basis of most statistical field theory
\cite{pbook,zinn,blz} : one assumes that {\it all} critical behavior in the
vicinity of $G_c$ is determined by one correlation length, which diverges
(in lattice units) at the critical point at $G_c$ in accordance with
Eq.~(\ref{eq:m_latt}).
As can be seen from Eq.~(\ref{eq:m_latt}), the scale $\xi$ is genuinely
non-perturbative, as in non-abelian gauge theories.
To determine the actual physical value of $\xi$ some physical input is needed,
as the underlying theory
{\it cannot} fix it : the ratio of the physical Newton's constant
to $\xi^2$ can be as small as one desires, provided the bare coupling $G$
is very close to its fixed point value $G_c$.

In conclusion, the above arguments and in particular the result of 
Eq.~(\ref{eq:wloop_curv}), suggest once more the identification of $\xi$
with the large scale curvature, the most natural
candidate being the (scaled) cosmological constant,

\beq
\lambda_{phys} \, \simeq \, { 1 \over \xi^2 } \;\;\;\; {\rm or} 
\;\;\;\; \xi \; = \; {1 \over \sqrt{\lambda_{phys}} } 
\label{eq:xi_lambda}
\eeq 
This relationship, taken at face value, implies a very large, cosmological value
for $\xi \sim 10^{28} cm$, given the present bounds on $\lambda_{phys}$ or
$H_0$.
Other closely related possibilities may exist,
such as an identification of $ \xi $ with the Hubble 
constant (as measured today) determining the macroscopic expansion rate of the
universe via the correspondence
\beq
\xi \; \simeq \; 1 / H_0 \;\; ,
\label{eq:hub}
\eeq
Since this quantity is presumably time-dependent, a possible scenario would
be one in which $\xi^{-1} = H_\infty = \lim_{t \rightarrow \infty} H(t) =
\sqrt{\Omega_\lambda}\, H_0$
with $H_\infty^2 = { 8 \pi G \over 3 } \lambda_0 = { \lambda \over 3 } $,
where $\lambda_0$ is the observed cosmological constant, and for which
the horizon radius is $R_{\infty} = H_{\infty}^{-1}$.

Should Newton's constant run with energy, and if so, according to what law?
Newton's constant enters the field equations, after multiplication by $G$
itself, as the coefficient of the $T_{\mu\nu}$ term

\beq
R_{\mu\nu} \, - \, \half \, g_{\mu\nu} \, R \, + \, \lambda \, g_{\mu\nu}
\; = \; 8 \pi \, G  \, T_{\mu\nu}
\eeq
In line with the previous discussion, the running of the (scaled)
cosmological term is according to the rule 
$\lambda \rightarrow 1 / \xi^2$, i.e. no scale dependence.
Since, as emphasized in Sec. 2.4, the gravitational path integral only depends
in a non-trivial way on the dimensionless combination $G \, \sqrt{\lambda_0}$
(see for example Eqs.~(\ref{eq:ac_cont1}) and (\ref{eq:z-cont}) 
and related discussion),
Newton's constant itself $G$ can be decomposed uniquely into non-running and
running parts, in the following way
\beq
G \; \equiv \; { 1 \over G \, \lambda_0 } \cdot 
\left ( G \, \sqrt{\lambda_0} \, \right )^2 \;\; \rightarrow \;\; 
{ 1 \over \xi^2 } \cdot 
\left [ \, (G \, \sqrt{\lambda_0}) \, (\mu^2) \, \right ]^2
\label{eq:g_decompose}
\eeq
where the running of the second term can be directly deduced from
either Eqs.~(\ref{eq:chising}) or (\ref{eq:rsing}) (both only depend
on the combination $ G \, \sqrt{\lambda_0} $), or Eq.~(\ref{eq:m_latt}),
which is related to the previous two by the scaling assumption
for the free energy $F$.

In conclusion, the modified Einstein equations, incorporating the
proposed quantum running of $G$, read
\beq
R_{\mu\nu} \, - \, \half \, g_{\mu\nu} \, R \, + \, \lambda \, g_{\mu\nu}
\; = \; 8 \pi \, G(\Box)  \, T_{\mu\nu}
\label{eq:field0}
\eeq
with $\lambda \simeq { 1 \over \xi^2 } $,
and only $G(\mu^2)$ on the r.h.s. scale-dependent.
The precise meaning of $G(\Box)$ is given in section 4. .

\subsection{Non-perturbative gravitational vacuum condensate}

In this section we will point out the deep relationship (well understood
in strongly coupled non-abelian gauge theories) between the non-perturbative
scale $\xi$ appearing in Eqs.~(\ref{eq:grun_latt}) and (\ref{eq:grun_m}), 
and the non-perturbative vacuum condensate of Eqs.~(\ref{eq:curvature})
and (\ref{eq:xi_lambda}), which is a measure of curvature.
The principal, and in our opinion inescapable, conclusion of
the results of Eqs.~(\ref{eq:curvature}) and (\ref{eq:wloop_curv}) is that
the scale $\xi$ appearing in Eqs.~(\ref{eq:grun_latt}) and (\ref{eq:grun_m})
is related to curvature, and must be {\it macroscopic} for the theory to be consistent.
How can quantum effects propagate to such large distances and give such drastic
modifications to gravity? 
The answer to this paradoxical question presumably lies in the fact
that gravitation is carried by a massless particle whose interactions
cannot be screened, on any length scale.

It is worth pointing out here that the gravitational vacuum condensate,
which only exists in the strong coupling phase $G>G_c$, and which is
proportional to the curvature, is genuinely non-perturbative.
If one uses a shorthand notation ${\cal R}$ for it, then one can summarize
the result of Eq.~(\ref{eq:xi_lambda}) as
\beq
{\cal R } \; \simeq \; (10^{-30} eV)^2 \, \sim \, \xi^{-2} 
\eeq
where the condensate is, according to Eqs.~(\ref{eq:m_latt}) and 
(\ref{eq:m_latt1}) (relating $\xi$ to $\vert G - G_c \vert$), 
and Eqs.~(\ref{eq:wloop_curv}) and (\ref{eq:xi_lambda}) (relating 
${\cal R}$ to $\xi$), non-analytic at $G=G_c$,
\beq
{\cal R } \; \sim \; \left | \, G \, - \, G_c \, \right |^{\, 2 \, \nu}
\label{eq:r_nonan}
\eeq
The non-perturbative curvature scale $\xi$ then corresponds
to a non-vanishing graviton vacuum condensate of order $\xi^{-1} \sim 10^{-30} eV$,
extraordinarily tiny compared to the QCD color condensate 
($\Lambda_{\overline{MS}} \simeq 220 \, MeV$) and the electro-weak Higgs condensate
($v \simeq 250 \, GeV$).
But as previously emphasized, the quantum gravity theory cannot provide,
at least in its present framework, a value for the non-perturbative curvature scale $\xi$, 
which ultimately can only be fixed by phenomenological input, either by
Eq.~(\ref{eq:grun_latt}) or, equivalently, by Eq.~(\ref{eq:xi_lambda}).
The main message here is that the scale in those two equations is one and the same.

Pursuing the analogy with strongly coupled Yang-Mills theories and QCD, we note
that there the non-perturbative gluon vacuum condensate depends in a
nontrivial way on the corresponding confinement scale parameter \cite{shifman},
\beq
\alpha_S <F_{\mu\nu} \cdot F^{\mu\nu}> \; \simeq \; (250 MeV)^4 \, \sim \, \xi^{-4} 
\eeq
with $\xi_{QCD}^{-1} \sim \Lambda_{\overline{MS}}$.
The above condensate is not the only one that appears in QCD,
another important non-vanishing vacuum condensate being the fermionic one \cite{hp}
\beq
(\alpha_S)^{4/\beta_0} <{\bar \psi} \, \psi > \; \simeq \; - \, (230 MeV)^3 \, \sim \, \xi^{-3} 
\eeq

\subsection{Quantum gravity near two dimensions }

The result of Eqs.~(\ref{eq:grun_latt}) and Eq.~(\ref{eq:beta_latt}) are
almost identical in overall structure to what one finds in $2+\epsilon$ dimensions, if
one allows for a different value of exponent $\nu$ as one transitions from
two dimensions to the physical case of four dimensions.
In this section we will explore their relationship, and
the lessons one learns from similar field theory models, such as the
non-linear sigma model above two dimensions.
But the second major ingredient which non-perturbative lattice studies
provide \cite{hw84,gra92}, besides the existence of a phase transition between
two geometrically rather distinct phases, is that the weakly coupled 
small $G$ phase is pathological, in
the sense that the theory becomes unstable, with the four-dimensional
lattice collapsing into a tree-like two-dimensional structure
for $G<G_c$.
Indeed the lattice theory close to
the transition at $G_c$ is ``on the verge'' of becoming two-dimensional \cite{gra92},
but only to the extent that the effects of higher derivative terms and conformal
anomaly contributions can be ignored at short distances
\footnote{
One way of determining coarse aspects of the underlying geometry
is to compute the effective dimension in the scaling regime,
for example by considering how the number
of points within a thin shell of geodesic distance between
$\tau$ and $\tau+\Delta$ scales with the geodesic distance itself \cite{gra92}.
For distances a few multiples of the average lattice spacing one finds
\beq
N(\tau) \mathrel{\mathop\sim_{\tau \rightarrow \, \infty}} \tau^{\; d_v} ,
\label{fractal}
\eeq
with $d_v =3.1(1)$ for $G>G_c$ (the smooth phase) and $d_v \simeq 1.6(2)$
for $G<G_c$ (the rough phase).
One concludes that in the rough phase the lattice tends to collapse into
a degenerate tree-like configuration,
whereas in the smooth phase the effective dimension of space-time
is consistent with four.
Higher derivative terms tend to affect these results at very short distances,
where they tend to make the geometry smoother \cite{hw84,hdqg}. 
}.

To one loop the $d=2+\epsilon$ result for the gravitational beta function 
was computed some time ago, and reads
\beq
\beta (G) \;  = \; (d-2) \, G \, - \, \beta_0 \, G^2 \, - \,  \dots \;\; ,
\label{eq:beta}
\eeq
It exhibits the celebrated non-trivial ultraviolet fixed point of $2+\epsilon$
quantum gravity at $G_c=(d-2)/\beta_0$ (one finds $\beta_0 >0$ for pure gravity).
Furthermore, the physics of it bears some striking similarity to the non-linear
sigma model, to be discussed below. 
The latter is also perturbatively non-renormalizable above two dimensions, but can
be constructed by a suitable double expansion in the coupling $g$ 
and $\epsilon=d-2$ \cite{sigma}.

In gravity the corresponding running of $G$ is obtained by integrating 
Eq.~(\ref{eq:beta})
\beq
G(k^2) \; = \; { G_c \over 1 \, \pm \, (m^2 / k^2 )^{(d-2)/2} } 
\label{eq:grun_cont} 
\eeq
The choice of $+$ or $-$ sign is determined from whether one is
to the left (+), or to right (-) of the fixed point, in which case
the effective $G(k^2)$ decreases or, respectively, increases as one flows away
from the ultraviolet fixed point at $G_c$.
Physically they represent a screening and an anti-screening solution.
It is noteworthy that the invariant mass scale $m$ arises as an arbitrary
integration constant of the renormalization group equations. 
While in the continuum both phases, and therefore both signs, seem
acceptable (giving rise to both a ``Coulomb'' or ``spin wave'' phase, and
a strong coupling phase), the euclidean lattice rules out the small $G<G_c$ phase
as pathological, in the sense that the lattice collapses into a
two-dimensional branched polymer \cite{hw84,critical}
\footnote{
The collapse stops at $d=2$ because the gravitational action becomes a
topological invariant.
}.

Thus the smooth phase with $G > G_c$ emerges as the only physically
acceptable phase in $d=3$ and $d=4$ \cite{hw3d,critical}.
These arguments suggest therefore that in the renormalization
group solution Eq.~(\ref{eq:grun_cont}) only the - sign is meaningful and
physical, corresponding to an infrared growth of the
coupling for $G>G_c$, the gravitational anti-screening solution given by
\beq
G(k^2) \; \simeq \; G_c \, \left [ 
\, 1 \, + \, \left ( {m^2 \over k^2 } \right )^{(d-2)/2} \, + \, \dots \right ]
\label{eq:grun_cont1} 
\eeq
The above expression has in fact exactly the same structure as the lattice 
result of Eq.~(\ref{eq:grun_latt}).
Even physically the result makes sense, as one would expect
that gravity cannot be screened (as would happen for the ``+'' sign choice). 
Repeating some of the general arguments presented in the previous sections,
and replacing $k^2 \rightarrow \Lambda^2$ in the above equation, 
where $\Lambda$ is the ultraviolet cutoff, one can solve, for $G > G_c $, 
for the genuinely non-perturbative, dynamically generated
mass scale $m$ in terms of the coupling at the cutoff (lattice) scale,
as in Eq.~(\ref{eq:m_cont}).
It should be noted here that Eq.~(\ref{eq:m_cont}) is essentially
the same as Eq.~(\ref{eq:m_latt}), and 
that Eq.~(\ref{eq:grun_cont}) is essentially
the same as Eq.~(\ref{eq:grun_latt}).

The derivative of the beta function at the fixed point gives the exponent $\nu$.
From Eq.~(\ref{eq:beta}) one has therefore close to two dimensions \cite{epsilon}
\beq
\beta ' (G_c) \, = \, - 1/ \nu  \; = \; (2-d) 
\eeq
Recently the above results have been extended to two loops, giving close to two dimensions
\footnote{
For a while there was considerable uncertainty, due in part to the kinematic
singularities which arise in gravity close to two dimensions, about the value
of the graviton contribution to $\beta_0$, which was quoted originally in 
\cite{tsao} as $38/3$, in \cite{epsilon} as $2/3$, and more recently
in \cite{epsilon2,epsilon3} as $50/3$.
As discussed in \cite{epsilon1}, the original expectation was that the graviton
contribution should be $d(d-3)/2 =-1$ times the scalar contribution close to $d=2$.
Direct numerical estimates in $d=3$ give $\nu^{-1} \simeq 1.67$ \cite{hw3d} and are 
therefore in much better agreement with the larger, more recent value for $\beta_0$.}
\beq
\beta (G) \, = \, (d-2) \, G \, - \, { 2 \over 3 } \, (25 - n_s) \, G^2 \,
- \, { 20 \over 3 } \, (25 - n_s) \, G^3 \, + \dots \;\; ,
\label{eq:betaeps}
\eeq
for $n_s$ massless real scalar fields minimally coupled to gravity.
After solving the equation $\beta (G_c) \, = \, 0 $ to determine the location
of the ultraviolet fixed point, one finds
\bea
G_c & = & {3 \over 2 \, ( 25 - n_s ) } \, (d-2) 
- {45 \over 2 ( 25 - n_s )^2 } \, (d-2)^2 + \dots
\nonumber \\
\nu^{-1} & = & - \, \beta ' (G_c) \, = \,
(d-2) \, + \, {15 \over 25 - n_s } \, (d-2)^2 \, + \, \dots 
\label{eq:nueps}
\eea
which gives, for pure gravity without matter ($n_s=0$) in four dimensions, to lowest order
$\nu^{-1} = 2$, and $\nu^{-1} \approx 4.4 $ at the next order.
Unfortunately in general the convergence properties of the $2+\epsilon$
expansion for some better understood field theories, such as the
non-linear sigma model, are not too encouraging, at least when compared for example
to well-established results obtained by other means directly in $d=3$
\cite{hikami,kleinert}.
This somewhat undesirable state of affairs is usually ascribed to the
suspected existence of renormalon-type singularities
$ \sim e^{-c/G}$ close to two dimensions, which could possibly 
arise in gravity as well.
At the quantitative level, the results of the $2+\epsilon$ expansion for gravity
remain therefore somewhat limited, and obtaining the three- or four-loop
corrections could represents a daunting task.
But one should not overlook the fact that there are quantum field theory
models which have in some qualitative respects astonishingly similar behavior in 
$2+\epsilon$ dimensions, and are at the same time very well understood.
Some analogies might therefore remain helpful.

The non-linear $O(N)$ sigma model is one example, studied extensively in the context
of the $2+\epsilon$ expansion, and solved exactly in the large $N$ limit \cite{zinn}.
The model is not perturbatively renormalizable above two dimensions.
Yet in both approaches it exhibits a non-trivial ultraviolet fixed point 
at $g_c$ (a phase transition in statistical mechanics language), 
separating a weak coupling massless ordered phase from a massive,
strong coupling phase.
Therefore the correct continuum limit has to be taken in the vicinity of
the non-trivial ultraviolet fixed point.
Perhaps one of the most striking aspects of the non-linear sigma model
above two dimensions is that all particles are massless in perturbation theory,
yet they all become massive in the strong coupling phase $g>g_c$, with
masses proportional to a non-perturbative scale $m$ \cite{bardeen}.

The second example of perturbatively non-renormalizable theory is the
chirally-invariant self-coupled fermion above two dimensions \cite{zinn}.
It too exhibits a non-trivial ultraviolet fixed point above two
dimensions, and can be studied perturbatively via a double $2+\epsilon$ expansion.
And it too can be solved exactly in the large $N$ limit.
With either method, one can show that the model is characterized
by two phases, a weak coupling phase where the fermions stay massless and chiral
symmetry is unbroken, and a strong coupling phase in which
chiral symmetry is spontaneously broken, a fermion condensate arises,
and a mass scale is generated non-perturbatively.

\subsection{Other determinations of the exponent $\nu$ }

Let us conclude this discussion by mentioning the remaining methods which have
been used to estimate $\nu$.

Recently some approximate renormalization group results have been obtained
in the continuum based on an Einstein-Hilbert action truncation.
In the limit of vanishing bare cosmological constant the result
$ \nu^{-1} \, = \, 2 d (d-2) / (d+2) = 2.667$ was given in $d=4$ \cite{litim}.
In the cited work the sensitivity of the numerical answer for the exponent to the choice
of gauge fixing term and to the specific shape of the momentum cutoff
were systematically investigated as well.
More details about the procedure can be found in the quoted references.
The more recent results extend earlier calculations for
the exponent $\nu$ done by similar operator truncation methods, described in
detail in \cite{reuter} and references therein.
A quantitative comparison of these continuum results with the lattice
answer for $\nu^{-1}$, and with the $2+\epsilon$ estimates, can be found in Figs. 1 and 2.
The overall the agreement seems reasonable, given the
uncertainties inherent in the various estimates.
There is also some connection between the just mentioned lattice and continuum renormalization
group results and the ideas in \cite{poly}, but in this last reference fractional
exponents, such as the ones in Eq.~(\ref{eq:grun_m}) and (\ref{eq:grun_cont}),
were not considered.

Both the lattice and the continuum renormalization group analysis can
be extended to dimensions greater than four.
In \cite{ttmodes} it was suggested, based on a simple geometric
argument, that $\nu = 1 / (d-1)$ for large $d$, and in
the large-$d$ limit one can prove that $\nu=0$ \cite{larged}.
Moreover, for the lattice theory in finite dimensions one finds no phase
transition in $d=2$ \cite{hw2d}, $\nu \approx 0.60$ in $d=3$ \cite{hw3d}
and $\nu \approx 0.33$ in $d=4$ \cite{critical,ttmodes},
which leads to the more or less constant sequence 
$(d-2) \nu$ = $1$, $0.60$ and $0.66$ in the three cases respectively, which
would suggest $\nu \sim 1/d$ for large $d$.
On the other hand, in the same large $d$ limit, in \cite{litim}
the value  $\nu \simeq 1/2d$ was obtained, again
performing a renormalization group analysis truncated to the Einstein-Hilbert
action with a cosmological term in the continuum.
One cannot help noticing some reasonable agreement between the estimates for
$\nu$ coming from the lattice, and the corresponding
results for $\nu$ using the truncated renormalization group expansion in
the continuum, as well as some degree of compatibility with the
$2+\epsilon$ expansion close to $d=2$.
For a more quantitative comparison, see again Figs. 1 and 2.

One possible advantage of the continuum renormalization group 
calculations is that, since they can be performed to a great extent
analytically, they allow greater flexibility in exploring various scenarios,
such as additional invariant operators in the action, measure
contributions, or varying dimensions.
At the same time the absence of a reliable estimate for the errors
involved in the truncation leaves the method with uncertainties, which
might be hard to quantify until an improved calculation is performed with
an extended operator basis.

%% Insert here Figures 1,2

\begin{center}
\epsfysize=8.0cm
\epsfbox{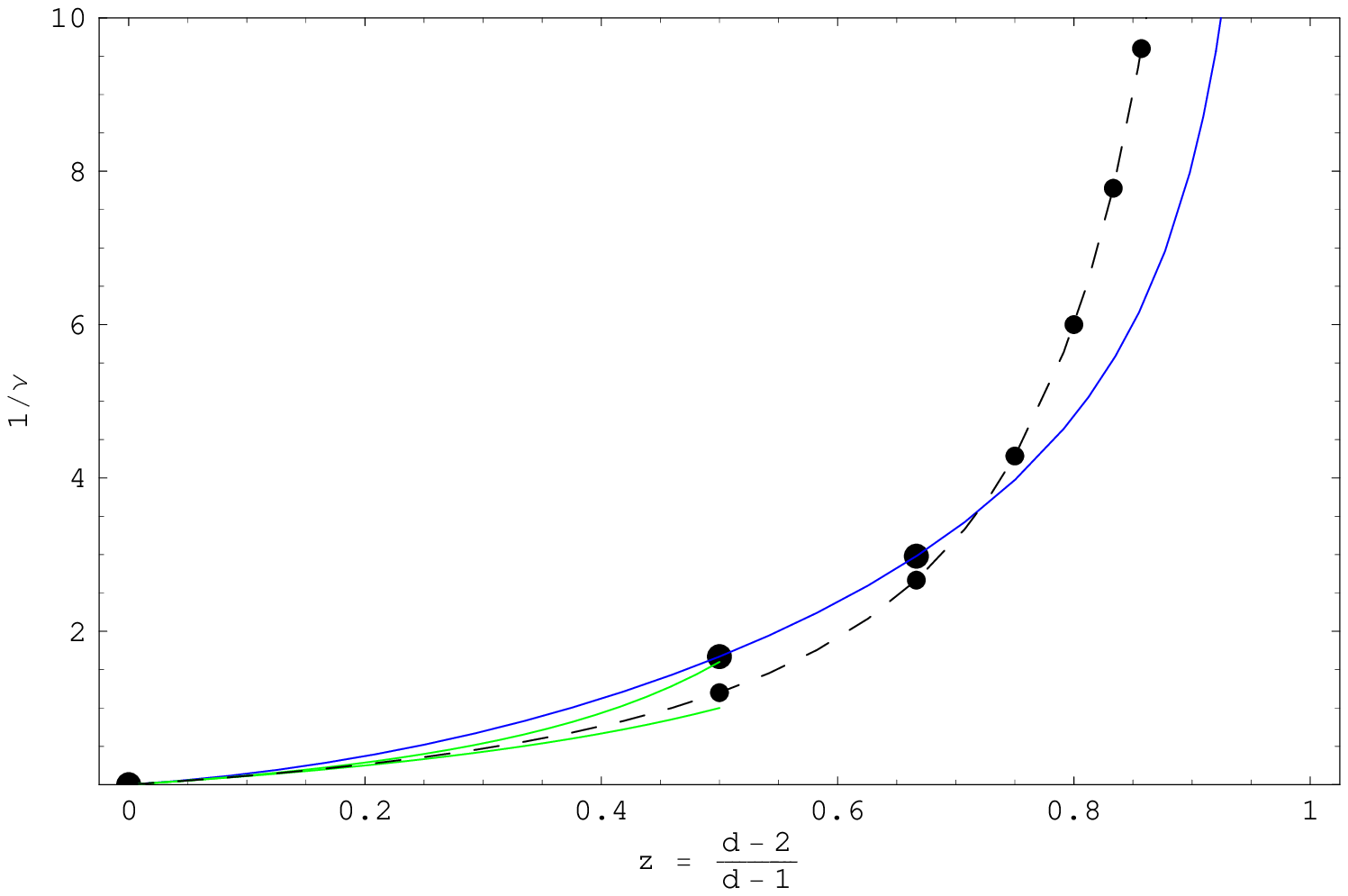}
\end{center}

\noindent{\small Fig. 1. Universal gravitational exponent $1/\nu$ 
of Eqs.~(\ref{eq:grun_latt}) and (\ref{eq:grun_m}) as a function of the dimension
(the abscissa here is $z=(d-2)/(d-1)$, where $d$ is the space-time dimension, 
which maps $d=2$ to $z=0$ and $d=\infty$ to $z=1$). 
The larger circles at $d=3$ and $d=4$ are the lattice gravity results of
\cite{hw3d,critical}, interpolated (continuous curve) using the exact
lattice results $1/\nu=0$ in $d=2$ \cite{hw2d}, and $\nu=0$ at $d=\infty$ \cite{larged}.
The smaller black dots (connected by the dashed curve) are the recent
results of \cite{litim,litim1}, obtained from a continuum renormalization group
study around the non-trivial fixed point of the gravitational action in $d$ dimensions
in the limit $\lambda \rightarrow 0$.
The two curves close to the origin are the $2+\epsilon$ expansion for $1/\nu$
to one loop (lower curve) and two loops (upper curve) \cite{epsilon3}.
\medskip}

% \newpage

% \begin{center}
% \epsfysize=8.0cm
% \epsfbox{exp1.eps}
% \end{center}

% \noindent{\small Fig. 2. Universal gravitational critical exponent $1/\nu$ 
% of Eq.~(\ref{eq:grun_latt}) as a function of dimension.
% The abscissa here is the space-time dimension $d$.
% The larger circles at $d=3$ and $d=4$ are the Regge lattice results of
% \cite{hw3d,critical}, interpolated (continuous curve) using the additional 
% exact lattice results $1/\nu=0$ in $d=2$ \cite{hw2d}, and $\nu=0$ at $d=\infty$ \cite{larged}.
% The smaller black dots (connected by the interpolating dashed curve) are the recent
% results of \cite{litim,litim1}, obtained from a continuum renormalization group
% study of the non-trivial fixed point of the gravitational action in $d$ dimensions.
% The two curves close to the origin are the $2+\epsilon$ expansion for $1/\nu$
% to one loop (lower curve) \cite{epsilon} and two loops (upper curve) \cite{epsilon3}.
% \medskip}

\newpage

\begin{center}
\epsfysize=8.0cm
\epsfbox{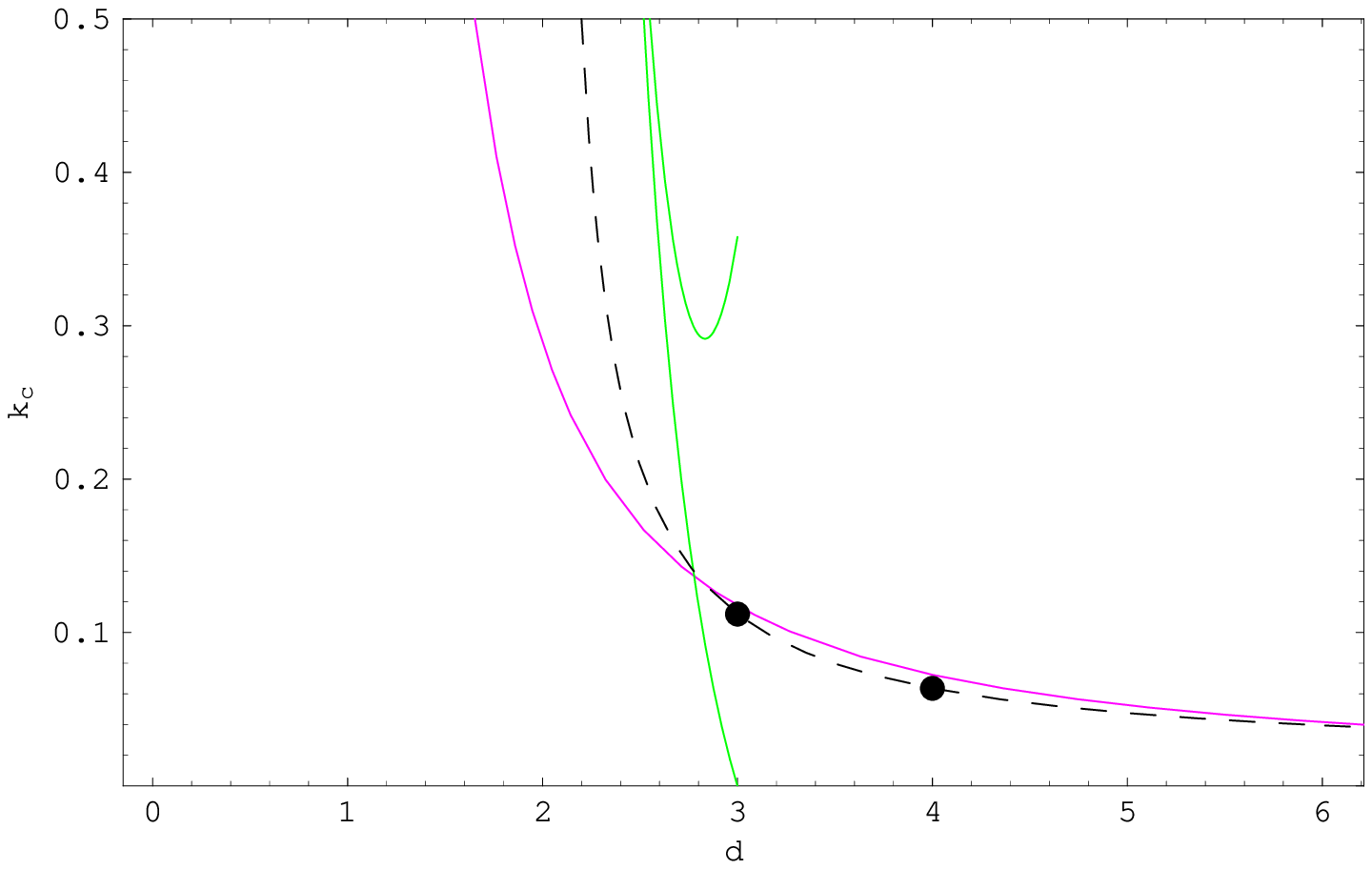}
\end{center}

\noindent{\small Fig. 2. Critical point $k_c=1/(8 \pi G_c)$ in units of the cutoff
(as it appears for ex. in Eqs.~(\ref{eq:m_latt}) and (\ref{eq:chising}))
as a function of dimension. The abscissa is the space-time dimension $d$.
The large circles at $d=3$ and $d=4$ are the Regge lattice results of
\cite{hw3d,critical}, interpolated (dashed curve) using the additional lattice
results $1/k_c=0$ in $d=2$ \cite{hw2d}.
The lower continuous curve is the large-$d$ lattice result of
Eq.~(\ref{eq:kcd}).
The two upper continuous curves are the two-loop $2+\epsilon$ expansion result
in the continuum of Eq.~(\ref{eq:nueps}) \cite{epsilon3}, whose prediction
becomes uncertain as $d$ approaches the value three.
\medskip}

\subsection{The running of $\alpha (\mu^2)$ in QED and QCD}

QED and QCD provide two invaluable illustrative cases where the running
of the gauge coupling with energy is not only theoretically well understood, but also
verified experimentally.
This section is intended to provide analogies and distinctions between the
two theories, and later gives justification, based on the structure of
infrared divergences in QCD, for the transition between the running of
the gravitational coupling as given in Eq.~(\ref{eq:grun_latt}), and its 
infrared regulated version of Eq.~(\ref{eq:grun_m}).
Most of the results found in this section are well known, but the purpose
here is provide a contrast (and in some instances, a close relationship)
with the gravitational case.

In QED the non-relativistic static Coulomb potential is affected by the vacuum
polarization contribution due to electrons (and positrons) of mass $m$.
To lowest order in the fine structure constant, the contribution is
from a single Feynman diagram involving a fermion loop.
One finds for the vacuum polarization contribution $\omega_R ({\bf k^2}) $
at small ${\bf k^2}$ the well known result \cite{iz}
\beq
{ e^2 \over {\bf k^2} } \; \rightarrow \;
{ e^2 \over {\bf k^2} [ \, 1 + \, \omega_R ({\bf k^2}) ] }
\; \sim \; 
{ e^2 \over {\bf k^2} } \, \left [ \, 1 \, + \, {\alpha \over 15 \, \pi } \,
{ {\bf k^2} \over m^2 } \, + \, O(\alpha^2) \, \right ] 
\label{eq:alpha_qed}
\eeq
which, for a Coulomb potential with a charge centered at the origin of strength
$-Ze$ leads to well-known Uehling \cite{uehling,kroll} $\delta$-function correction 
\beq
V(r) \; = \; \left ( \, 1 \, - \, {\alpha \over 15 \, \pi } \,
{ \Delta \over m^2 } \, \right ) \, { - Z \, e^2 \over 4 \, \pi \, r} \; = \;
{ - Z \, e^2 \over 4 \, \pi \, r} \, - \, {\alpha \over 15 \, \pi } \, { - Z \, e^2 \over m^2 } 
\, \delta^{(3)} ({\bf x})
\label{eq:pot_uel}
\eeq
It is not necessary though to resort to the small-${\bf k^2}$ approximation,
and in general a static charge of strength $e$ at the origin will
give rise to a modified potential
\beq
{ e \over 4 \, \pi \, r } \; \rightarrow \;
{ e \over 4 \, \pi \, r } \, Q(r)
\eeq
with 
\beq
Q(r) \; = \; 1 \, + \, {\alpha \over 3 \, \pi } \, \ln { 1 \over m^2 \, r^2 } \, + \, \dots 
\;\;\;\;\; m \, r \ll 1
\label{eq:qed_s}
\eeq
for small $r$, and
\beq
Q(r) \; = \; 1 \, + \, {\alpha \over 4 \, \sqrt{\pi} \, (m r)^{3/2} } \, 
e^{- \, 2 \, m \, r  } \, + \, \dots 
\;\;\;\;\; m \, r \gg 1
\label{eq:qed_l}
\eeq
for large $r$.
Here the normalization is such that the potential at infinity has $Q(\infty)=1$
\footnote{
The running of the fine structure constant has recently been verified
experimentally at LEP, the scale dependence of the vacuum polarization effects gives
a fine structure constant changing from $\alpha (0) \sim 1/137.036$
at atomic distances
to about $ \alpha (m_{Z_0}) \sim 1/128.978 $ at energies comparable to
the $Z^0$ boson mass, in good agreement with the theoretical prediction \cite{lep}.}.
The reason we have belabored on this example is to show that the screening
vacuum polarization contribution would have dramatic effects in QED if for some
reason the particle running through the fermion loop diagram had a much
smaller (or even close to zero) mass.

There are two interesting aspects of the (one-loop) result of
Eqs.~(\ref{eq:qed_s}) and (\ref{eq:qed_l}).
The first one is that the exponentially small size of the correction at large $r$
is linked with the fact that the electron mass $m_e$ is not too small:
the range of the correction term is 
$\xi = 2 \hbar / m c = 0.78 \times 10^{-10} cm$, but would have been much
larger if the electron mass had been a lot smaller.
In fact in the limit of zero electron mass the correction progressively 
extends out all the way to infinity.
Ultimately the fact that the Uehling correction is not important in atomic physics
is largely due to the fact that its value is already quite small on atomic scales,
comparable to the Bohr radius $ a_0 = \hbar / m e^2 = 0.53 \times 10^{-8}$.

The second interesting aspect is that the correction is divergent at small $r$ due
to the log term, giving rise to a large unscreened charge close to the origin; and 
of course the smaller the electron mass $m$, the larger the correction.
Also, it should be noted that the observed laboratory value for
the effective electromagnetic charge is normalized so that $Q(r)$ is $Q(r=\infty)=1$,
whereas in gravity the laboratory value corresponds to the ``short distance'' limit,
$G(r=0)$, due to the very large size of $\xi$ in Eq.~(\ref{eq:grun_m}).

In QCD (and related Yang-Mills theories) radiative corrections are also known to alter 
significantly the behavior of the static potential at short distances.
The changes in the potential are best expressed in terms 
of the running strong coupling constant $\alpha_S (\mu) $, whose scale
dependence is determined by the celebrated beta function of $SU(3)$ QCD
with $n_f$ light fermion flavors \cite{pdg}
\beq
\mu \, { \partial \, \alpha_S \over \partial \, \mu } \; = \; 
2 \, \beta ( \alpha_S ) \; = \; 
- \, { \beta_0 \over 2 \, \pi } \, \alpha_S^2 
\, - \, { \beta_1 \over 4 \, \pi^2 } \, \alpha_S^3 
\, - \, { \beta_2 \over 64 \, \pi^3 } \, \alpha_S^4 \, - \, \dots
\label{eq:beta_qcd}
\eeq
with $\beta_0 = 11 - {2 \over 3} n_f $, $\beta_1 = 51 - {19 \over 3} n_f $, 
and $\beta_2 = 2857 - {5033 \over 9} n_f + {325 \over 27} n_f^2 $.
The solution of the renormalization group equation  
Eq.~(\ref{eq:beta_qcd}) then gives for the running of $\alpha_S (\mu )$
\beq
\alpha_S (\mu ) \; = \; 
{ 4 \, \pi \over \beta_0 \ln { \mu^2 / \Lambda_{\overline{MS}}^2  } }
\, \left [ 1 \, - \, { 2 \beta_1 \over \beta_0^2 } \,
{ \ln \, [ \ln { \mu^2 / \Lambda_{\overline{MS}}^2 ] } 
\over \ln { \mu^2 / \Lambda_{\overline{MS}}^2  } }
\, + \, \dots \right ]
\label{eq:alpha_qcd}
\eeq
The non-perturbative scale $ \Lambda_{\overline{MS}} $ appears as an
integration constant of the renormalization group equations, and
is therefore - by construction - scale independent.
The physical value of $ \Lambda_{\overline{MS}} $
cannot be fixed from perturbation theory alone, and needs to be determined
by experiment
\footnote{
Alternatively, the integration constant in Eq.~(\ref{eq:beta_qcd})
can be fixed by specifying the
value of $\alpha_S$ at some specific energy scale, usually chosen to be the 
$Z^0$ mass.}.

In QCD one determines experimentally from Eq.~(\ref{eq:alpha_qcd})
(for example, from the size of scaling violations in deep inelastic scattering)
$ \Lambda_{\overline{MS}} \simeq 220 MeV$, which not surprisingly is
close to a typical hadronic scale.
But from a purely theoretical standpoint one could envision other scenarios,
where this scale would take on a completely different value. 
The point is that nothing in QCD itself determines the absolute magnitude
of this scale: only its size in relation to other physical
observables such as hadron masses is theoretically calculable within QCD.

In principle one can solve for $ \Lambda_{\overline{MS}} $ in terms of the 
coupling at any scale, and in particular at the cutoff scale $\Lambda$,
obtaining 
\beq
\Lambda_{\overline{MS}} \; = \; 
\Lambda \, \exp \left ( 
{ - \int^{\alpha_S (\Lambda)} \, {d \alpha_S' \over 2 \, \beta ( \alpha_S') } }
\right )
\; = \;
\Lambda \, \left ( { \beta_0 \, \alpha_S ( \Lambda ) \over 4 \, \pi } 
\right )^{\beta_1 / \beta_0^2 }
\, e^{ - { 2 \, \pi \over \beta_0 \, \alpha_S (\Lambda ) } }
\, \left [ \, 1 \, + \, O( \alpha_S ( \Lambda) ) \, \right ] 
\label{eq:lambda_qcd}
\eeq
It should be clear by now that this expression is the analog of Eqs.~(\ref{eq:m_latt})
and (\ref{eq:m_cont}) in the gravitational case. 
In lattice QCD this is usually taken as the definition of
the running strong coupling constant $\alpha_S (\mu) $.
It then leads to an effective potential between quarks and anti-quarks of the form
\beq
V ( {\bf k^2} ) \; = \; - \, {4 \over 3} \, { \alpha_S ( {\bf k^2} ) 
\over {\bf k^2} }
\label{eq:qcd_pot}
\eeq
and the leading logarithmic correction makes the potential appear
softer close to the origin, $V(r) \sim 1 / ( r \ln r) $.

When the QCD result is contrasted with the QED answer of
Eqs.~(\ref{eq:alpha_qed}) and (\ref{eq:pot_uel}) it appears
that the infrared small ${\bf k^2}$ singularity in 
Eqs.~(\ref{eq:qcd_pot}) is quite serious.
An analogous conclusion is reached when examining 
Eqs.~(\ref{eq:alpha_qcd}) : the coupling strength
$ \alpha_S ( {\bf k^2} )$ diverges in the infrared due to the 
singularity at $k^2=0$.
In order to avoid a meaningless divergent answer,
the uncontrolled growth in $ \alpha_S ( {\bf k^2} )$
needs therefore to be regulated (or self-regulated) by the 
dynamically generated QCD infrared cutoff $\Lambda_{\overline{MS}}$ \cite{richardson}
(as indeed happens in simpler theories, such as the non-linear sigma model,
which can be solved exactly in the large $N$ limit, where this 
mechanism can therefore be explicitly demonstrated \cite{zinn}).
To lowest order in the coupling this implies
\beq
V ( {\bf k^2} ) \; = \; - \, {4 \over 3} \, 
{ 4 \, \pi \over \beta_0 {\bf k^2} \, 
\ln \left ( {\bf k^2} / \Lambda_{\overline{MS}}^2 \right ) }
\; \rightarrow \;
- \, {4 \over 3} \, 
{ 4 \, \pi \over \beta_0 {\bf k^2} \, 
\ln \left (
( {\bf k^2} \, + \, \Lambda_{\overline{MS}}^2 ) / \Lambda_{\overline{MS}}^2
\right ) }
\label{eq:qcd_pot_rich}
\eeq
The resulting potential can then be evaluated for small 
$ {\bf k^2} $ 
\beq
V ( {\bf k^2} ) \;
\mathrel{\mathop\sim_{ {\bf k^2} / \Lambda_{\overline{MS}}^2 \rightarrow 0 }} \;
- \, {4 \over 3} \, 
{ 4 \, \pi \over \beta_0 {\bf k^2} \, 
( {\bf k^2} / \Lambda_{\overline{MS}}^2 ) }
\label{eq:qcd_pot_sing}
\eeq
giving, after Fourier transforming back to real space, for large $r$,
\beq
V ( r ) \;
\mathrel{\mathop\sim_{ r \rightarrow \, \infty }} \;
- \, {4 \over 3} \, 
{ 4 \, \pi \over \beta_0 } \, \Lambda_{\overline{MS}}^2 \,
\cdot \, \left ( - \, { 1 \over 8 \, \pi } \, r \right ) \, + \, O( r^0) 
\; \simeq \; + \, \sigma \, r 
\label{eq:qcd_pot_conf}
\eeq
The desired linear potential, here with string tension $\sigma$,
is then indeed recovered at large distances.
In fact the interpolating potential of Eq.~(\ref{eq:qcd_pot_rich})
is remarkably successful in describing non-relativistic
QCD bound states, as discussed in \cite{richardson,eichten},
incorporating correctly to some extent the leading short- and large-distance
QCD corrections
\footnote{
A similar infrared regularization was done in 
Eq.~(\ref{eq:grun_m}), 
$ ( m^2 / {\bf k^2} )^{1/2 \nu} = 
\exp \left [ - {1 \over 2 \nu } \ln ( {\bf k^2} / m^2 )  \right ] 
\; \rightarrow \; \exp 
\left \{ - {1 \over 2 \nu } \ln [ ( {\bf k^2} + m^2 ) / m^2 ] \right \} $
}.

What is rather remarkable in this context is that the removal of a serious
infrared divergence in $ \alpha_S ( {\bf k^2} ) $, by the replacement
of Eq.~(\ref{eq:qcd_pot_rich}), has in fact caused an even 
stronger infrared divergence in $V ( {\bf k^2} ) $ itself, which
leads though, in this instance, to precisely the desired
result of Eq.~(\ref{eq:qcd_pot_conf}), namely a confining potential
at large distances, as expected on the basis of non-perturbative arguments
\cite{wilson-lgt}.
From $\alpha_S ( {\bf k^2} ) $ given in Eq.~(\ref{eq:qcd_pot_rich})
one can calculate the corresponding beta function
\beq
\beta ( \alpha_S ) \; = \; 
- \, { \beta_0 \over 4 \, \pi } \, \alpha_S^2 \,
\left ( 1 \, - \, e^{ - \, { 4 \, \pi \over \beta_0 \, \alpha_S } } \right )
\, + \, O( \alpha_S^3 ) 
\label{eq:beta_rich}
\eeq
which now exhibits a non-analytic (renormalon) contribution at $\alpha_S =0$, not
detectable to any order in perturbation theory \cite{dok,beneke}.

The very fruitful analogy between strongly coupled non-abelian gauge theories and strongly
coupled gravity can be pushed further \cite{larged}, by
developing other aspects of the correspondence which naturally lend themselves
to such a comparison.
In gravity the analog to the Wilson loop $W(\Gamma)$ of non-abelian gauge theories
\beq
W(\Gamma) \, \sim \, \Tr \, {\cal P} \, \exp \, \left [
\int_C \, A_{\mu} \, dx^{\mu}  \right ]
% \label{eq:wloop}
\eeq
exists as well (see Eq.~(\ref{eq:wloop})), defined in the gravitational
case as the path-ordered exponential of the affine connection 
$\Gamma_{\mu\nu}^{\lambda}$ around a closed planar loop.

Furthermore, it is known that in QCD, due to the non-trivial strong
coupling dynamics, there arise several non-perturbative condensates.
Thus the gluon condensate is related to the confinement scale via \cite{shifman}
\beq
\alpha_S <F_{\mu\nu} \cdot F^{\mu\nu}> \; \simeq \; (250 MeV)^4 \, \sim \, \xi^{-4} 
\eeq
with $\xi_{QCD}^{-1} \sim \Lambda_{\overline{MS}}$.
Another important non-vanishing non-perturbative condensate in QCD is the fermionic
one \cite{veneziano,hp}
\beq
(\alpha_S)^{4/\beta_0} <{\bar \psi} \, \psi > \; \simeq \; - (230 MeV)^3 \, \sim \, \xi^{-3} 
\eeq
and represents a key quantity in discussing the size of chiral symmetry
breaking, and the light quark masses.

\vskip 30pt
\newsection{Poisson's Equation with Running G }
\hspace*{\parindent}

Given the running of $G$ from either Eq.~(\ref{eq:grun_m}), or Eq.~(\ref{eq:grun_latt}) 
in the large $\bf k$ limit, the next step is naturally a solution
of Poisson's equation with a point source at the origin, in order to determine
the structure of the quantum corrections to the gravitational potential in real space.
The more complex solution of the fully relativistic problem will then
be addressed in the following sections.
In the limit of weak fields the relativistic field equations
\beq
R_{\mu\nu} \, - \, \half \, g_{\mu\nu} \, R \, + \, \lambda \, g_{\mu\nu}
\; = \; 8 \pi G \, T_{\mu\nu}
\label{eq:naive_t}
\eeq
give for the $\phi$ field (with $g_{00}(x) \simeq  -(1+2 \phi(x)\,)$
\beq
( \Delta \, - \, \lambda ) \, \phi(x) \; = \; 
4 \, \pi \, G \, \rho (x) \, - \, \lambda
\label{eq:pois}
\eeq
which would suggest that the scaled cosmological constant $\lambda$ acts like a mass term
$m = \sqrt{\lambda}$.
For a point source at the origin, the first term on the r.h.s is just
$4 \pi M G \, \delta^{(3)} ( {\bf x}) $.
The solution for $\phi(r)$ can then be obtained simply by Fourier transforming
back to real space Eq. (\ref{eq:pot_def}), and, up to an additive constant,
one has 
\beq
\phi (r) \; = \; - M \, G \, { e^{- m \, r } \over r } \, - \, 
{ a_0 \, m \, M \, G  \, \over
2^{{1 \over 2} ( {1 \over \nu} -1 )} \, \Gamma(1 \, + \, {1 \over 2 \, \nu} ) \sqrt{\pi} }
\, ( m \, r )^{{1 \over 2} ( {1 \over \nu} -1 )} 
\, K_{{1 \over 2} ( {1 \over \nu} -1 )} ( m \, r ) 
\eeq
where $K_n(x)$ is the modified Bessel function of the second type.
The behavior of $\phi(r)$ would then be Yukawa-like 
$\phi(r) \sim {\rm const.} \;  e^{- m \, r } / r $ and thus rapidly
decreasing for large $r$.

But the reason why both of the above results are in fact {\it incorrect} (assuming
of course the validity of general coordinate invariance at very large 
distances $r \gg 1/\sqrt{\lambda}$) is that the
exact solution to the field equations in the static isotropic case with a $\lambda$ term
gives
\beq
- \, g_{00} \; = \; B(r) \; = \; 1 - { 2 \, M \, G \over r} 
\, -  \, {\lambda \over 3} \, r^2 
\eeq
showing that the $\lambda$ term definitely does {\it not} act like
a mass term in this context.

Therefore the zeroth order contribution to the potential should be taken to
be proportional to $4 \pi /( {\bf k^2} + \mu^2 ) $ with $\mu \rightarrow 0$, as
already indicated in fact in Eq.~(\ref{eq:pot_m}).
Also, proper care has to be exercised in providing an appropriate infrared regulated
version of $G({\bf k^2})$, and therefore $V({\bf k^2})$, which from
Eq.~(\ref{eq:pot_m}) reads
\beq
{ 4 \pi \over ( {\bf k^2} \, + \, \mu^2 ) } \, \left [ \; 1 \, 
+ \, a_0 \left ( { m^2 \over {\bf k^2} \, + \, m^2 } \right )^{1 \over 2 \nu} \, \right ]
\label{eq:pot_m1}
\eeq
and where the limit $\mu \rightarrow 0$ is intended to be taken
at the end of the calculation.

There are in principle two equivalent ways to compute the potential $\phi(r)$,
either by inverse Fourier transform of the above expression, or by solving Poisson's
equation $\Delta \phi = 4 \pi \rho$ with $\rho (r)$ given by the inverse Fourier
transform of the correction to $G(k^2)$, as given later in Eq.~(\ref{eq:rho_vac}).
Here we will first use the first, direct method.

\subsection{Large r limit}

The zeroth order term gives the standard Newtonian $-M G/r$ term, while the correction
in general is given by a rather complicated hypergeometric function.
But for the special case $\nu=1/2$ one has for the Fourier transform of the correction
to $ \phi (r) $
\beq
a_0 \, m^{1/\nu} \, { 4 \pi \over {\bf k^2} \, + \, \mu^2 } \, 
{ 1 \over ( {\bf k^2} \, + \, m^2 )^{1 \over 2 \nu} } \; \rightarrow \; 
a_0 \, m^2 \, { e^{- \mu \, r} \, - \, e^{- m \, r } \over r \, (m^2 \, - \, \mu^2 ) }   
\; \mathrel{\mathop\sim_{ \mu  \, \rightarrow \, 0  }} \; 
a_0 \, m^2 \, { 1 \, - \, e^{- m \, r } \over m^2 \, r }   
\eeq
giving for the complete quantum-corrected potential
\beq
\phi(r) \; = \; - \, { M \, G \over r } \, \left [ \, 
1 \, + \, a_0 \, \left ( \, 1 \, - \, e^{- m \, r } \, \right ) \, \right ]   
\label{eq:phi_half}
\eeq
For this special case the running of $G(r)$ is particularly transparent,
\beq
G(r) \; = \; G_{\infty} \left ( 
\, 1 \, - \, { a_0 \over 1 \, + \, a_0 } \, e^{- m r} \, \right )
\label{eq:grun_half}
\eeq
with $G_{\infty} \equiv (1+a_0) \, G $ and $G \equiv G(0)$.
$G$ therefore increases slowly from its value $G$ at small $r$ to the larger
value $(1+a_0)\, G$ at infinity.
Figure 3. provides a schematic illustration of the behavior of $G$ as a function
of $r$.

%% Insert here Figure 3.

% \newpage

\begin{center}
\epsfysize=8.0cm
\epsfbox{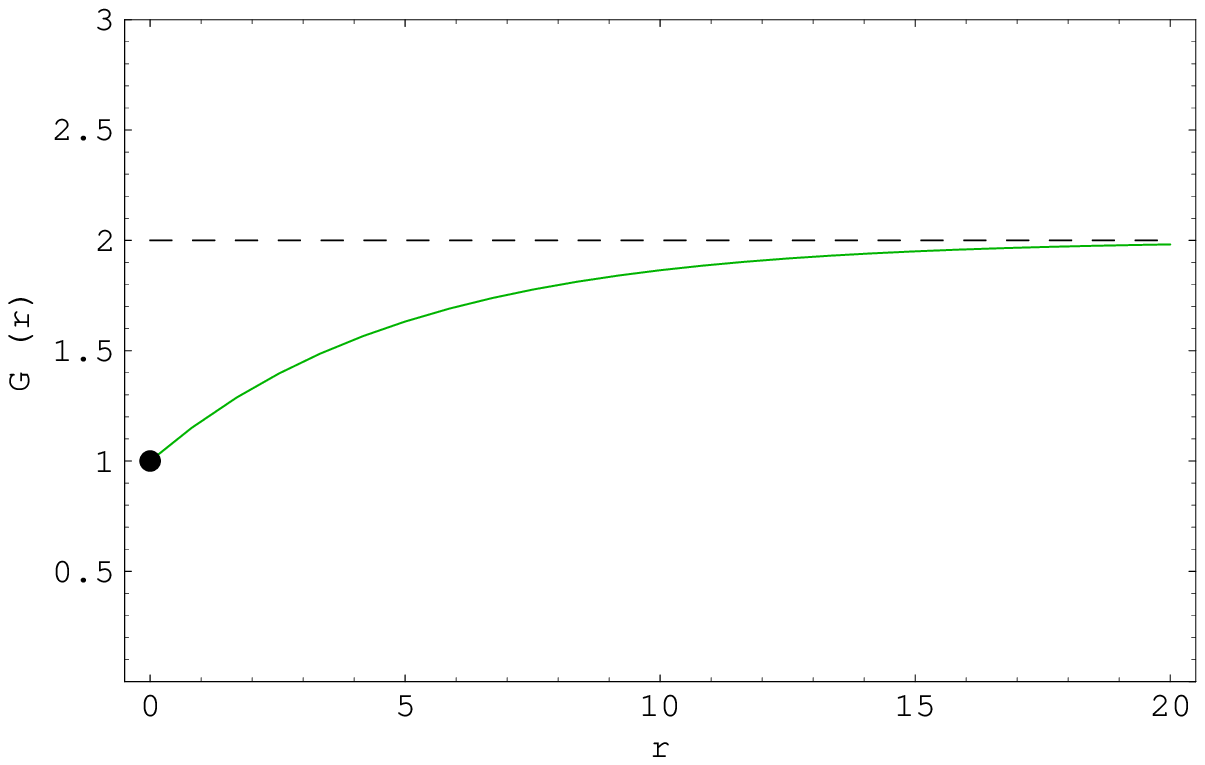}
\end{center}

\noindent{\small Fig. 3. Schematic scale dependence of the gravitational coupling
$G(r)$ from Eq.~(\ref{eq:grun_half}), here for $\nu=1/2$.
The gravitational coupling rises initially like a power of $r$, and then
approaches the asymptotic value $G_{\infty}=(1+a_0) G$ for large $r$.
The behavior for other values of $\nu>1/3$ is similar.
\medskip}

% \newpage

Returning to the general $\nu$ case, one can expand for small ${\bf k}$ to get
the correct large $r$ behavior,
\beq
{ 1 \over ( {\bf k^2} \, + \, \mu^2 ) \, ( {\bf k^2} \, + \, m^2 )^{1 \over 2 \nu} }
\; \simeq \; 
{ 1 \over m^{1 \over \nu} } \, { 1 \over ( {\bf k^2} \, + \, \mu^2 ) } \, - \,
{ 1 \over 2 \, \nu \, m^2 } \, 
{ 1 \over ( {\bf k^2} \, + \, m^2 )^{1 \over 2 \nu} } \, + \, \dots
\eeq
After Fourier transform, one obtains the previous answer for $\nu=1/2$,
whereas for $\nu=1/3$ one finds
\beq
- \, { M \, G \over r } \, \left [ \, 1 \, + \, a_0 \, 
\left ( 1 \, - \, { 3 \, m \, r \over \pi } \, K_0 ( m \, r )  \right ) \,
\right ]
\eeq
and for general $\nu$
\beq
- \, { M \, G \over r } \, \left [ 1 \, + \, a_0 \, 
\left ( 1 \, - \, 
{ 2^{ {1 \over 2} ( 3 - {1 \over \nu} ) } \, m \, r \,
\over 2 \, \nu \, \sqrt{\pi} \, \Gamma( {1 \over 2 \, \nu} ) } \, 
( m \, r )^{- {1 \over 2} (3 - {1 \over \nu}) } \,
\, K_{ {1 \over 2} ( 3 - {1 \over \nu} ) } ( m \, r ) \, \right ) \, \right ]
\label{eq:phi_large}
\eeq
Using the asymptotic expansion of the modified Bessel function $K_n(x)$
for large arguments,
$K_n (z) \sim \sqrt{\pi/2} \, z^{-1/2} \, e^{-z} \, ( 1+ O(1/z) )$, 
one finally obtains in the large $r$ limit
\beq
\phi(r) \; \mathrel{\mathop\sim_{ r  \, \rightarrow \, \infty  }} \; 
- \, { M \, G \over r } \, \left [ 1 \, + \, a_0 \, 
\left ( 1 \, - \, c_l \, ( m \, r )^{ {1 \over 2 \nu} - 1 } \,
e^{- m r } \right ) \, \right ]
\label{eq:phi_large1}
\eeq
with $c_l = 1 / ( \nu \, 2^{1 \over 2 \nu} \, \Gamma ( {1 \over 2 \nu } )) $.

\subsection{Small r limit}

In the small $r$ limit one finds instead, using again Fourier transforms,
for the correction for $\nu=1/3$
\beq
( - \, M \, G ) \, a_0 \, m^2 \,  
{ r^2 \over 3 \, \pi } \left [ \, \ln ( { m \, r \over 2 } ) \, + \, 
\gamma - { 5 \over 6 } \, \right ] \, + \, O(r^3) 
\label{eq:phi_small}
\eeq
In the general case the complete leading correction to the potential
$\phi(r)$ for small $r$ (and $\nu > 1/3 $) has the structure 
$ ( - {\rm const.} ) (- M G) a_0 \, m^{1 \over\nu } \,  r^{ {1\over\nu} - 1 } $.
Note that the quantum correction always vanishes at short distances
$r \rightarrow 0$, as expected from the original result of
Eqs.~(\ref{eq:grun_latt}) or (\ref{eq:grun_m}) for $k^2 \rightarrow \infty$
\footnote{
At very short distances $r \sim l_P$ other quantum corrections come into
play, which are not properly encoded in Eq.~(\ref{eq:grun_latt}), which 
after all is supposed to describe the universal running in the {\it scaling region}
$ l_P \ll r \ll \xi $.
Furthermore, higher derivative terms could also have important effects at very short
distances.}.

The same result can be obtained via a 
different, but equivalent, procedure, in which one solves directly the radial Poisson
equation for $\phi(r)$.
First, for a point source at the origin, $4 \pi M G \, \delta^{(3)} ( {\bf x}) $,
with
\beq
\delta^{(3)} ({\bf x}) \; = \; {1 \over 4 \, \pi} \, 2 \, { \delta (r) \over r^2 }
\eeq
one sets $\Delta \phi(r) \rightarrow r^{-1} d^2 / d r^2 [ r \phi(r) ] $
in radial coordinates.
In the $a_0 \neq 0$ case one then needs to solve $\Delta \phi = 4 \pi \rho$,
or in the radial coordinate for $ r >0 $
\beq
{ 1 \over r^2} \, { d \over d \, r } \, 
\left ( \, r^2 \, {d \, \phi \over d \, r } \, \right ) \; = \;
4 \, \pi \, G \, \rho_m (r)
\eeq
with the source term $\rho_m $ determined from the inverse Fourier
transform of the correction term in Eq.~(\ref{eq:grun_m}), namely
\beq
a_0 \, M \, \left ( { m^2 \over {\bf k^2} \, + \, m^2 } \right )^{1 \over 2 \nu}
\eeq
One finds
\beq
\rho_m (r) \; = \; { 1 \over 8 \pi } \, c_{\nu} \, a_0 \, M \, m^3 \,
( m \, r )^{ - {1 \over 2} (3 - {1 \over \nu}) }  
\, K_{ {1 \over 2} ( 3 - {1 \over \nu} ) } ( m \, r ) 
\label{eq:rho_vac}
\eeq
with 
\beq
c_{\nu} \; \equiv \; { 2^{ {1 \over 2} (5 - {1 \over \nu}) }
\over \sqrt{\pi} \, \Gamma( {1 \over 2 \, \nu} ) }
\label{eq:rho_vac1}
\eeq
The vacuum polarization density $\rho_m$ has the property 
\beq
4 \, \pi \, \int_0^\infty \, r^2 \, d r \, \rho_m (r) \; = \;  a_0 \, M 
\label{eq:rho_vac2}
\eeq
where the standard integral 
$\int_0^\infty dx \, x^{2-n} \, K_n (x) = 
2^{-n} \sqrt{\pi} \Gamma \left ( {3 \over 2} -n \right ) $ has been used. 
Note that the vacuum polarization distribution is singular close to $r=0$, just as
in QED, Eq.~(\ref{eq:qed_s}).

The $r \rightarrow 0 $ result for $\phi (r) $ 
(discussed in the following, as an example, for $\nu=1/3$) can then
be obtained by solving the radial equation for $\phi(r)$,
\beq
{ 1 \over r} \, { d^2 \over d \, r^2 } \, \left [ \, r \, \phi (r) \, \right ] \; = \;
{ 2 \, a_0 \, M \, G \, m^3 \over \pi } \, K_0 ( m \, r ) 
\eeq
where the (modified) Bessel function is expanded out 
to lowest order in $r$, 
$K_0 ( m \, r ) = - \gamma - \ln \, \left ( { m \, r \over 2 } \right ) + O(m^2 \, r^2) $,
giving
\beq
\phi (r) \; = \; - \, { M \, G \over r } \, + \,
a_0 \, M \, G \, m^3 \,  
{ r^2 \over 3 \, \pi } \left [ - \, \ln ( { m \, r \over 2 } ) \, - \, 
\gamma + { 5 \over 6 } \, \right ] \, + \, O(r^3) 
\label{eq:phi_small1}
\eeq
where the two integration constants are matched to the large $r$ solution
of Eq.~(\ref{eq:phi_large}).
Note again that the vacuum polarization
density $\rho_m (r)$ has the expected normalization property
\beq
4 \, \pi \, \int_0^\infty \, r^2 \, d r \, 
{  a_0 \, M \,  m^3 \over 2 \, \pi^2 } \, K_0 ( m \, r )
\; = \;  
{ 2 \, a_0 \, M \,  m^3 \over \pi } \cdot { \pi \over 2 \, m^3 } \; = \; a_0 \, M 
\eeq
so that the total enclosed additional ``charge'' is indeed just $a_0 M$,
and $G_\infty = G_0 (1 + a_0) $ (see for comparison
also Eq.~(\ref{eq:phi_large})).
Using then the same method for general $\nu > {1 \over 3} $, one finds for small $r$
(using the expansion of the modified Bessel function $K_n (x)$ for small
arguments as given later in Eq.~(\ref{eq:bessel})) 
\beq
\rho_m (r) \; \mathrel{\mathop\sim_{ r \, \rightarrow \, 0 }} \; 
{ \vert \sec \left ( { \pi \over 2 \nu } \right ) \vert 
\over 4 \pi \, \Gamma \left ( {1 \over \nu} -1 \right ) }
\, a_0 \, M \, m^{1 \over \nu} \; r^{ {1 \over \nu} - 3 }
\; \equiv \; A_0 \; r^{ {1 \over \nu} - 3 } 
\eeq
and from it the general result
\beq
\phi(r) \; 
\; \mathrel{\mathop\sim_{ r \, \rightarrow \, 0 }} \; 
- \, { M \, G \over r } \, + \,
a_0 \, M \, G \, c_s \, m^{1 \over \nu} \, r^{ { 1 \over \nu } - 1 } \, + \, \dots
\label{eq:phi_small2}
\eeq
with $c_s = \nu \vert \sec \left ( { \pi \over 2 \nu } \right ) 
\vert / \Gamma ( {1 \over \nu } ) $.

\vskip 30pt
\newsection{Relativistic Field Equations with Running G}
\hspace*{\parindent}

Solutions to Poisson's equation with a running $G$ provide some insights
into the structure of the quantum corrections, but a complete analysis
requires the study of the full relativistic field equations, which will
be discussed next in this section.
A set of relativistic field equations incorporating the running of
$G$ is obtained by doing the replacement \cite{effective} 
\beq
G (k^2) \;\; \rightarrow \;\; G( \Box )
\label{eq:gbox0}
\eeq
with the d'Alembertian $\Box$ intended to correctly represent invariant
distances, and incorporating the running of $G$ as expressed in 
either Eqs.~(\ref{eq:grun_latt}) or (\ref{eq:grun_m}),
\beq
G \; \rightarrow \;  G( \Box) \; = \; G \, \left [ 1 \, + \, 
a_0 \, \left ( { m^2 \over - \, \Box + m^2 } \right )^{ 1 \over 2 \nu} \, 
+ \, \dots \, \right ]
\label{eq:gbox}
\eeq
For the use of $\Box$ to express the running of couplings in gauge theories
the reader is referred to the references in \cite{vilko}.
Here the $\Box$ operator is defined through 
the appropriate combination of covariant derivatives
\beq
\Box \; \equiv \; g^{\mu\nu} \, \nabla_\mu \nabla_\nu 
\eeq
and whose explicit form depends on the specific tensor nature of
the object it is acting on, as in the case of the energy-momentum tensor
\beq
\Box \; T^{\alpha\beta \dots}_{\;\;\;\;\;\;\;\; \gamma \delta \dots}
\; = \; g^{\mu\nu} \, \nabla_\mu \left( \nabla_\nu \;
T^{\alpha\beta \dots}_{\;\;\;\;\;\;\;\; \gamma \delta \dots} \right)
\eeq
Thus on scalar functions one obtains the fairly simple result
\beq
\Box \, S(x) \; = \; 
{1 \over \sqrt{g} } \, \partial_\mu \, g^{\mu\nu} \sqrt{g} \, \partial_\nu
\, S(x)
\eeq
whereas on second rank tensors one has the significantly
more complicated expression
$\Box \, T_{\alpha\beta} \, \equiv \, 
g^{\mu\nu} \, \nabla_\mu (\nabla_\nu T_{\alpha\beta}) $.
In general the invariant operator appearing in the above expression, namely
\beq
A (\Box) \; = \; a_0 \left( { m^2 \over - \Box \;\; } \right)^{1/2\nu} 
\label{eq:abox0}
\eeq
or its infrared regulated version
\beq
A (\Box) \; = \; a_0 \left( { m^2 \over - \Box \, + \, m^2 } \right)^{1/2\nu} 
\label{eq:abox}
\eeq
has to be suitably defined by analytic continuation from positive
integer powers; the latter can be often be done by computing $\Box^n$
for positive integer $n$, and then analytically continuing to 
$n \rightarrow -1/2\nu$.
In the following the above analytic continuation from positive
integer $n$ will always be understood.
Usually it is easier to work with the expression in Eq.~(\ref{eq:abox0}), and
then later amend the final result to include the infrared regulator, if needed.

One is therefore lead to consider the effective field equations of Eq.~(\ref{eq:field0}),
namely
\beq
R_{\mu\nu} \, - \, \half \, g_{\mu\nu} \, R \, + \, \lambda \, g_{\mu\nu}
\; = \; 8 \, \pi \, G  \, \left( 1 + A( \Box ) \right) \, T_{\mu\nu}
\label{eq:field}
\eeq
with $A( \Box ) $ given by Eq.~(\ref{eq:abox}) and $\lambda \simeq 1 / \xi^2$,
as well as the trace equation
\beq
R \, - \, 4 \, \lambda \; = \; - \, 8 \, \pi \, G \, \left( 1 + A( \Box ) \right) \, T
\label{eq:field_tr}
\eeq
Being manifestly covariant, these expressions at least satisfy some
of the requirements for a set of consistent field equations
incorporating the running of $G$, and can then be easily re-cast in a form
similar to the classical field equations
\beq
R_{\mu\nu} \, - \, \half \, g_{\mu\nu} \, R \, + \, \lambda \, g_{\mu\nu}
\; = \; 8 \pi G  \, {\tilde T_{\mu\nu}}
\eeq
with $ {\tilde T_{\mu\nu}} \, = \, \left( 1 + A( \Box ) \right) \, T_{\mu\nu}$
defined as an effective, gravitationally dressed, energy-momentum tensor.
Just like the ordinary Einstein gravity case,
in general ${\tilde T_{\mu\nu}}$ might not be covariantly conserved a priori,
$\nabla^\mu \, {\tilde T_{\mu\nu}} \, \neq \, 0 $, but ultimately the
consistency of the effective field equations demands that it
be exactly conserved in consideration of the Bianchi identity satisfied
by the Einstein tensor \cite{effective}.
The ensuing new covariant conservation law
\beq
\nabla^\mu \, {\tilde T_{\mu\nu}} \; \equiv \; 
\nabla^\mu \, \left [ \left ( 1 + A( \Box ) \right ) \, T_{\mu\nu} 
\right ] \;  = \;  0
\label{eq:continuity}
\eeq
can be then be viewed as a constraint
on ${\tilde T_{\mu\nu}}$ (or $T_{\mu\nu}$) which, for example,
in the specific case of a perfect fluid, 
implies a definite relationship between the density $\rho(t) $,
the pressure $p(t)$ and the metric components \cite{effective}.

From now on, we will set the cosmological constant $\lambda=0$, and its
contribution can then be added at a later stage.
As long as one is interested in static isotropic solutions, one
takes for the metric the most general form
\beq
ds^2 \; = \; - \, B(r) \, dt^2 \, + \, A(r) \, d r^2 \, 
+ \, r^2 \, ( d \theta^2 \, + \, \sin^2 \theta \, d \varphi^2 )
\label{eq:metric}
\eeq
and for the energy momentum tensor the perfect fluid form 
\beq
T_{\mu \nu} \; = \; {\rm diag} \,
[ \, B(r) \, \rho(r), \, A(r) \, p(r) ,
\, r^2 \, p(r), \, r^2 \, \sin^2 \theta \, p(r) \, ]
\label{eq:perfect_si}
\eeq
with trace $T = 3 \, p - \rho $.
The trace equation then only involves the (simpler)
scalar d'Alembertian, acting on the trace of the energy-momentum tensor

To the order one is working here, the
above effective field equations should be equivalent to
\beq
{ 1 \over 1 \, + \, A( \Box ) } \,  
\left( R_{\mu\nu} - \half \, g_{\mu\nu} \, R \right) \; = \;
\left( 1 \, - \, A( \Box )  \, + \, \dots \, \right) 
\left( R_{\mu\nu} - \half \, g_{\mu\nu} \, R \right)
\; = \; 8 \pi G \, T_{\mu\nu}
\label{eq:naive_r}
\eeq
where the running of $G$ has been moved over to the ``gravitational'' side,
and later expanded out, assuming the correction to be small.
For the vacuum solutions, the r.h.s. is zero for $ r \neq 0$, 
and one can re-write the last equation simply as
\beq
{ 1 \over 8 \pi G \left ( 1 \, + \, A( \Box ) \right ) } \,  
\left( R_{\mu\nu} - \half \, g_{\mu\nu} \, R \right)
\; = \; 0
\label{eq:field_rev}
\eeq

\subsection{Dirac delta function source}

A mass point source is most suitably described by a Dirac delta function.
The delta function at the origin can be represented for example as
\beq
\delta (r) \; = \; \lim_{\epsilon \rightarrow 0}
\, \frac{\epsilon }{ \pi \left( r^2+\epsilon ^2 \right) }
\eeq
As an example, the derivative operator $(d / d \, r)^n$ acting on the delta
function would give, for small $\epsilon$,
\beq
(-1)^n \, \Gamma (n+2) \, \pi^{-1} \, \epsilon \, r^n \, \left(r^2+\epsilon ^2\right)^{-n-1} 
\eeq
which can be formally analytically continued to a fractional value $n=-1/\nu$
\beq
(-1)^{-1/\nu } \, \Gamma \left( 2-\frac{1}{\nu } \right) \, \pi^{-1} 
\, \epsilon  \, r^{-1/\nu } \, \left(r^2+\epsilon ^2\right)^{\frac{1}{\nu }-1} 
\eeq
and then re-expressed as an overall factor in front of the original $\delta$ function
\beq
(-1)^{-1/\nu } r^{-1/\nu } \left(r^2+\epsilon ^2\right)^{\frac{1}{\nu }} 
\Gamma \left( 2-\frac{1}{\nu } \right) \; \delta (r) 
\; \rightarrow \; 
(-1)^{-1/\nu } r^{1/\nu } \, \Gamma \left( 2-\frac{1}{\nu } \right) \; \delta (r) 
\eeq
The procedure achieves in this case the desired result, namely the multiplication 
of the original delta function by a power $r^{1/\nu}$.
Note also that if one just takes the limit $\epsilon \rightarrow 0$ at
fixed $r$ one always gets zero for any $r>0$, so close to the origin
$r$ has to be sent to zero as well to get a non-trivial result.

The above considerations suggests that one should be able to write
for the point source at the origin 
\beq
T_{\mu \nu} (r) \; = \; {\rm diag} [ \, B(r) \, \rho(r), \, 0, \, 0, \, 0 \, ]
\label{eq:enmom_point}
\eeq
with (in spherical coordinates)
\beq
\rho \; = \; M \, \delta^{(3)} ({\bf x}) \; \rightarrow \; 
M \, {1 \over 4 \, \pi} \, 2 \, { \delta (r) \over r^2 }
\eeq
and the delta function defined as a suitable limit of a smooth function, and
with vanishing pressure $p(r)=0$.

Next we will consider as a warmup the trace equation
\beq
R \; = \; - 8 \, \pi \, G  (\Box) \, T 
\; \equiv \;  - \, 8 \, \pi \, G \, \left ( 1 \, + \, A( \Box ) \right ) \, T 
\; = \; + 8 \, \pi \, G \, \left ( 1 \, + \, A( \Box ) \right ) \, \rho 
\eeq
where we have used the fact that the point source at the origin
is described just by the density term.
One then computes the repeated action of the invariant d'Alembertian on $T$,
\beq
\Box \, (- 8 \pi G \, T ) \; = \; \Box (+ 8 \pi G \, \rho ) \; = \;
\frac{16\,G\,\pi \,\rho '}{r\,A} - 
  \frac{4\,G\,\pi \,A'\,\rho '}{{A}^2} + 
  \frac{4\,G\,\pi \,B'\,\rho '}{A\,B} + 
  \frac{8\,G\,\pi \,\rho ''}{A}
\eeq
or, using the explicit form for $\rho(r)$,
\beq
\Box \, (- 8 \pi G \, T ) \; = \;
- \frac{8 G M \epsilon 
\left(-6 A B-r A' B+r A B'\right)}
{\pi  ( r^2+\epsilon^2)^3 A^2 B} \; + \; O(\epsilon^3 )
\eeq
In view of the rapidly escalating complexity of the problem 
it seems sensible to expand around the Schwarzschild solution, and set therefore 
\beq
A(r)^{-1} \; = \; 1 \, - \, { 2 \, M \, G \over r} \, + \, { \sigma (r) \over r }
\label{eq:sigma}
\eeq
and
\beq
B(r) \; = \; 1 \, - \, { 2 \, M \, G \over r} \, + \, { \theta (r) \over r }
\label{eq:theta}
\eeq
where the correction to the standard solution are 
parametrized here by the two functions $\sigma(r)$ and $\theta(r)$, both
assumed to be ``small'', i.e. proportional to $a_0$ as in
Eqs.~(\ref{eq:abox0}) or (\ref{eq:abox}), with $a_0$ considered a small parameter.
Then for the scalar curvature, to lowest order in $\sigma$ and $\theta$, one has
\beq
{ G\,M\, \left ( \sigma - \theta  \right ) -  \left( 2\,G\,M - r \right) 
\,\left( G\,M\,\theta ' +  \left( 3\,G\,M - 2\,r \right) \,\sigma ' 
+ \left( 2\,G\,M - r \right) \,r\,\theta '' \right)
\over r^2 \, ( r \, - \, 2 \, M \, G)^2 }
\label{eq:scalar_a0_sp}
\eeq
To simplify the problem even further, we will assume that for 
$ 2 M G \ll r \ll \xi $
(the ``physical'' regime) one can set
\beq
\sigma (r) \; = \; - \, a_0 \, M \, G \, c_\sigma \, r^\alpha 
\label{eq:sigma_p}
\eeq
and
\beq
\theta (r) \; = \; - \, a_0 \, M \, G \,  c_\theta \, r^\beta 
\label{eq:theta_p}
\eeq
This assumption is in part justified by the form of the non-relativistic
correction of Eqs.~(\ref{eq:phi_small}).
Then for $\alpha=\beta$ (the equations seem impossible to satisfy if
$\alpha$ and $\beta$ are different) one obtains for the scalar curvature
\beq
R \; = \; 0 \, + \, \alpha \, \left ( 2 \, c_\sigma + ( \alpha - 1 ) \, c_\theta \right ) \,
a_0 \, M \, G \, r^{\alpha-3} \, + \, + O( a_0^2 )
\label{eq:scalar_a0}
\eeq
A first result can be obtained in the following way.
Since in the ordinary Einstein case one has
for a perfect fluid $R = - 8 \pi G T = + 8 \pi G ( \rho - 3 p)$, and 
since $\rho_m (r) \sim r^{{1\over \nu} -3}$
from Eq.~(\ref{eq:rho_vac}) in the same regime, one concludes that a
solution is given by
\beq
\alpha \; = \; { 1 \over \nu } 
\label{eq:alpha_nu}
\eeq
which also seems consistent with the Poisson equation result of Eq.~(\ref{eq:phi_small2}).

Next one needs the action of $\Box^n$ on the point source (here hidden in $T$).
To lowest order one has for the source term
\beq
8 \pi G \, \rho \; = \; M \, G \, { 4 \over \pi} \, { \epsilon \over r^2 ( r^2 + \epsilon^2 ) }
\eeq
The d'Alembertian then acts on the source term and gives
\beq
\Box^n \, ( 8 \pi G \, \rho ) \; \rightarrow \; { ( 2 n \, + \, 2)! \over 2 }
 \, M \, G \, { 4 \over \pi } \, r^{- 2 n - 3} \, 
\left ( {\epsilon \over r} \right ) 
\eeq
which can then be analytically continued to $n=- {1 \over 2 \nu}$, resulting in
\beq
(\Box)^{- {1 \over 2 \nu}} \, ( 8 \pi G \, \rho ) \; \rightarrow \; 
{ \Gamma ( 3 - { 1 \over \nu } ) \over 2 } 
\, M \, G \, { 4 \over \pi } \, r^{ {1 \over \nu} - 3} \, 
\left ( {\epsilon \over r} \right ) 
\eeq
After multiplying the above expression by $a_0$, consistency with l.h.s. of
the trace equation, Eq.~(\ref{eq:scalar_a0}), is achieved to lowest order in $a_0$
provided again $\alpha = 1 / \nu$.
To zeroth order in $a_0$, the correct solution is of course already built into
the structure of Eqs.~(\ref{eq:sigma}) and (\ref{eq:theta}).
Also note that setting $\epsilon =0$ would give nonsensical results, and in
particular the effective density would be zero for $r \neq 0$, in disagreement
with the result of Eq.~(\ref{eq:rho_vac}), $\rho_m (r) \sim r^{ {1 \over \nu} - 3} $
for small $r$.

The next step up would be the consideration of the action of $\Box$ on the point source,
as it appears in the full effective field equations of Eq.~(\ref{eq:field}), 
with again $T_{\mu\nu}$ described by Eq.~(\ref{eq:enmom_point}).
One perhaps surprising fact is the generation of an effective
pressure term by the action of $\Box$, suggesting that both terms should arise
in the correct description of vacuum polarization effects, 
\bea
\left( \Box \, T_{\mu\nu} \right )_{tt} \; & = & \; 
  - \frac{ \rho \,{B'}^2 }{2\,A\,B} + 
  \frac{2\,B\,\rho '}{r\,A} - 
  \frac{B\,A'\,\rho '}{2\,{A}^2} + 
  \frac{B'\,\rho '}{2\,A} + \frac{B\,\rho ''}{A}
\nonumber \\
\left( \Box \, T_{\mu\nu} \right )_{rr} \; & = & \; 
- \frac{ \rho \,{B'}^2 }{2\,{B}^2}
\label{eq:boxont}
\eea
and $ \left( \Box \, T_{\mu\nu} \right )_{\theta\theta} = 
\left( \Box \, T_{\mu\nu} \right )_{\varphi\varphi} = 0 $.
A similar effect, namely the generation of an effective vacuum pressure
term in the field equations by the action of $\Box$, is seen also
in the Robertson-Walker metric case \cite{effective}.

\subsection{Effective trace equation}

To check the overall consistency of the approach, consider the set of
effective field equations that
are obtained when the operator $(1+A(\Box))$ appearing in 
Eqs.~(\ref{eq:field}) and (\ref{eq:field_tr}) is moved over to the
gravitational side, as in Eq.~(\ref{eq:field_rev}).
Since the r.h.s of the field equations then vanishes for $r \neq 0$, one 
has apparently reduced the problem to one of finding vacuum solutions
of a modified, non-local field equation.

Let us first look at the simpler trace equation, valid again for $r \neq 0$.
If we denote by $\delta R$ the lowest order variation (that is, of order $a_0$)
in the scalar curvature over the ordinary vacuum solution $R=0$, then one 
needs to find solutions to
\beq
{ 1 \over 8 \pi G \, A( \Box ) } \; \delta R \; = \; 0
\label{eq:field_tr_rev1}
\eeq
On a generic scalar function $F(r)$ one has the following action of the
covariant d'Alembertian $\Box$ :
\beq
\Box \, F(r) \; = \; 
-\frac{A' F'}{2 A^2}+\frac{B' F'}
{2 A B}+\frac{2 F'}{r A}+\frac{F''}{A}
\eeq
The Ricci scalar is complicated enough, even in this simple case, and equal to
\beq
\frac{B'^2}{2 A B^2}+\frac{A' B'}{2 A^2 B}
-\frac{2 B'}{r A B}+\frac{2 A'}{r A^2}
-\frac{B''}{A B}-\frac{2}{r^2 A}+\frac{2}{r^2}
\eeq
The action of the covariant d'Alembertian on it produces the rather formidable
expression
\bea
&& \frac{5 B'^4}{2 A^2 B^4}+\frac{3 A' B'^3}{A^3 B^3}-
\frac{5 B'^3}{r A^2  B^3}+\frac{3 A'^2 B'^2}{A^4 B^2}
-\frac{6 A' B'^2}{r A^3 B^2} -\frac{5 A'' B'^2}{4 A^3 B^2}
-\frac{6 B'' B'^2}{A^2 B^3}+\frac{B'^2}{r^2 A^2 B^2}
\nonumber \\
&& +\frac{7 A'^3 B'}{2 A^5 B}-\frac{9 A'^2 B'}{r A^4 B}
-\frac{A' B'}{r^2 A^3 B}-\frac{13 A' A'' B'}{4 A^4 B}
+\frac{4 A'' B'}{r A^3 B}-\frac{23 A' B'' B'}{4 A^3 B^2}
+\frac{9 B'' B'}{r A^2 B^2}+\frac{A^{(3)} B'}{2 A^3 B}
\nonumber \\
&& +\frac{5 B^{(3)} B'}{2 A^2 B^2}-\frac{2 B'}{r^3 A B}
+\frac{2 B'}{r^3 A^2 B}+\frac{14 A'^3}{r A^5}
-\frac{4 A'^2}{r^2 A^4}+\frac{2 B''^2}{A^2 B^2}
+\frac{2 A'}{r^3 A^2}-\frac{6 A'}{r^3 A^3}
+\frac{2 A''}{r^2 A^3}
\nonumber \\
&& -\frac{19 A'^2 B''}{4 A^4 B}
+\frac{8 A' B''}{r A^3 B}+\frac{2 A'' B''}{A^3 B}
+\frac{2 A^{(3)}}{r A^3}+\frac{3 A' B^{(3)}}{A^3 B}
-\frac{4 B^{(3)}}{r A^2 B}-\frac{B^{(4)}}{A^2 B}
+\frac{4}{r^4 A}-\frac{4}{r^4 A^2}
\eea
To lowest order in the functions $\sigma$ and $\theta$, from Eq.~(\ref{eq:scalar_a0_sp}),
the scalar curvature is given by
\beq
\left ( G M \left ( \sigma - \theta  \right ) -  \left( 2 G M - r \right) 
\left( G M \theta ' +  \left( 3 G M - 2 r \right) \sigma ' 
+ \left( 2 G M - r \right) r \theta '' \right) \right )
/ \left ( r^2 ( r - 2 M G)^2 \right )
\eeq
After having the d'Alembertian $\Box$ act on this expression, one obtains the still formidable
(here again to lowest order in $\sigma$ and $\theta$) result
\bea
&& ( \theta ^{(4)} r^7+2 \sigma ^{(3)} r^6-8 G M \theta
   ^{(4)} r^6-4 \sigma '' r^5+G M \theta ^{(3)} r^5-15 G M \sigma ^{(3)} r^5+24
   G^2 M^2 \theta ^{(4)} r^5+4 \sigma ' r^4
\nonumber \\
&& +3 G M \theta '' r^4 +31 G M 
   \sigma '' r^4-6 G^2 M^2 \theta ^{(3)} r^4+42 G^2 M^2 \sigma ^{(3)} r^4-32 G^3
   M^3 \theta ^{(4)} r^4-12 G M \theta ' r^3
\nonumber \\
&& -28 G M \sigma ' r^3-14 G^2 M^2
   \theta '' r^3-94 G^2 M^2 \sigma '' r^3+12 G^3 M^3 \theta ^{(3)} r^3
-52 G^3 M^3 \sigma ^{(3)} r^3+16 G^4 M^4 \theta ^{(4)} r^3
\nonumber \\
&& +12 G M \theta 
   r^2-12 G M \sigma  r^2+48 G^2 M^2 \theta ' r^2+96 G^2 M^2 \sigma ' r^2+20 G^3
   M^3 \theta '' r^2+132 G^3 M^3 \sigma '' r^2
\nonumber \\
&& -8 G^4 M^4 \theta ^{(3)} r^2 
 +24 G^4 M^4 \sigma ^{(3)} r^2-24 G^2 M^2 \theta  r+24 G^2 M^2 \sigma  r-64
   G^3 M^3 \theta ' r-160 G^3 M^3 \sigma ' r
\nonumber \\
&& -8 G^4 M^4 \theta '' r-72 G^4
   M^4 \sigma '' r+16 G^3 M^3 \theta -16 G^3 M^3 \sigma +32 G^4 M^4 \theta '
  +96 G^4 M^4 \sigma '  ) / 
\nonumber \\
&& \;\;\;\;\;\; \left ( (2 G M-r)^3 r^5 \right )
\eea
Higher powers of the d'Alembertian $\Box$ then lead to even more
complicated expressions, with increasingly higher derivatives of 
$\sigma(r)$ and $\theta(r)$.
Assuming a power law correction, as in Eqs.~(\ref{eq:sigma_p}) 
and (\ref{eq:theta_p}), with $\alpha=\beta$, as in Eq.~(\ref{eq:scalar_a0}),
\beq
R \; = \; a_0 \, M \, G \, \alpha \, 
\left( \, 2 \, c_\sigma + c_\theta \, ( \alpha -1 )  \, \right) 
\, r^{\alpha-3} \, + \, O (a_0^2 )
\eeq
and then
\bea
\Box \,\; R \; & \rightarrow & \; a_0 \, M \, G \, r^{\alpha-5} \, 
\left( \, 2 \, c_\sigma + c_\theta \, ( \alpha - 1 ) \, \right ) \, 
\alpha \, (  \alpha - 2 ) \, ( \alpha -3 ) 
\nonumber \\
\Box^2 \, R \; & \rightarrow & \; a_0 \, M \, G \, r^{\alpha - 7} \, 
\left( \, 2 \, c_\sigma + c_\theta \, ( \alpha - 1 )  \, \right) \, 
\alpha \, ( \alpha  - 2 ) \, ( \alpha - 3 ) \, ( \alpha - 4 )
\, ( \alpha - 5 )
\eea
and so on, and for general $n \rightarrow + { 1 \over 2 \nu} $
\beq
\Box^n \, R \; \rightarrow \;
a_0 \, M \, G \, \left ( \, 2 \, c_\sigma + c_\theta \, ( \alpha - 1 ) \, \right) \, 
{ \alpha \, \Gamma ( 2 + {1 \over \nu } - \alpha ) \over \Gamma ( 2 - \alpha ) }
\; r^{\alpha - {1 \over \nu } - 3 } 
\eeq
Therefore the only possible power solution for $r \gg M G $
is $\alpha= 0, \, 2 \dots {1 \over \nu} + 1$, with $c_\sigma$ and $c_\theta$
unconstrained to this order.

\subsection{Full effective field equations}

Next we examine the full effective field equations (as opposed
to just their trace part) as in Eq.~(\ref{eq:field_rev}) with $\lambda=0$,
\beq
{ 1 \over 8 \pi G \left ( 1 \, + \, A( \Box ) \right ) } \,  
\left( R_{\mu\nu} - \half \, g_{\mu\nu} \, R \right) \; = \; 0
\label{eq:field_vac}
\eeq
valid for $r \neq 0$.
If one denotes by 
$\delta G_{\mu\nu} \equiv \delta \left( R_{\mu\nu} - \half \, g_{\mu\nu} \, R \right)$
the lowest order variation (that is, of order $a_0$)
in the Einstein tensor over the ordinary vacuum solution $G_{\mu\nu}=0$, then one has
\beq
{ 1 \over 8 \pi G \, A( \Box ) } \;\,
\delta \left( R_{\mu\nu} - \half \, g_{\mu\nu} \, R \right) \; = \; 0
\label{eq:field_vac1}
\eeq
again for $r \neq 0$.
Here the covariant d'Alembertian operator
\beq
\Box \; = \; g^{\mu\nu} \, \nabla_\mu \nabla_\nu 
\eeq
acts on a second rank tensor,
\bea
\nabla_{\nu} T_{\alpha\beta} \, = \, \partial_\nu T_{\alpha\beta} 
- \Gamma_{\alpha\nu}^{\lambda} T_{\lambda\beta} 
- \Gamma_{\beta\nu}^{\lambda} T_{\alpha\lambda} \, \equiv \, I_{\nu\alpha\beta}
\nonumber
\eea
\beq 
\nabla_{\mu} \left( \nabla_{\nu} T_{\alpha\beta} \right)
= \, \partial_\mu I_{\nu\alpha\beta} 
- \Gamma_{\nu\mu}^{\lambda} I_{\lambda\alpha\beta} 
- \Gamma_{\alpha\mu}^{\lambda} I_{\nu\lambda\beta} 
- \Gamma_{\beta\mu}^{\lambda} I_{\nu\alpha\lambda} 
\eeq
and would thus seem to require the calculation of as many as 1920 terms,
of which many fortunately vanish by symmetry.
In the static isotropic case the components of the Einstein tensor are given by 

\bea
G_{tt} \; & = & \; \frac{A' B}{r A^2}-\frac{B}{r^2 A}+\frac{B}{r^2}
\nonumber \\
G_{rr} \; & = & \; -\frac{A}{r^2}+\frac{B'}{r B}+\frac{1}{r^2}
\nonumber \\
G_{\theta\theta} \; & = & \; -\frac{B'^2 r^2}{4 A B^2}-
\frac{A' B' r^2}{4 A^2 B}
+\frac{B'' r^2}{2 A B}
-\frac{A' r}{2 A^2}+\frac{B' r}{2 A B}
\nonumber \\
G_{\varphi\varphi} \; & = & \; \sin^2 \theta \, G_{\theta\theta} 
\eea
Using again the expansion of Eqs.~(\ref{eq:sigma}) and (\ref{eq:theta}) one obtains for the
$tt$, $rr$ and $\theta\theta$ components of the
Einstein tensor, to lowest order in $\sigma$ and $\theta$,
\bea
G_{tt} \; & \simeq & \; 
\frac{ \left( 2\,G\,M - r \right) \,\sigma '}{r^3}
\nonumber \\
G_{rr} \; & \simeq & \; 
\frac{-\theta  + \sigma  
+ \left( -2\,G\,M + r \right) \,\theta ' }
{r\,{\left( -2\,G\,M + r \right) }^2}
\nonumber \\
G_{\theta\theta} \; & \simeq & \; 
\frac{ \left( -\,G\,M + \,r \right) \, \left( \theta  
- \sigma  + \left( 2\,G\,M - r \right) \, \left( \theta ' 
- \sigma ' \right)  \right) 
+  \,r\,{\left( -2\,G\,M + r \right) }^2\,\theta '' }
{2\, {\left( -2\,G\,M + r \right) }^2}
\nonumber \\
G_{\varphi\varphi} \; & = & \; \sin^2 \theta \, G_{\theta\theta} 
\eea
After acting with $\Box$ on this expression one finds a rather
complicated result.
Here we will only list $ (\Box G)_{tt} $:  
\bea
&& \frac{6 B A'^3}{r A^5}+\frac{2 B A'^2}{r^2 A^4}
-\frac{4 B' A'^2}{r A^4}-\frac{2 B A'}{r^3 A^3}
-\frac{6 B A'' A'}{r A^4}+\frac{B'' A'}{r A^3}
+\frac{6 B}{r^4 A}
\nonumber \\
&& -\frac{6 B}{r^4 A^2}
-\frac{4 B'}{r^3 A}+\frac{4 B'}{r^3 A^2}
-\frac{B A''}{r^2 A^3}+\frac{2 B' A''}{r A^3}
+\frac{B''}{r^2 A}-\frac{B''}{r^2 A^2}
+\frac{B A^{(3)}}{r A^3}
\eea
To lowest order in $\sigma(r)$ and $\theta(r)$ one finds the slightly simpler expressions
\bea
(\Box G)_{tt} \; = &&
        - ( -2\,G^2M^2\,\theta  + 
        2\,G^2M^2\,\sigma  - 
        4\,G^3M^3\,\theta ' + 
        2\,G^2M^2\,r\,\theta ' - 
        28\,G^3M^3\,\sigma ' + 38\,G^2M^2\,r\,\sigma '
\nonumber \\
      &&  - 16\,GM\,r^2\,\sigma ' + 2\,r^3\,\sigma '
        + 24\,G^3M^3\,r\,\sigma '' - 
        32\,G^2M^2\,r^2\,\sigma '' + 
        14\,GM\,r^3\,\sigma '' 
\nonumber \\
      && - 2\,r^4\,\sigma '' - 
        8\,G^3M^3\,r^2\,\sigma ^{(3)} + 
        12\,G^2M^2\,r^3\,\sigma ^{(3)} - 
        6\,GM\,r^4\,\sigma ^{(3)} + r^5\,\sigma ^{(3)}
        ) /  
\nonumber \\
     && \left ( r^6 \, \left ( -2\,GM + r \right ) \right )
\eea

\bea
(\Box G)_{rr} \; = &&
      ( 6\,G^2M^2\,\theta  - 
      4\,GM\,r\,\theta  - 6\,G^2M^2\,\sigma  + 
      4\,GM\,r\,\sigma  + 12\,G^3M^3\,\theta ' - 
      14\,G^2M^2\,r\,\theta ' 
\nonumber \\
   && + 4\,GM\,r^2\,\theta ' - 
      12\,G^3M^3\,\sigma ' + 6\,G^2M^2\,r\,\sigma '
      + 4\,GM\,r^2\,\sigma ' - 2\,r^3\,\sigma ' + 
      8\,G^3M^3\,r\,\theta ''
\nonumber \\
   && - 12\,G^2M^2\,r^2\,\theta '' + 
      6\,GM\,r^3\,\theta '' - 
      r^4\,\theta '' + 4\,G^2M^2\,r^2\,\sigma '' - 
      4\,GM\,r^3\,\sigma '' + r^4\,\sigma '' 
\nonumber \\
   && -8\,G^3M^3\,r^2\,\theta ^{(3)} + 
      12\,G^2M^2\,r^3\,\theta ^{(3)} - 
      6\,GM\,r^4\,\theta ^{(3)} + 
      r^5\,\theta ^{(3)} ) / 
\nonumber \\
   && \left ( r^4\, { \left ( -2\,GM + r \right) }^3 \right )
\eea

\bea
(\Box G)_{\theta\theta} \; = &&
      ( 24\,G^3M^3\,\theta  - 
      36\,G^2M^2\,r\,\theta  + 
      16\,GM\,r^2\,\theta  - 
      24\,G^3M^3\,\sigma  + 36\,G^2M^2\,r\,\sigma  - 
      16\,GM\,r^2\,\sigma
\nonumber \\
   && + 48\,G^4M^4\,\theta ' - 
      96\,G^3M^3\,r\,\theta ' 
      +68\,G^2M^2\,r^2\,\theta ' - 
      16\,GM\,r^3\,\theta ' + 
      16\,G^4M^4\,\sigma ' - 32\,G^3M^3\,r\,\sigma '
\nonumber \\
   && +  28\,G^2M^2\,r^2\,\sigma ' - 
      16\,GM\,r^3\,\sigma ' + 4\,r^4\,\sigma ' + 
      8\,G^4M^4\,r\,\theta '' 
      -12\,G^3M^3\,r^2\,\theta '' + 
      10\,G^2M^2\,r^3\,\theta '' 
\nonumber \\
   &&  - 5\,GM\,r^4\,\theta '' + 
      r^5\,\theta '' - 24\,G^4M^4\,r\,\sigma '' + 
      52\,G^3M^3\,r^2\,\sigma '' - 
      46\,G^2M^2\,r^3\,\sigma '' 
      +19\,GM\,r^4\,\sigma '' 
\nonumber \\
   &&  - 3\,r^5\,\sigma '' - 
      24\,G^4M^4\,r^2\,\theta ^{(3)} + 
      44\,G^3M^3\,r^3\,\theta ^{(3)} - 
      30\,G^2M^2\,r^4\,\theta ^{(3)} + 
      9\,GM\,r^5\,\theta ^{(3)} - 
      r^6\,\theta ^{(3)}
\nonumber \\
   && +  8\,G^4M^4\,r^2\,\sigma ^{(3)} - 
      20\,G^3M^3\,r^3\,\sigma ^{(3)} 
      + 18\,G^2M^2\,r^4\,\sigma ^{(3)} - 
      7\,GM\,r^5\,\sigma ^{(3)} + r^6\,\sigma ^{(3)}
\nonumber \\
   && + 16\,G^4M^4\,r^3\,\theta ^{(4)} - 
      32\,G^3M^3\,r^4\,\theta ^{(4)}
     + 24\,G^2M^2\,r^5\,\theta ^{(4)} - 
      8\,GM\,r^6\,\theta ^{(4)} + 
      r^7\,\theta ^{(4)} ) / 
\nonumber \\
   &&  \left ( 2\,r^3\, {\left( -2\,GM + r \right) }^3 \right )
\eea
with the $\varphi\varphi$ component proportional to the 
$\theta\theta$ component.
If one again assumes that the corrections are given by a power, as in
Eqs.~(\ref{eq:sigma_p}) and (\ref{eq:theta_p}), with $\alpha=\beta$,
then one has to zeroth order
\bea
G_{tt} \; & = & \; 
a_0 \, M \, G \, c_\sigma \, \alpha \, r^{\alpha -3} 
\nonumber \\
G_{rr} \; & = & \; 
- \, a_0 \, M \, G \, (c_\sigma+c_\theta (\alpha -1)) \, r^{\alpha -3} 
\nonumber \\
G_{\theta\theta} \; & = & \; 
- \frac{1}{2} \, a_0 \, M \, G \, (c_\sigma+c_\theta (\alpha -1))
\, (\alpha -1) \, r^{\alpha -1} 
\eea
with the $\varphi\varphi$ component again proportional to the 
$\theta\theta$ component.
Applying $\Box$ on the above Einstein tensor one then gets
\bea
\left ( \Box \, G \right )_{tt} \; & = & \; 
a_0 \, G\,M\, c_\sigma \, \alpha  ( \alpha - 2) \, ( \alpha - 3) \, r^{\alpha - 5}
\nonumber \\
\left ( \Box \, G \right )_{rr} \; & = & \; 
-  \, a_0 \,G \,M \, \left( c_\sigma + c_\theta \, ( \alpha -1 )  \right) \,
\alpha \, ( \alpha - 3 ) \, r^{\alpha - 5 }
\nonumber \\
\left ( \Box \, G \right )_{\theta\theta} \; & = & \; 
- \, \frac{1}{2} \, a_0 \,G \,M \, \left( c_\sigma + c_\theta \, ( \alpha -1 )  \right) \,
\alpha \, ( \alpha - 3 )^2 \, r^{\alpha - 3 }
\eea
again with the $\varphi\varphi$ component proportional to the 
$\theta\theta$ component.
Applying $\Box$ again one obtains
\bea
\left ( \Box^2 \, G \right )_{tt} \; & = & \; 
a_0 \,G\,M\, c_\sigma \, \alpha  \, ( \alpha - 2) \, ( \alpha - 3) ( \alpha - 4)
\, ( \alpha -5) \, r^{\alpha - 7}
\nonumber \\
\left ( \Box^2 \, G \right )_{rr} \; & = & \; 
-  a_0 \,G \,M \, \left( c_\sigma + c_\theta \, ( \alpha -1 )  \right) \,
\alpha \, ( \alpha - 2 ) \, ( \alpha - 3 ) \, ( \alpha - 5 ) 
\, r^{\alpha - 7 }
\nonumber \\
\left ( \Box^2 \, G \right )_{\theta\theta} \; & = & \; 
- \frac{1}{2} \, a_0 \,G \,M \, \left( c_\sigma + c_\theta \, ( \alpha -1 )  \right) \,
\alpha \, ( \alpha - 2 ) \, ( \alpha - 3 ) \, ( \alpha - 5 )^2
\, r^{\alpha - 5 }
\eea
and so on, and for general $n \rightarrow + { 1 \over 2 \nu} $
\bea
\left ( \Box^n \, G \right )_{tt} \; & \rightarrow & \; a_0 \,G\,M\, c_\sigma \, 
\frac{\Gamma (2 + {1 \over \nu} - \alpha )}{\left( \alpha -1 \right) \, \Gamma (-\alpha )}
\, r^{\alpha - 3 - {1 \over \nu}}
\nonumber \\
\left ( \Box^n \, G \right )_{rr} \; & \rightarrow & \; -  a_0 \,G \,M \, 
\left( c_\sigma + c_\theta \, ( \alpha -1 )  \right) \,
\frac{ \Gamma (2 + {1 \over \nu} - \alpha )}{\left( \alpha -1 \right) \,
\left( \alpha - {1 \over \nu} \right) \, \Gamma(-\alpha )}
\, r^{\alpha - 3 - {1 \over \nu}  }
\nonumber \\
\left ( \Box^n \, G \right )_{\theta\theta} \; & \rightarrow & \; 
- \frac{1}{2} \, a_0 \,G \,M \, \left( c_\sigma + c_\theta \, ( \alpha -1 )  \right) \,
\frac{\left( \alpha - 1 - {1 \over \nu} \right) \, \Gamma (2 + {1 \over \nu} - \alpha )}
{\left( \alpha -1 \right) \, \left( \alpha - {1 \over \nu} \right) \, \Gamma(-\alpha )}
\, r^{\alpha - 1 - {1 \over \nu} }
\eea
Inspection of the above results reveals a common factor $1/\Gamma(-\alpha)$,
which would allow only integer powers $\alpha=0,1,2 \dots$, but the additional
factor of $1/(\alpha-1)$ excludes $\alpha=1$ from being a solution.
Even for $\alpha$ close to $1 / \nu$ (as expected on the basis of the non-relativistic
expression of Eq.~(\ref{eq:phi_small2}), as well as from
Eq.~(\ref{eq:alpha_nu})) $\nu \sim 1/\alpha - \epsilon$ 
only integer values $\alpha=2,3,4 \dots $ are allowed.
For the covariant divergences $ \nabla^\mu ( \Box^n \,  G )_{\mu\nu} $ one
has
\beq
\nabla^\mu \left ( \delta \, G \right )_{\mu r} \; = \; 0 
\eeq
and, at the next order,
\beq
\nabla^\mu \left ( \Box \, G \right )_{\mu r} \; = \; 
2 \, a_0 \, G^2 \,M^2 \, \alpha  \, ( \alpha - 3) \, 
\left ( c_\sigma \, + \, c_\theta ( 1 - \alpha ) \right ) \, r^{\alpha - 7}
\eeq
with the other components vanishing identically, and 
\beq
\nabla^\mu \left ( \Box^2 \, G \right )_{\mu r} \; = \; 
4 \, a_0 \, G^2 \,M^2 \, \alpha  \, ( \alpha - 3) \, ( \alpha - 5)^2 \, 
\left ( c_\sigma \, + \, c_\theta ( 1 - \alpha ) \right ) \, r^{\alpha - 9}
\eeq
again with the other components of the divergence vanishing identically.

In general the problem of finding a complete general solution to the effective
field equations by this method lies in the difficulty of computing
arbitrarily high powers of $\Box$ on general functions such as
$\sigma(r)$ and $\theta(r)$, which eventually involve a large number
of derivatives. 
Assuming for these functions a power law dependence on $r$ simplifies
the problem considerably, but also restricts the kind of solutions
that one is likely to find.
More specifically, if the solution involves (say for small $r$, but still 
with
$r \gg 2 M G$) a term of the type $r^\alpha \ln m r$, as in
Eqs.~(\ref{eq:phi_small1}), (\ref{eq:a_small_r3}) and (\ref{eq:b_small_r3})
for $\nu \rightarrow 1/3$, then this method will have to be dealt
with very carefully. 
This is presumably the reason why in some of the $\Gamma$-function
coefficients encountered here one finds a power solution (in fact $\alpha=3$)
for $\nu$ close to a third, but one gets indeterminate expression
if one sets exactly $\alpha=1/\nu=3$.

\vskip 30pt
\newsection{The Quantum Vacuum as a Fluid}
\hspace*{\parindent}

The discussion of Sec. 3 suggests that the quantum correction due
to the running of $G$ can be described, at least in the non-relativistic
limit of Eq.~(\ref{eq:grun_m}) as applied to Poisson's equation,
in terms of a vacuum energy density $\rho_m(r)$, distributed around
the static source of strength $M$ in accordance with the result of 
Eqs.~(\ref{eq:rho_vac}) and Eq.~(\ref{eq:rho_vac2}).
These expressions, in turn, can be obtained by Fourier transforming back 
to real space the original result for $G(k^2)$ of Eq.~(\ref{eq:grun_m}).

Furthermore, it was shown in Sec. 4 (and was discovered in \cite{effective}
as well, see for example Eq.~(\ref{eq:p_eff}) later in this paper)
that a manifestly covariant implementation of the running of $G$, 
via the $G(\Box)$ given in Eqs.~(\ref{eq:gbox}) and (\ref{eq:abox}),
will induce a non-vanishing effective pressure term.
This result can be seen clearly, in the case of the static isotropic metric,
for example from the result of Eq.~(\ref{eq:boxont}).

We will therefore, in this section, consider a relativistic perfect fluid, 
with energy-momentum tensor
\beq
T_{\mu\nu} \;  = \; 
\left [ \, p \, + \, \rho \, \right ] \, u_\mu \, u_\nu
\, + \, g_{\mu \nu} \, p 
\label{eq:perfect}
\eeq
which in the static isotropic case reduces to Eq.~(\ref{eq:perfect_si}),
\beq
T_{\mu \nu} \; = \; {\rm diag} \,
[ \, B(r) \, \rho(r), \, A(r) \, p(r) ,
\, r^2 \, p(r), \, r^2 \, \sin^2 \theta \, p(r) \, ]
% \label{eq:perfect_si}
\eeq
and gives a trace $T = 3 \, p - \rho $.

The $tt$, $rr$ and $\theta\theta$ components of the field equations then read 
\beq
-\lambda  B(r)+\frac{A'(r) B(r)}{r A(r)^2}
-\frac{B(r)}{r^2 A(r)}+\frac{B(r)}{r^2} \; = \; 8 \pi G  B(r) \rho (r)
\eeq

\beq
\lambda  A(r)-\frac{A(r)}{r^2}+\frac{B'(r)}{r B(r)}+\frac{1}{r^2}
\; = \; 8 \pi G  A(r) p (r)
\eeq

\beq
-\frac{B'(r)^2 r^2}{4 A(r) B(r)^2}+\lambda  r^2
-\frac{A'(r) B'(r) r^2}{4 A(r)^2 B(r)}
+\frac{B''(r) r^2}{2 A(r) B(r)}
-\frac{A'(r) r}{2 A(r)^2}
+\frac{B'(r) r}{2 A(r) B(r)} \; = \; 8 G \pi  r^2 p(r)
\eeq
with the $\varphi\varphi$ component equal to $\sin^2 \theta$ times the
$\theta\theta$ component.

Energy conservation $\nabla^{\mu} \, T_{\mu\nu} =0 $ implies
\beq
\left [ \, p(r)+\rho (r) \, \right ] \frac{ B'(r)}{2 B(r)} + p'(r) \; = \; 0
\eeq
and forces a definite relationship between $B(r)$, $\rho(r)$ and $p(r)$.
The three field equations and the energy conservation equation
are, as usual, not independent, because of the Bianchi identity.

It seems reasonable to attempt to solve the above equations (usually
considered in the context of relativistic stellar structure \cite{mtw})
with the density $\rho(r)$ given by the $\rho_m (r)$ of
Eqs.~(\ref{eq:rho_vac}), (\ref{eq:rho_vac1}) and (\ref{eq:rho_vac2}).

This of course raises the question of how the relativistic pressure $p(r)$
should be chosen, an issue that the non-relativistic calculation
did not have to address.
We will argue below that covariant energy conservation completely
determines the pressure in the static case, leading to
consistent equations and solutions (note that in particular it would
not be consistent to take $p(r)=0$). 

Since the function $B(r)$ drops out of the $tt$ field equation,
the latter can be integrated immediately, giving

\beq
A(r)^{-1} \; = \; 1 \, + \, { c_1 \over r } \, - \, { \lambda \over 3 } \, r^2 
\, - \, { 8 \pi G \over r } \, \int_0^r d x \, x^2 \, \rho (x)
\label{eq:ar}
\eeq

which suggests the introduction of a function $m(r)$
\beq
m(r) \; \equiv \; 4 \, \pi \, \int_0^r d x \, x^2 \, \rho (x)
\label{eq:mr}
\eeq
It also seems natural in our case to identify $c_1 = - 2 M G $,
which of course corresponds to the solution with $a_0=0$ ($p=\rho=0$)
(equivalently, the point source at the origin of strength $M$ 
could be included as an additional $\delta$-function contribution to
$\rho(r)$).

Next, the $rr$ field equation can be solved for $B(r)$,

\beq
B(r) \; = \; \exp \left \{ c_2 \, - \, \int_{r_0}^r \, d y \, 
\frac{1 + A(y) \left( \lambda \, y^2 - 8 \pi G \, y^2 \, p(y) - 1 \right)}{y} 
\right \}
\label{eq:br}
\eeq
with the constant $c_2$ again determined by the requirement that the
above expression for $B(r)$ reduce to the standard Schwarzschild solution
for $a_0=0$ ($p=\rho=0$), giving 
$c_2=\ln (1 - 2 M G / r_0 - \lambda r_0^2 / 3 )$.
The last task left therefore is the determination of the pressure $p(r)$.

Using the $rr$ field equation, $B'(r)/B(r)$ can be expressed in
term of $A(r)$ in the energy conservation equation, which results in
\beq
2 r \, p'(r) - \left [ 1 + A(r) \left (\lambda \,  r^2 - 8 \pi G \, r^2 \, p(r)
- 1 \right) \right ] ( \, p(r)+\rho (r)\, )  \; = \; 0
\label{eq:pr}
\eeq

Inserting the explicit expression for $A(r)$, from Eq.~(\ref{eq:ar}), one obtains

\beq
p'(r) + \frac{\left ( 8 \pi G \, r^3 \, p(r) \, + \, 2 M G \, 
- \, { 2 \over 3} \lambda r^3 \, 
+ \, 8 \pi G \, \int_{r_0}^r dx \, x^2 \rho (x) \right ) \, (p(r)+\rho (r))}
{2 \, r \left( r \, - \, 2 M G \, - \, {\lambda \over 3}\, r^3  
- 8 \pi G \, \int_0^r dx \, x^2 \, \rho (x) \right) } \; = \; 0
\label{eq:pres1}
\eeq
which is usually referred to as the equation of hydrostatic equilibrium.
From now on we will focus only the case $\lambda=0$.
Then
\beq
p'(r) + \frac{\left( 8 \pi G \, r^3 \, p(r) \, + \, 2 M G \, 
+ \, 8 \pi G \int_0^r dx \, x^2 \rho (x) \right) (p(r)+\rho (r))}
{2 \, r \left( r \, - \, 2 M G \, - \, 
8 \pi G \, \int_0^r dx \, x^2 \, \rho (x) \right) } \; = \; 0
\label{eq:pres2}
\eeq

The last equation, a non-linear differential equation for $p(r)$, can
be solved to give the desired solution $p(r)$, which then,
by equation Eq.~(\ref{eq:br}), determines the remaining function $B(r)$.

In our case though it will be sufficient to solve the above equation
for small $a_0$, where $a_0$
(see Eq.~(\ref{eq:grun_m}) and Eq.~(\ref{eq:rho_vac}))
is the dimensionless parameter which,
when set to zero, makes the solution revert back to the classical one.

It will also be convenient to pull out of $A(r)$ and $B(r)$ the
Schwarzschild solution part, by introducing the small corrections
$\sigma(r)$ and $\theta(r)$ (already defined before in Eqs.~(\ref{eq:sigma})
and (\ref{eq:theta})), both of which are expected to be proportional
to the parameter $a_0$.
One has
\beq
\sigma (r) \; = \; - 8 \pi G \, \int_0^r dx \, x^2 \, \rho (x) 
\; \equiv \; - \, 2 \, m(r) \, G
\label{eq:sigma_sol}
\eeq
and
\beq
\theta (r) \; = \; 
\exp \left \{ c_2 + \int_{r_0}^r dy \, { 1 + 8 \pi G \, y^2 \, p(y) \over
y - 2 M G - 8 \pi G \int_0^y dx \, x^2 \, \rho (x) } \right \}
\, + \, 2 M G \, - \, r 
\label{eq:theta_sol}
\eeq
Again, the integration constant $c_2 $ needs to be chosen here so that the normal
Schwarzschild solution is recovered for $p=\rho=0$.  

To order $a_0$ the resulting equation for $p(r)$, from Eq.~(\ref{eq:pres2}), is
\beq
\frac{M G \, ( p(r) \, + \, \rho (r))} {r \, ( r \, - \, 2 M G )} 
\, + \, p'(r) \; \simeq \; 0
\label{eq:pres0}
\eeq
Note that in regions where $p(r)$ is slowly varying, $p'(r) \sim 0$,
one has $p \simeq -\rho $, i.e. the fluid contribution is
acting like a cosmological constant term with 
$ \sigma (r) \sim \theta (r) \sim - (\rho /3) \, r^3$.

The last differential equation can then be solved for $p(r)$,
\beq
p_m (r) \; = \; {1 \over \sqrt{1 - \frac{2 M G}{r} } } \,
\left ( c_3 - \int_{r_0}^r dz \, 
\frac{M G \rho (z)} {z^2 \, \sqrt{1-\frac{2 M G}{z}} }
\right )
\label{eq:pres0_sol}
\eeq
where the constant of integration has to be chosen so that
when $\rho(r)=0$ (no quantum correction) one has $p(r)=0$ as well.
Because of the singularity in the integrand at $r= 2 M G$, we will
take the lower limit in the integral to be $r_0 = 2 M G + \epsilon$,
with $\epsilon \rightarrow 0$.

To proceed further, one needs the explicit form for $\rho_m(r)$, which
was given in Eqs.~(\ref{eq:rho_vac}), (\ref{eq:rho_vac1}) and (\ref{eq:rho_vac2}).
The required integrands involve for general $\nu$ the modified Bessel function $K_n(x)$,
and can be therefore a bit complicated.
But in some special cases the general form of the density $\rho_m$ of Eq.~(\ref{eq:rho_vac})
\beq
\rho_m (r) \; = \; {1 \over 8 \pi } \, c_{\nu} \, a_0 \, M \, m^3 \,
( m \, r )^{ - {1 \over 2} (3 - {1 \over \nu}) }  
\, K_{ {1 \over 2} ( 3 - {1 \over \nu} ) } ( m \, r ) 
\eeq
reduces to a relatively simple expression, which we will list here.
For $\nu=1$ one has
\beq
\rho_m (r) \; = \; {1 \over 2 \pi^2 } \, a_0 \, M \, m^3 \, {1 \over m \, r}
\, K_1 ( m \, r ) 
\eeq
whereas for $\nu=1/2$ one has
\beq
\rho_m (r) \; = \; {1 \over 4 \pi } \, a_0 \, M \, m^3 \, {1 \over m \, r}
\, e^{- \, m \, r} 
\eeq
and for $\nu=1/3$ 
\beq
\rho_m (r) \; = \; {1 \over 2 \pi^2 } \, a_0 \, M \, m^3 \, 
\, K_0 ( m \, r ) 
\label{eq:rho_vac_3}
\eeq
and finally for $\nu=1/4$
\beq
\rho_m (r) \; = \; {1 \over 8 \pi } \, a_0 \, M \, m^3 \, 
\, e^{- \, m \, r} 
\eeq
Note that $\rho_m(r)$ diverges at small $r$ for $\nu \geq 1/3$

Here we will limit our investigation to the small $r$ ($ m r \ll 1$)
and large $r$ ($m r \gg 1$) behavior.
Since $m=1/\xi$ is very small, the first limit appears to be of greater physical
interest.

\subsection{Small r limit}

For small $r$ the density $\rho_m(r)$ has the following behavior
(see Eq.~(\ref{eq:rho_vac})),
\beq
\rho_m (r) \; \mathrel{\mathop\sim_{ r \, \rightarrow \, 0 }} \; 
A_0 \; r^{ {1 \over \nu} - 3 } 
\label{eq:rho_small}
\eeq
for $\nu > 1/3$, with 
\beq
A_0 \; \equiv \; { c_k \, c_\nu \over 8 \pi } \, a_0 \, M \, m^{1 \over \nu} 
\; = \; 
{ \vert \sec \left ( { \pi \over 2 \nu } \right ) \vert 
\over 4 \pi \, \Gamma \left ( {1 \over \nu} -1 \right ) } \, a_0 \, M \, m^{1 \over \nu}
\eeq
where the dimensionless positive constant $c_k$ 
is determined from the small $x$ behavior of the modified Bessel function
$K_n(x)$,
\beq
x^{\frac{1}{2} \left( \frac{1}{\nu }-3\right)} 
\, K_{\frac{1}{2} \left(3-\frac{1}{\nu }\right)} (x)
\; \mathrel{\mathop\sim_{ x \, \rightarrow \, 0 }} \; 
- \frac{ 2^{\frac{1}{2} \left(1-\frac{1}{\nu }\right)} 
\, \pi \, \sec \left(\frac{\pi }{2 \nu }\right) }
{ \Gamma \left( \frac{1}{2 \nu }-\frac{1}{2} \right) } 
\, x^{\frac{1}{\nu }-3}  \; \equiv \; c_k \, x^{\frac{1}{\nu }-3}
\label{eq:bessel}
\eeq
valid for $\nu > 1/3$, and $c_\nu$ is given in Eq.~(\ref{eq:rho_vac1}). 
For $\nu < 1/3$ $\rho_m (r) \, \sim \, {\rm const.} \, a_0 \, M \, m^3 $, 
independent of $r$.
For $\nu=1/3$ the expression for $\rho_m(r)$ in Eq.~(\ref{eq:rho_vac_3})
should be used instead.

Therefore in this limit, with ${1 \over 3} < \nu < 1 $, one has 
\beq
m (r) \; \simeq \; 4 \, \pi \, \nu \, A_0 \,  r^{ {1 \over \nu} } 
\eeq
and, from the definition of $\sigma (r)$,
\beq
\sigma (r) \; \simeq \; - \, 2 \, m(r) \, G \, = \, 
- \, 8 \, \pi \, \nu \, G \, A_0 \,  r^{ {1 \over \nu} } 
\eeq
and finally
\beq
A^{-1}(r) \; = \; 1 \, - \, { 2 \, M \, G \over r} \, - \, 
2 \, a_0 \, M \, G \, c_s \, m^{1 \over \nu} \, r^{{1\over\nu}-1} \, + \, \dots
\label{eq:a_small_r}
\eeq
with the constant $c_s = \nu \vert \sec \left ( { \pi \over 2 \nu } \right )
\vert / \Gamma ( {1 \over \nu } - 1 ) $.
For $\nu=1/3$ the last contribution is indistinguishable from
a cosmological constant term $- {\lambda \over 3} r^2$, except 
for the fact that the coefficient here is quite different, 
being proportional to $\sim a_0 \, M \, G \, m^3$.

To determine the pressure, we suppose that it as well has a power
dependence on $r$ in the regime under consideration, 
$p_m (r) = c_p \, A_0 \, r^\gamma $, where $c_p$ is a numerical constant,
and then substitute $p_m (r)$ into the pressure equation Eq.~(\ref{eq:pres0}).
This gives, past the horizon $r \gg 2 M G$,
\beq
( 2 \gamma \, - \,1 ) \, c_p \, M \, G \, r^{\gamma - 1} \, 
- \, c_p \, \gamma \, r^\gamma \, - \, M \, G \, r^{1/\nu -4} \, \simeq 0
\eeq
giving the same power $\gamma=1/\nu-3$ as for $\rho(r)$, $c_p=-1$
and surprisingly also $\gamma=0$, implying that in this regime
only $\nu=1/3$ gives a consistent solution.
Again, the resulting correction is quite similar to what one would expect
from a cosmological term, with an effective 
$\lambda_m / 3 \simeq  8 \pi \, \nu \, a_0 \, M \, G \, m^{1 \over \nu} $.
One then has for $\nu$ near $1/3$
\beq
p_m (r) \; = \;  A_0 \, c_p \; r^{ {1 \over \nu} - 3 }  \, + \, \dots
\label{eq:p_small_r}
\eeq
and thus from Eq.~(\ref{eq:pres0}) 
\beq
B(r) \; = \; 1 \, - \, { 2 \, M \, G \over r} \, - \, 
2 \, a_0 \, M \, G \, c_s \, m^{1 \over \nu} \, r^{{1\over\nu}-1} \, + \, \dots
\label{eq:b_small_r}
\eeq
Both the result for $A(r)$ in Eq.~(\ref{eq:a_small_r}), and the above result for $B(r)$
are, for $r \gg 2 \, M \, G$, consistent with
a gradual slow increase in $G$ in accordance with the formula
\beq
G \; \rightarrow \; G(r) \; = \; 
G \, \left ( 1 \, + \, a_0 \, c_s \, m^{1 \over \nu} \, r^{1\over\nu}
\, + \, \dots
\right )
\label{eq:grun_small_r}
\eeq
We note here that both expressions for $A(r)$ and $B(r)$ have some similarities
with the approximate non-relativistic (Poisson equation) result of Eq.~(\ref{eq:phi_small2}),
with the correction proportional to $a_0$ agreeing roughly in magnitude (but not in sign).

The case $\nu=1/3$ requires a special treatment, since the coefficient $c_k$
in Eq.~(\ref{eq:bessel}) diverges as $\nu \rightarrow 1/3$.
Starting from the expression for $\rho_m(r)$ for $\nu=1/3$
in Eq.~(\ref{eq:rho_vac_3}),
\beq
\rho_m (r) \; = \; {1 \over 2 \pi^2 } \, a_0 \, M \, m^3 \, 
\, K_0 ( m \, r ) 
% \label{eq:rho_vac_3}
\eeq
one has for small $r$ 
\beq
\rho_m (r) \; = \; - \, {a_0 \over 2 \pi^2 } \, M \, m^3 \, 
\left ( \ln { m \, r \over 2 } \, + \, \gamma \, \right ) \, + \, \dots
\label{eq:rho_vac_3_s}
\eeq
and therefore from Eq.~(\ref{eq:theta_sol}),
\beq
\sigma (r) \; = \; {4 \, a_0 \, M \, G \, m^3 \over 3 \, \pi } 
\, r^3 \, \ln \, ( m \, r ) \, + \, \dots
\eeq
and consequently
\beq
A^{-1} (r) \; = \; 1 \, - { 2 \, M \, G \over r } \, + \, 
{4 \, a_0 \, M \, G \, m^3 \over 3 \, \pi } \, r^2 \, \ln \, ( m \, r ) 
\, + \, \dots
\label{eq:a_small_r3}
\eeq
From Eq.~(\ref{eq:pres0}) one can then obtain an expression for the
pressure $p_m(r)$, and one finds
\beq
p_m (r) \; = \; 
\frac{a_0 M m^3 \log (m r) }{2 \pi ^2}
-\frac{a_0 M m^3 \log \left (r + r \sqrt{1 - { 2 M G \over r } } - M G \right) }
{2 \pi ^2 \sqrt{1- { 2 M G \over r } }}
+\frac{a_0 M m^3}{\pi ^2}
+\frac{a_0 M m^3 c_3 }{2 \pi^2 \sqrt{1- {2 M G \over r } }}
\eeq
where $c_3$ is again an integration constant.
Here we will be content with the $r \gg 2 M G $ limit of the above expression,
which we shall write therefore as 
\beq
p_m (r) \; = \; {a_0 \over 2 \pi^2 } \, M \, m^3 \, \ln \, (  m \, r ) 
\, + \, \dots
\label{eq:p_vac_3}
\eeq
After performing the required $r$ integral in Eq.~(\ref{eq:theta_sol}),
and evaluating the resulting expression in the limit $r \gg 2 M G$, one obtains
an expression for $\theta (r)$, and consequently from it
\beq
B(r) \; = \; 1 \, - \, { 2 \, M \, G \over r} \, + \, 
{4 \, a_0 \, M \, G \, m^3 \over 3 \, \pi } \, r^2 \, \ln \, ( m \, r ) 
\, + \, \dots
\label{eq:b_small_r3}
\eeq
The expressions for $A(r)$ and $B(r)$ are, for $r \gg 2 \, M \, G$, consistent with
a gradual slow increase in $G$ in accordance with the formula
\beq
G \; \rightarrow \; G(r) \; = \; 
G \, \left ( 1 \, + \, 
{ a_0 \over 3 \, \pi } \, m^3 \, r^3 \, \ln \, { 1 \over  m^2 \, r^2 }  
\, + \, \dots
\right )
\label{eq:g_small_r3}
\eeq
and therefore consistent as well with the original result of Eqs.~(\ref{eq:grun_latt})
or (\ref{eq:grun_m}), namely that the classical laboratory value of $G$ 
is obtained for $ r \ll \xi $.
In fact it is reassuring that the renormalization properties of $G(r)$
as inferred from $A(r)$ are the same as what one finds from $B(r)$.
Note that the correct relativistic small $r$ correction of Eq.~(\ref{eq:g_small_r3})
agrees roughly in magnitude (but not in sign) with the approximate
non-relativistic, Poisson equation result of Eq.~(\ref{eq:phi_small1}).

One further notices some similarities, as well as some rather substantial differences,
with the corresponding QED result of Eq.~(\ref{eq:qed_s}).
In the gravity case, the correction vanishes as $r$ goes to zero: in this
limit one is probing the bare mass, unencumbered by its virtual graviton vacuum
polarization cloud.
On the other hand, in the QED case, as one approaches the source
one is probing the bare charge, unscreened by the electron's vacuum polarization cloud,
and whose magnitude diverges logarithmically for small $r$.

It should be recalled here that neither function $A(r)$ or $B(r)$
are directly related
to the relativistic potential for particle orbits,
which is given instead by the combination
\beq
V_{eff} (r) \; = \; {1 \over 2 \, A(r)} 
\left [ \, \frac{l^2}{r^2} \, - \, \frac{1}{B(r)} \, + \, 1 \, \right ]
\label{eq:veff}
\eeq
where $l$ is proportional to the orbital angular momentum of the test
particle \cite{hartle-book}.

Furthermore, from the metric of Eqs.~(\ref{eq:a_small_r}) and (\ref{eq:b_small_r})
one finds for $\nu \rightarrow 1/3$ the following results for the curvature
invariants
\bea
R^2 \; & = & \; 1024 A_0^2 G^2 \pi ^2
\nonumber \\
R_{\mu\nu} \, R^{\mu\nu} \; & = & \; 256 A_0^2 G^2 \pi ^2
\nonumber \\
R_{\mu\nu\lambda\sigma} \, R^{\mu\nu\lambda\sigma} \; & = & \;
16 G^2 \left( \frac{32 \pi ^2 A_0^2}{3} \, +\, \frac{3 M^2}{r^6}\right)
\label{eq:r2terms}
\eea
which are non-singular at $r = 2 \, M \, G$, and again consistent with an
effective mass density around the source $m(r) \propto r^3$.

\subsection{Large r limit}

For large $r$ one has instead, from Eq.~(\ref{eq:rho_vac}) for $\rho_m(r)$,
\beq
\rho_m (r) \; \mathrel{\mathop\sim_{ r \, \rightarrow \, \infty }} \; 
A_0 \; r^{ {1 \over 2 \, \nu} - 2 } \, \, e^{- \, m \, r} 
\label{eq:rho_large}
\eeq
with $A_0 = 1 / \sqrt{128 \pi} \, c_{\nu} \, a_0 \, M \, m^{1 + { 1 \over 2 \nu }} $.
In the same limit, the integration constants is chosen so that the
solution for $A(r)$ and $B(r)$ at large $r$ corresponds to a mass $M' = (1+a_0) M$
(see the expression for the integrated density in Eq.~(\ref{eq:rho_vac2})),
or equivalently
\beq
\sigma (r) \; \sim \; \theta (r) 
\; \mathrel{\mathop\sim_{ r \, \rightarrow \, \infty }} \; - 2 \, a_0 \, M \, G
\eeq
On then recovers a result similar to the non-relativistic expression of
Eqs.~(\ref{eq:phi_half}), (\ref{eq:grun_half}) and (\ref{eq:phi_large1}),
with $G(r)$ approaching the constant value
$G_{\infty} = (1 + a_0) G$, up to exponentially small corrections 
in $ m r $ at large $r$.

In conclusion, it appears that a solution to relativistic static isotropic
problem of the running gravitational constant can be found, provided that
the exponent $\nu$ in either Eq.~(\ref{eq:grun_m}) or Eq.~(\ref{eq:field})
is close to one third.
This last result seems to be linked with the fact that the running coupling
term acts in some way like a local cosmological constant term, for which the
$r$ dependence of the vacuum solution for small $r$ is fixed by the nature
of the Schwarzschild solution with a cosmological constant term
\footnote{
In $d \ge 4$ dimensions the Schwarzschild solution to Einstein
gravity with a cosmological term is \cite{perry}
$A^{-1}(r) = B(r) = 1 - 2 M G c_d \, r^{3-d} - { 2 \lambda \over (d-2)(d-1) } \, r^2$,
with $c_d = 4 \pi \Gamma({d-1 \over 2}) / (d-2) \pi^{d-1 \over 2}$,
which would suggest, in analogy with the results for $d=4$ given in this section,
that in $d \ge 4 $ dimensions only $\nu=1/(d-1)$ is possible.
This last result would also be in agreement with the exact value
$\nu=0$ found at $d=\infty$ \cite{larged} }.

\vskip 30pt
\newsection{Distortion of the Gravitational Wave Spectrum}
\hspace*{\parindent}

A scale-dependent gravitational constant $G(k^2)$ will cause slight distortions
in the spectrum of gravitational radiation at extremely low frequencies,
to some extent irrespective of the nature of the perturbations that cause them.
From the field equations with $\lambda=0$
\beq
R_{\mu\nu} \, - \, \half \, g_{\mu\nu} \, R \, 
\; = \; 8 \pi \, G  \, T_{\mu\nu}
\eeq
one obtains in the weak field limit with harmonic gauge condition
\beq
\Box \, h_{\mu\nu} \; = \; 8 \pi \, G  \, \bar T_{\mu\nu}
\eeq
with as usual 
\beq
\bar T_{\mu\nu} \; \equiv \; T_{\mu\nu} \, - \, {1 \over 2} 
\, \eta_{\mu\nu} \, T^{\;\;\lambda}_\lambda
\eeq
Density perturbations $\delta \rho ( {\bf x} , t)$ will enter the r.h.s.
of the field equations and give rise to gravitational waves
with Fourier components
\beq
h_{\mu\nu} (k) \; = \; - 8 \pi \, G  \, { 1 \over k^2 } \, \bar T_{\mu\nu} (p,\rho) (k)
\eeq
giving for the power spectrum of transverse traceless (gravitational wave) modes
\beq
P_{TT} (k^2) \; \simeq \; k^3 \,
\vert \, h_{TT} (k) \, \vert^2 \; = \; ( 8 \pi )^2 \, G^2 \,  { 1 \over k } 
\, \vert \, \bar T (p,\rho) (k) \, \vert^2
\eeq
A scale dependent gravitational constant, with variation in accordance
with Eq.~(\ref{eq:grun_m}), 
\beq
G \; \rightarrow \; G ( k^2 )
\eeq
would affect the spectrum of very long wavelength modes via
\beq
P_{TT} (k^2) \; \simeq \; 
k^3 \, \vert h_{\mu\nu} (k) \vert^2 \; = \; ( 8 \pi )^2 \, G^2(k^2) \, { 1 \over k } 
\, \vert \, \bar T_{\mu\nu} (p,\rho) \, \vert^2
\eeq
Specifically, according to the expression in Eq.~(\ref{eq:grun_m}) for the running 
of the gravitational constant, 
\beq
{ G(k^2) \over G } \; \simeq \; 1 \, + \, a_0 \,
\left ( { m^2 \over k^2 \, + \, m^2 } \right )^{1 \over 2 \nu} \, + \, \cdots
\eeq
one has for the tensor power spectrum
\beq
P_{TT} (k^2) \; \simeq \; 
k^3 \, \vert h_{\mu\nu} (k) \vert^2 \; = \; ( 8 \pi )^2 \, 
G^2 \, { 1 \over k } \, \left [ \, 1 \, + \, a_0 \, 
\left ( { m^2 \over k^2 + m^2 } \right )^{1 \over 2 \nu} \; \right ]^2
\, \vert \, \bar T_{\mu\nu} (p,\rho) \, \vert^2
\label{eq:power}
\eeq
with the expression in square brackets varying perhaps by as much
as an order of magnitude from short wavelengths $k \gg 1/\xi $, to 
very long wavelengths $k \sim 1/\xi$.

\vskip 30pt
\newsection{Quantum Cosmology - An Addendum}
\hspace*{\parindent}

In this section we will discuss briefly what modifications are expected
when one uses Eq.~(\ref{eq:grun_m}) instead of Eq.~(\ref{eq:grun_latt})
in the effective field equations.
In \cite{effective} cosmological solutions within the 
Friedmann-Robertson-Walker (FRW) framework were discussed,
starting from the quantum effective field equations
of Eq.~(\ref{eq:field0}), 
\beq
R_{\mu\nu} \, - \, \half \, g_{\mu\nu} \, R \, + \, \lambda \, g_{\mu\nu}
\; = \; 8 \pi G  \, \left( 1 + A( \Box ) \right) \, T_{\mu\nu}
\eeq
with $A(\Box)$ defined in either Eq.~(\ref{eq:abox0}) or Eq.~(\ref{eq:abox}), 
and applied to the standard Robertson-Walker metric
\beq
ds^2 \; = \; - dt^2 + a^2(t) \, \left \{ { dr^2 \over 1 - k\,r^2 } 
+ r^2 \, \left( d\theta^2 + \sin^2 \theta \, d\varphi^2 \right)  \right \}
\eeq
It should be noted that there are {\it two} quantum contributions to this set of equations. 
The first one arises because of the presence of a non-vanishing 
cosmological constant $\lambda \simeq 1 / \xi^2 $, as in Eq.~(\ref{eq:xi_lambda}),
originating in the non-perturbative vacuum condensate of the curvature. 
As in the case of standard FRW cosmology, this is the dominant contributions
at large times $t$, and gives an exponential expansion of the scale factor.

The second contribution arises because of the running of $G$ for $t \ll \xi$ in 
the effective field equations,
\beq
G(\Box)  \; = \; G \, \left( 1 + A( \Box ) \right) \; = \; 
G \, \left [ \, 1 + a_0 \left ( \xi^2 \Box \right )^{-{ 1 \over 2 \nu }} \, + \, \dots \, \right ]
\eeq
with $\nu \simeq 1/3$ and $a_0$ a calculable coefficient of order one
(see Eqs.~(\ref{eq:grun_latt}) and (\ref{eq:grun_m})).

In the simplest case, namely for a universe filled with non-relativistic matter ($p$=0),
the effective Friedmann equations then have the following appearance \cite{effective}
\bea
{ k \over a^2 (t) } \, + \,
{ \dot{a}^2 (t) \over a^2 (t) }  
& \; = \; & { 8 \pi G(t) \over 3 } \, \rho (t) \, + \, { 1 \over 3 \, \xi^2 }
\nonumber \\
& \; = \; & { 8 \pi G \over 3 } \, \left [ \,
1 \, + \, c_\xi \, ( t / \xi )^{1 / \nu} \, + \, \dots \, \right ]  \, \rho (t)
\, + \, { 1 \over 3 \, \xi^2 } 
\label{eq:fried_tt}
\eea
for the $tt$ field equation, and
\bea
{ k \over a^2 (t) } \, + \, { \dot{a}^2 (t) \over a^2 (t) }
\, + \, { 2 \ddot{a}(t) \over a(t) } 
& \; = \; & - \, { 8 \pi G \over 3 } \, \left [ \, c_\xi \, ( t / \xi )^{1 / \nu} 
\, + \, \dots \, \right ] \, \rho (t) 
\, + \, { 1 \over \xi^2 }
\label{eq:fried_rr}
\eea
for the $rr$ field equation.
The running of $G$ appropriate for the RW metric, and appearing explicitly in
the first equation, is described by
\beq
G(t) \; = \; G \, \left [ \; 1 \, + \, c_\xi \, 
\left ( { t \over \xi } \right )^{1 / \nu} \, + \, \dots \, \right ]
\label{eq:grun_frw}
\eeq
(with $c_\xi$ or the same order as $a_0$ of Eq.~(\ref{eq:grun_latt}) \cite{effective}).
Note that the running of $G(t)$ induces as well an effective pressure term in the second 
($rr$) equation.
\footnote{
We wish to emphasize that we are {\it not} talking here about models with a
time-dependent value of $G$.
Thus, for example, the value of $G \simeq G_c$ at laboratory scales should be taken to
be constant throughout most of the evolution of the universe.} 

One can therefore talk about an effective density
\beq
\rho_{eff} (t) \; = \; { G(t) \over G } \, \rho (t)
\label{eq:rho_eff}
\eeq
and an effective pressure
\beq
p_{eff} (t) \; = \; 
{ 1 \over 3 } \, \left ( { G(t) \over G } \, - \, 1 \right ) \, \rho (t) 
\label{eq:p_eff}
\eeq
with $ p_{eff} (t) / \rho_{eff} (t) = \third ( G(t) - G ) / G(t) $
\footnote{
Strictly speaking, the above results can only be proven if one assumes that
the pressure's time dependence is given by a power law, as discussed in detail
in \cite{effective}.
In the more general case, the solution of the above equations for various choices of
$\xi$ and $a_0$ has to be done numerically.
}.
Within the FRW framework, the gravitational vacuum polarization term behaves therefore
in some ways (but not all) like a positive pressure term, with $p(t) = \omega \, \rho(t)$ and
$\omega=1/3$, which is therefore characteristic of radiation.
One could therefore visualize the gravitational vacuum polarization contribution
as behaving like ordinary radiation, in the form of a dilute virtual
graviton gas: a radiative fluid with an equation of state $p={1 \over 3} \rho$.
It should be emphasized though that the relationship between density $\rho (t)$
and scale factor $a(t)$ is very different from the classical case.

The running of $G(t)$ in the above equations follows directly
from the basic result of Eq.~(\ref{eq:grun_latt}) 
(with the dimensionless constant $c_\xi$ proportional to $a_0$, with a numerical
coefficient of order one given in magnitude in \cite{effective}), but transcribed,
by explicitly computing the action of the covariant d'Alembertian
$ \Box \; \equiv \; g^{\mu\nu} \, \nabla_\mu \nabla_\nu $ on $T_{\mu\nu}$,
for the RW metric.
In other words, following the more or less unambiguously defined sequence
$G(k^2) \rightarrow G(\Box) \rightarrow G(t)$.
At the same time, the discussion of Sec. 1 underscores the fact that
for large times $t \gg \xi$ the form of Eq.~(\ref{eq:grun_latt}), and therefore
Eq.~(\ref{eq:grun_frw}), is no longer appropriate, due
to the spurious infrared divergence of Eq.~(\ref{eq:grun_latt}) at
small $k^2$.
Indeed from Eq.~(\ref{eq:grun_m}), the infrared regulated version of 
the above expression should read instead
\beq
G(t) \; \simeq \; G \, \left [ \; 1 \, 
+ \, c_\xi \, \left ( { t^2 \over t^2 \, + \, \xi^2 } \right )^{1 \over 2 \nu} \, 
+ \, \dots \; \right ]
\label{eq:grun_frw_m}
\eeq
with $\xi=m^{-1}$ the (tiny) infrared cutoff.
Of course it reduces to the expression in Eq.~(\ref{eq:grun_frw}) in the limit
of small times $t$, but for very large times $t \gg \xi$ the gravitational
coupling, instead of unphysically diverging, approaches a constant, finite value 
$G_\infty = ( 1 + a_0 + \dots ) \, G_c $, independent of $\xi$.
The modification of Eq.~(\ref{eq:grun_frw_m}) should apply whenever one considers
times for which $t \ll \xi$ is not valid. 
But since $\xi \sim 1 \sqrt{\lambda}$ is of the order the size of the visible
universe, the latter regime is largely of academic interest, and was therefore 
not discussed much in \cite{effective}. 

It should be noted that the effective Friedman equations of Eqs.~(\ref{eq:fried_tt}) 
and (\ref{eq:fried_rr}) also bear a superficial degree of resemblance to what
might be obtained in some scalar-tensor theories of gravity, where the
gravitational Lagrangian is postulated to be some singular function of the scalar
curvature \cite{inverse,moffat}.
Indeed in the FRW case one has, for the scalar curvature in terms of the scale factor,
\beq
R \; = \; 6 \left( k \, + \, {\dot{a}}^2(t) \, + \, a(t) \, \ddot{a}(t) \right) / a^2(t)
\eeq
and for $k=0$ and $a(t) \sim t^{\alpha}$ one has
\beq
R  \; = \; { 6 \, \alpha ( 2 \, \alpha - 1 ) \over t^2 } 
\eeq
which suggests that the quantum correction in Eq.~(\ref{eq:fried_tt}) is,
at this level, nearly indistinguishable from an inverse curvature term of the type
$ ( \xi^2 \, R )^{-1 / 2 \nu }$, 
or $ 1/( 1 \, + \, \xi^2 R )^{1 / 2 \nu }$ if one uses the infrared regulated version.
The former would then correspond the to an effective gravitational action

\beq
I_{eff}  \; \simeq \; { 1 \over 16 \pi G } \int dx \, \sqrt{g} \,
\left ( \, R \, + \, 
{ f \, \xi^{- {1\over\nu} } \over \left ( R \right )^{{1\over 2 \nu} -1} } 
\, - \, 2 \, \lambda \, \right )
\label{eq:scalar}
\eeq
with $f$ a numerical constant of order one, and $\lambda \simeq 1 / \xi^2 $.
But this superficial resemblance is seen here more as an artifact, due to the 
particularly simple form of the RW metric, with the coincidence
of several curvature invariants not expected to be true in general.

\vskip 30pt
\newsection{Conclusions}
\hspace*{\parindent}

In this paper we have examined a number of basic issues connected with the
renormalization group running of the gravitational coupling.
The scope of this paper was to explore the overall consistency of the picture 
obtained from the lattice, by considering a number of basic issues, one of
which is the analogy, or contrast, with a much better understood class of
theories, namely QED on the one hand, and non-abelian gauge theories and QCD
on the other.

The starting point for our discussion of the renormalization group
running of $G$ (Sec. 2) is Eq.~(\ref{eq:grun_latt}) (valid at short distances $k \gg m$,
or, equivalently $r \ll \xi$), and its improved infrared regulated version
of Eq.~(\ref{eq:grun_m}).
The scale dependence for $G$ obtained from the lattice is remarkably similar to
the result of the $2+\epsilon$ expansion in the continuum,
as in Eq.~(\ref{eq:grun_cont}), with two important differences: only the
strong coupling phase $G>G_c$ is physical, and for the exponent one has
$\nu \simeq 1/3$ in four dimensions.
The similarity between the two results in part also originates from the
fact that in both cases the renormalization group properties of $G$
are inferred (implicitly, in the $2+\epsilon$ case) from the requirement that
the non-perturbative scale of Eq.~(\ref{eq:m_latt}) be treated as an invariant.

Inspection of the quantum gravitational functional integral $Z$ of Eq.~(\ref{eq:zc})
reveals that its singular part can only depend of the dimensionless combination
$\lambda_0 G^2$, up to an overall factor which cannot affect the non-trivial scaling
behavior around the fixed point, since it is analytic in the couplings.
This then leaves the question open of which coupling(s) run and which
ones do not.

The answer in our opinion is possibly quite simple, and is perhaps best inferred from
the nature of the Wilson loop of Eq.~(\ref{eq:wloop_curv}): 
the appropriate renormalization scheme for quantum gravity is one in which $G$ runs
with scale according to the prediction Eq.~(\ref{eq:grun_m}),
and the scaled cosmological constant $\lambda$ is kept fixed, 
as in Eqs.~(\ref{eq:xi_lambda}) and (\ref{eq:g_decompose}).
Since the scale $\xi$ is related to the observable curvature at large scales,
it is an almost inescapable conclusion of these arguments that it must be macroscopic.
Furthermore, it is genuinely non-perturbative and non-analytic in $G$, as 
seen for example from Eq.~(\ref{eq:r_nonan}), and represents the
effects of the gravitational vacuum condensate which makes its appearance
in the strongly coupled phase $G>G_c$.

Another aspect we have investigated in this paper is the nature of the
quantum corrections to the gravitational potential $\phi (r)$ in real space, 
arising from the scale dependence of Newton's constant $G$.
The running is originally formulated in momentum space (see Eq.~(\ref{eq:grun_m})), 
since it originates in the momentum dependence of $G$ as it arises on the lattice,
or in the equivalent renormalization group equations, Eqs.~(\ref{eq:beta_latt})
or (\ref{eq:beta_mu}).
The solution $\phi(r)$ to the non-relativistic Poisson equation for a point source
is given in Eq.~(\ref{eq:phi_small1}) of Sec. 3 for various values of the exponent $\nu$.
The solution is obtained by first computing the effective vacuum polarization
density $\rho_m(r)$ of Eq.~(\ref{eq:rho_vac}), and then using it as a source term in
Poisson's equation.
Already in the non-relativistic case, the value $\nu=1/3$ appears to stand out,
since it leads to logarithmic corrections at short distances $r \ll \xi$.

A relativistic generalization of the previous results was worked out in 
Secs. 4 and 5.
First it was shown that the scale dependence of $G$ can be consistently embedded
in a relativistic covariant framework using the d'Alembertian $\Box$ operator,
leading to a set of nonlocal effective field equations, Eq.~(\ref{eq:field}).
The consequences can then be worked out in some detail for the static isotropic
metric (Sec. 4), at least in a regime where $ 2 M G \ll r \ll \xi$, and under the assumption
of a power law correction (otherwise the problem becomes close to intractable).
One then finds that the structure of the leading quantum correction
severely restricts the possible values for the exponent $\nu$, in the sense that 
no consistent solution to the effective non-local
field equations, incorporating the running of $G$, can be found unless $\nu^{-1}$
is an integer.

A somewhat different approach to the solution of the static isotropic
metric was then discussed in Sec. 5, in terms of the effective vacuum density
of Eq.~(\ref{eq:rho_vac}), and a vacuum pressure chosen so as to satisfy a
covariant energy conservation for the vacuum polarization contribution.
The main result is the derivation from the relativistic field equations of
an expression for the metric coefficients $A(r)$ and $B(r)$, given in 
Eqs.~(\ref{eq:a_small_r3}) and (\ref{eq:b_small_r3}).
For $\nu=1/3$ it implies for the running of $G$ in the region $ 2 M G \ll r \ll \xi$
the result of Eq.~(\ref{eq:g_small_r3}),
\beq
G(r) \; = \; 
G \, \left ( 1 \, + \, 
{ a_0 \over 3 \, \pi } \, m^3 \, r^3 \, \ln \, { 1 \over  m^2 \, r^2 }  
\, + \, \dots
\right )
% \label{eq:g_small_r3}
\eeq
indicating therefore a gradual, very slow increase in $G$ from the ``laboratory''
value $G \equiv G(r=0)$.
For the actual values of the parameters appearing in the above expression
one expects that $m$ is related to the curvature on the largest scales,  
$m^{-1} = \xi \sim 10^{28} cm$, and that $a_0 \sim O(10)$.
From the nature of the solution for $A(r)$ and $B(r)$ one finds again 
that unless the exponent $\nu$ is close to $1/3$, a consistent solution
of the field equations cannot be found.
Note that for very large $r \gg \xi$ the growth in $G(r)$ saturates and the value 
$G_{\infty} = (1 + a_0 ) G$ is obtained, in accordance with the original
formula of Eq.~(\ref{eq:grun_m}) for $k^2 \simeq 0$.
A natural comparison is with the QED result of Eq.~(\ref{eq:qed_s}).

At the end of the paper we have added some remarks on
the solution of the gravitational wave equation with a running $G$.
We find that a running Newton's constant will slightly distort the gravitational
wave spectrum at very long wavelengths (Sec. 6), according to 
Eq.~(\ref{eq:power}).
Regarding the problem of finding solutions of the effective
non-local field equations in a cosmological context \cite{effective}, wherein
quantum corrections to the Robertson-Walker metric and the basic
Friedman equations (Eqs.~(\ref{eq:fried_tt}) and (\ref{eq:fried_rr}))
are worked out, we have discussed some of the simplest
and more plausible scenarios for the growth (or lack thereof)
of the coupling at very large distances, past the de Sitter horizon.

\vspace{20pt}

{\bf Acknowledgements}

The authors wish to thank Luis \'Alvarez Gaum\'e, Gabriele Veneziano and the 
Theory Division at CERN for their warm hospitality.
The work of Ruth M. Williams was supported in part by the UK Particle
Physics and Astronomy Research Council.

% \newpage

\vfill

\newpage


\vskip 30pt



\begin{thebibliography}{99}


\bibitem{feylec}
  R.~P.~Feynman, {\sl `Lectures on Gravitation'}, 1962-1963,
  edited by F.~B.~Morinigo and W.~G.~Wagner, California Institute
  of Technology (Pasadena, 1971).

\bibitem {thooft}
  G.~'t Hooft and M.~Veltman, 
  %``One Loop Divergencies In The Theory Of Gravitation,''
  {\it Ann. Inst. Poincar\'e} {\bf 20} 69 (1974).

\bibitem{deser}
  S.~Deser and P.~van Nieuwenhuizen, {\it Phys. Rev.}{\bf D10} 401,410 (1974);
  %``One Loop Divergences Of Quantized Einstein-Maxwell Fields,''
  %``Nonrenormalizability Of The Quantized Dirac - Einstein System,''
  S.~Deser, H.~S.~Tsao and P.~van Nieuwenhuizen,
  %``One Loop Divergences Of The Einstein Yang-Mills System,''
  {\it Phys.\ Rev.} {\bf D 10}, 3337 (1974).

%% General Renormalization Group

\bibitem{wilson}
  K.~G.~Wilson, {\it Rev.\ Mod.\ Phys.} {\bf 47} 773 (1975);
  {\bf 55} 583 (1983);
  K.~G.~Wilson and M.~Fisher, {\it Phys.\ Rev.\ Lett.} {\bf 28} 240 (1972).

%% Non-Renormalizable theories

\bibitem{wilseps}
  K.~G.~Wilson, {\it Phys.\ Rev.} {\bf D 7}, 2911 (1973).

\bibitem {parisi}
  G.~Parisi, {\it Lett.\ Nuovo Cim.} {\bf 6} 450 (1973);
  {\it Nucl.\ Phys.} {\bf B100} 368 (1975), {\it ibid.} 
  {\bf 254} 58 (1985); {\sl `On Non-renormalizable Interactions'}, 
  in {\it New Development in Quantum Field Theory and Statistical Mechanics},
  (Cargese 1976, M.~Levy and P.~Mitter eds. , Plenum Press, New York 1977).

\bibitem{sym}
  K.~Symanzik, {\it Comm.\ Math.\ Phys.} {\bf 45} 79-98 (1975).

% \bibitem{nonren-wein}
%  S.~Weinberg, {\it Phys.\ Rev.} {\bf D56} 2303 (1997).

\bibitem{epsilon1}
  S.~Weinberg, in {\sl `General Relativity - An Einstein Centenary
  Survey'}, edited by S.~W.~Hawking and W.~Israel,
  (Cambridge University Press, 1979).

\bibitem {hawking}
  S.~W.~Hawking, in {\sl `General Relativity - An Einstein Centenary
  Survey'}, edited by S.~W.~Hawking and W.~Israel,
  (Cambridge University Press, 1979);

\bibitem{harhawk}
  J.~B.~Hartle and S.~W.~Hawking,
  %``Wave Function Of The Universe,''
  {\it Phys.\ Rev.} {\bf D 28}, 2960 (1983).

\bibitem{harcosm}
  J.~B.~Hartle,
  %``Quantum Cosmology,''
  lectures delivered at the {\it Theoretical Advanced Study Institute}
  in Elementary Particle Physics, Yale University, 1985, vol. 2, p. 471-566. 

\bibitem {effective}
  H.~W.~Hamber and R.~M.~Williams, {\it Phys.\ Rev.} {\bf D 72} 044026 (2005);
  Mod.\ Phys.\ Lett.\ A {\bf 21}, 735 (2006).

\bibitem {critical}
  H.~W.~Hamber, {\it Phys.\ Rev.} {\bf D45} 507 (1992);
  {\it Phys.\ Rev.} {\bf D61} 124008 (2000).

\bibitem {ttmodes}
  H.~W.~Hamber and R.~M.~Williams, 
  {\it Phys.\ Rev.} {\bf D 70}, 124007 (2004).

\bibitem{larged}
  H.~W.~Hamber and R.~M.~Williams,
  % ``Quantum gravity in large dimensions,''
  {\it Phys.\ Rev.} {\bf D73}, 044031 (2006).


%% Regge lattice

\bibitem {regge}
  T.~Regge, {\it Nuovo Cimento} {\bf 19} 558 (1961).

\bibitem {rowi}
  M.~Ro\u cek and R.~M.~Williams, {\it Phys.\ Lett.} {\bf 104B} 31 (1981);
  {\it Z.\ Phys.} {\bf C21} 371 (1984).

\bibitem {cms1}
  J.~Cheeger, W.~M\"uller and R.~Schrader, 
  %``Lattice Gravity Or Riemannian Structure On Piecewise Linear Spaces,''
  in {\sl `Unified Theories Of Elementary Particles'},  
  {\it Heisenberg Symposium}, M\"unchen 1981, p. 176-188, 
  (Springer, New York, 1982).

\bibitem {lee}
  T.~D.~Lee, in {\sl `Discrete Mechanics'}, 
  1983 Erice International School of Subnuclear Physics,
  vol. 21 (Plenum Press, New York 1985), and references therein.

\bibitem {hw84}
  H.~W.~Hamber and R.~M.~Williams, {\it Nucl.\ Phys.} {\bf B248} 392 (1984);
  {\bf B260} 747 (1985); {\it Phys.\ Lett.} {\bf 157B} 368 (1985);
  {\it Nucl.\ Phys.} {\bf B269} 712 (1986).

\bibitem {hartle}
  J.~B.~Hartle, {\it J.\ Math.\ Phys.} {\bf 26} 804 (1985);
  {\bf 27} 287 (1986); {\bf 30} 452 (1989).

\bibitem {monte}
  B.~Berg,  {\it Phys.\ Rev.\ Lett.} {\bf 55} 904 (1985);
  {\it Phys.\ Lett.} {\bf B176} 39 (1986).

\bibitem {hw2d}
  H.~W.~Hamber and R.~M.~Williams, {\it Nucl.\ Phys.} {\bf B267} 482 (1986).

\bibitem {hw3d}
   H.~W.~Hamber and R.~M.~Williams, {\it Phys.\ Rev.} {\bf D47} 510 (1993).

\bibitem {corr}
  H.~W.~Hamber, {\it Phys.\ Rev.} {\bf D50} 3932 (1994).


\bibitem{gra92}
  H.~W.~Hamber,
  % ``Phases of simplicial quantum gravity in four-dimensions: Estimates for the
  % critical exponents,''
  Nucl.\ Phys.\ B {\bf 400}, 347 (1993).


%% gravitational measure 

\bibitem {dewitt}
  B.~DeWitt, {\it Phys.\ Rev.} {\bf 160} 1113 (1967);
  in {\sl `General Relativity - An Einstein Centenary Survey'},
  edited by S.~W.~Hawking and W.~Israel, Cambridge University Press (1979).

\bibitem {misner}
  C.~W.~Misner, {\it Rev. Mod. Phys.} {\bf 29} 497 (1957);
  L.~D.~Faddeev and V.~N.~Popov, {\it Sov.\ Phys.\ Usp.} 
  {\bf 16} 777-788 (1974);
  {\it Usp.\ Fiz.\ Nauk.} {\bf 111} 427-450 (1973).


%% Wilson loop

\bibitem{fadpop}
  V.~N.~Popov and L.~D.~Faddeev,
  %``Perturbation theory for gauge-invariant fields,''
  Fermilab-pub-72-057-T, in 
  {\sl ``50 years of Yang-Mills theory''}, edited by G. 't Hooft, 
  (World Scientific, Singapore, 2004) pp. 40-64.

\bibitem{caselle}
  M.~Caselle, A.~D'Adda and L.~Magnea,
  %``Regge Calculus As A Local Theory Of The Poincare Group,''
  {\it Phys.\ Lett.} {\bf B 232}, 457 (1989).

\bibitem{froh}
  J.~Fr\"ohlich, {\sl `Regge Calculus and Discretized Gravitational Functional
  Integrals'}, I.~H.~E.~S.~preprint 1981 (unpublished);
  {\it Non-Perturbative Quantum Field Theory: Mathematical Aspects and
  Applications}, Selected Papers (World Scientific, Singapore, 1992),
  pp. 523-545. 

\bibitem{bianchi}
  H.~W.~Hamber and G.~Kagel, {\it Class.\ and Quant.\ Grav.} 
  {\bf 21}, 5915 (2004), and references therein.

\bibitem{moda}
  G.~Modanese,
  %``Geodesic round trips by parallel transport in quantum gravity,''
  {\it Phys.\ Rev.} {\bf D 47}, 502 (1993);
  %``Wilson loops in four-dimensional quantum gravity,''
  {\it Phys.\ Rev.} {\bf D 49}, 6534 (1994).

\bibitem {lines}
  H.~W.~Hamber and R.~M.~Williams, {\it Nucl.\ Phys.} {\bf B435} 361 (1995);
  {\it Phys.\ Rev.} {\bf D59} 064014 (1999).

\bibitem{wilson-lgt}
  K.~G.~Wilson,
  %``Confinement Of Quarks,''
  {\it Phys.\ Rev.} {\bf D 10}, 2445 (1974);
  {\sl `Quarks And Strings On A Lattice'},
  in {\it New Phenomena In Subnuclear Physics}, Erice 1975 
  (Plenum Press, New York, 1977).

\bibitem {peskin}
  See, for example, M.~E.~Peskin and D.~V.~Schroeder, 
  {\sl `An Introduction to Quantum Field Theory'}, sec. 22.1 
  (Addison Wesley, Reading, Massachusetts, 1995).



%% Universality of critical behavior


\bibitem {pbook}
  G.~Parisi, {\sl `Statistical Field Theory'},
  Addison Wesley, New York (1988).

\bibitem{zinn}
  See, for example, J.~Zinn-Justin, {\sl `Quantum Field Theory and Critical Phenomena'},
  (Oxford University Press, 3rd edition, 1996).

\bibitem{blz}
  E.~Brezin, J.~C.~Le Guillou and J.~Zinn-Justin,
  {\sl `Field Theoretical Approach To Critical Phenomena'},
  in ``Phase Transitions and Critical Phenomena'', 
  edited by C. Domb and M. Green (Academic Press, London 1976), 
  Vol.6, pp. 125-247. 



%% Condensates in QCD

\bibitem{shifman}
   M.A. Shifman, A.I. Vainshtein and V.I. Zakharov, Nucl. Phys. B147 (1979), 
   p. 385; p. 448; p. 519;
   V.A. Novikov, M.A. Shifman, A.I. Vainshtein and V.I. Zakharov preprint ITEP-42 Moscow (1981).

\bibitem{veneziano}
  A.~Armoni, M.~Shifman and G.~Veneziano,
  %``QCD quark condensate from SUSY and the orientifold large-N expansion,''
  Phys.\ Lett.\ B {\bf 579}, 384 (2004);
  A.~Armoni, G.~Shore and G.~Veneziano,
  %``Quark condensate in massless QCD from planar equivalence,''
  Nucl.\ Phys.\ B {\bf 740}, 23 (2006).

\bibitem{hp}
  H.~W.~Hamber and G.~Parisi,
  % ``Numerical Estimates For The Spectrum Of Quantum Chromodynamics,''
  Phys.\ Rev.\ D {\bf 27}, 208 (1983); Phys.\ Rev.\ Lett.\  {\bf 47}, 1792 (1981).
%  H.~W.~Hamber,
%  % ``Hadron Spectroscopy On A Large Lattice,''
%  Phys.\ Lett.\ B {\bf 178}, 277 (1986).



% 2 + epsilon for gravity

\bibitem{tsao}
  H.~S.~Tsao,
  %``Conformal Anomalies In A General Background Metric,''
  Phys.\ Lett.\ B {\bf 68}, 79 (1977).

\bibitem {epsilon}
  R.~Gastmans, R.~Kallosh and C.~Truffin, 
  {\it Nucl.\ Phys.} {\bf B133} 417 (1978);  
  S.~M.~Christensen and M.~J.~Duff, {\it Phys.\ Lett.} {\bf B79} 213 (1978).

\bibitem {epsilon2}
  H.~Kawai and M.~Ninomiya,  {\it Nucl.\ Phys.} {\bf B336} 115 (1990);  
  H.~Kawai, Y.~Kitazawa and M.~Ninomiya,  {\it Nucl.\ Phys.} 
  {\bf B393} 280 (1993) and {\bf B404} 684 (1993);  
  Y.~Kitazawa and M.~Ninomiya, {\it Phys.\ Rev.} {\bf D55} 2076 (1997).

\bibitem {epsilon3}
  T.~Aida and Y.~Kitazawa, {\it Nucl.\ Phys.} {\bf B491} 427 (1997).

\bibitem {hdqg}
  E.~S.~Fradkin and A.~A.~Tseytlin, {\it Phys.\ Lett.} {\bf 104B} 377 (1981)
  and {\bf 106B} 63 (1981); {\it Nucl.\ Phys.} {\bf B201} 469 (1982);
  I.~G.~Avramidy and A.~O.~Barvinsky,{\it Phys.\ Lett.} {\bf 159B} 269 (1985).


% 2 + epsilon for sigma model

\bibitem {sigma}
  E.~Brezin and J.~Zinn-Justin, {\it Phys.\ Rev.\ Lett.} {\bf 36} 691 (1976); 
  {\it Phys.\ Rev.} {\bf D14} 2615 (1976); 
  {\it Phys.\ Rev.} {\bf B14}, 3110 (1976);
  E.~Brezin and S.~Hikami, LPTENS-96-64 (Dec 1996) [cond-mat/9612016].

\bibitem{bardeen}
  W.~A.~Bardeen, B.~W.~Lee and R.~E.~Shrock,
  % ``Phase Transition In The Nonlinear Sigma Model In Two + Epsilon Dimensional Continuum,''
  Phys.\ Rev.\ D {\bf 14}, 985 (1976).

\bibitem{david}
  F.~David,
  % ``Cancellations Of Infrared Divergences In The Two-Dimensional Nonlinear
  % Sigma Models,''
  Commun.\ Math.\ Phys.\  {\bf 81}, 149 (1981).

\bibitem{hikami}
  S.~Hikami and E.~Brezin,
  %``Three Loop Calculations In The Two-Dimensional Nonlinear Sigma Model,''
  {\it J.\ Phys.} {\bf A 11}, 1141 (1978).

\bibitem{kleinert}
  H.~Kleinert,
  %``Variational resummation of epsilon-expansions of critical exponents of
  %nonlinear O(N)-symmetric sigma-model in 2+epsilon dimensions,''
  {\it Phys.\ Lett.} {\bf A 264}, 357 (2000).


% ERG Truncation method

\bibitem{litim}
  D.~Litim, CERN-Th-2003-299, {\it Phys.\ Rev.\ Lett.} {\bf 92} 201301 (2004);
  CERN-Th-2005-256 (hep-th/0606044) (2006).

\bibitem{litim1}
  P.~Fischer and D.~F.~Litim,
  % ``Fixed points of quantum gravity in higher dimensions,''
  CERN-Th-2005-257 (hep-th/0606135) (Feb. 2006); CERN-Th-2006-066
  (hep-th/0602203) (June 2006) .

\bibitem{reuter}
  M.~Reuter and F.~Saueressig, {\it Phys.\ Rev.} {\bf D65} 065016 (2002);
  O.~Lauscher and M.~Reuter, {\it Class.\ Quant.\ Grav.} {\bf 19} 483 (2002).

\bibitem{poly}
  A.~M.~Polyakov,
  % ``A Few projects in string theory,''
  in the proceedings of the Les Houches Summer School
  on {\sl Gravitation and Quantizations}, Session 57, Aug. 1992, 
  J.~Zinn-Justin and B.~Julia eds. (North-Holland, 1995) 
  (hep-th/9304146).  

% \bibitem{ambjorn}
%  J.~Ambjorn, J.~Jurkiewicz and R.~Loll,
%  % ``Reconstructing the universe,''
%  Phys.\ Rev.\ D {\bf 72}, 064014 (2005), and references therein.


% Running in QED and QCD

\bibitem{iz}
  See, for example, C.~Itzykson and J.~B.~Zuber, {\sl `Quantum Field Theory'},
  (McGraw-Hill, New York, 1980).

\bibitem{uehling}
   E.~A.~Uehling, {\it Phys.\ Rev.} {\bf 48}, 55-63 (1935).

\bibitem{kroll}
   E.~Wichmann and N.~Kroll, {\it Phys.\ Rev.} {\bf 96}, 232-234 (1954).

\bibitem{lep}
  A.~D.~Martin, J.~Outhwaite and M.~G.~Ryskin,
  % ``A New Determination Of The QED Coupling Alpha(M(Z)**2) Lets The Higgs  Off
  %The Hook,''
  Phys.\ Lett.\ B {\bf 492}, 69 (2000).

\bibitem{pdg}
  We follow the conventions of the Particle Data Group, pdg.lbl.gov,
  S.~Eidelman et al., {\it Phys. Lett.} {\bf B} 592, 1 (2004). 



% IR renormalons in QCD

\bibitem{dok}
  Y.~L.~Dokshitzer, G.~Marchesini and B.~R.~Webber,
  %  ``Dispersive Approach to Power-Behaved Contributions in QCD Hard Processes,''
  Nucl.\ Phys.\ B {\bf 469}, 93 (1996);
  A.~V.~Manohar and M.~B.~Wise,
  % ``Power suppressed corrections to hadronic event shapes,''
  Phys.\ Lett.\ B {\bf 344}, 407 (1995);
  R.~Akhoury and V.~I.~Zakharov,
  % ``On the universality of the leading, 1/Q power corrections in QCD,''
  Phys.\ Lett.\ B {\bf 357}, 646 (1995);
  B.~R.~Webber,
  % ``Renormalon phenomena in jets and hard processes,''
  Nucl.\ Phys.\ Proc.\ Suppl.\  {\bf 71}, 66 (1999).

\bibitem{beneke}
  M.~Beneke,
  % ``Renormalons,''
  Phys.\ Rept.\  {\bf 317}, 1 (1999);
  M.~Beneke and V.~M.~Braun,
  % ``Power corrections and renormalons in Drell-Yan production,''
  Nucl.\ Phys.\ B {\bf 454}, 253 (1995).

\bibitem{richardson}
  J.~L.~Richardson,
  % ``The Heavy Quark Potential And The Upsilon, J / Psi Systems,''
  Phys.\ Lett.\ B {\bf 82}, 272 (1979).

\bibitem{eichten}
  E.~Eichten and F.~Feinberg,
  %  ``Spin Dependent Forces In QCD,''
  Phys.\ Rev.\ D {\bf 23}, 2724 (1981);
  W.~Buchmuller and S.~H.~H.~Tye,
  % ``Quarkonia And Quantum Chromodynamics,''
  Phys.\ Rev.\ D {\bf 24}, 132 (1981);
  U.~Ellwanger, M.~Hirsch and A.~Weber,
  % ``The heavy quark potential from Wilson's exact renormalization group,''
  Eur.\ Phys.\ J.\ C {\bf 1}, 563 (1998).

\bibitem{vilko}
G.~A.~Vilkovisky,
%``The Unique Effective Action In Quantum Field Theory,''
{\it Nucl.\ Phys.} {\bf B 234} (1984) 125;
A.~O.~Barvinsky and G.~A.~Vilkovisky, {\it Nucl. Phys.} {\bf B 282}, 163 (1987);
{\bf B 333}, 471 (1990); {\bf B 333}, 512 (1990); 
A.~O.~Barvinsky, Y.~V.~Gusev, V.~V.~Zhytnikov and G.~A.~Vilkovisky,
Print-93-0274 (Manitoba);
A.~O.~Barvinsky and G.~A.~Vilkovisky,
%``The Generalized Schwinger-DeWitt Technique In Gauge Theories And Quantum
%Gravity,''
{\it Phys.\ Rept.} {\bf 119}, 1 (1985).

\bibitem {mtw}
  C.~W.~Misner, K.~S.~Thorne and J.~A.~Wheeler,
  {\sl Gravitation}, 
  (W.~H.~Freeman and Co. Publishers, New York, 1973).

\bibitem{hartle-book}
  J.~B.~Hartle, ``Gravity: an Introduction to Einstein's General Relativity'',
  (Addison-Wesley, New York, 2002).

\bibitem{perry}
  R.~C.~Myers and M.~J.~Perry,
  %``Black Holes In Higher Dimensional Space-Times,''
  Annals Phys.\  {\bf 172}, 304 (1986);
  D.~Y.~Xu,
  % ``EXACT SOLUTIONS OF EINSTEIN AND EINSTEIN-MAXWELL EQUATIONS IN HIGHER
  % DIMENSIONAL SPACE-TIME,''
  Class.\ Quant.\ Grav.\  {\bf 5} (1988) 871.

\bibitem{inverse}
  S.~Capozziello, S.~Carloni and A.~Troisi, in 
  ``{\sl Recent Research Developments in Astronomy and Astrophysics}''
  -RSP/AA/21-2003, [astro-ph/0303041];
   S.~M.~Carroll, V.~Duvvuri, M.~Trodden and M.~S.~Turner,
  {\it Phys.\ Rev.} {\bf D 70}, 043528 (2004);
  E.~E.~Flanagan, {\it Phys.\ Rev.\ Lett.} {\bf 92}, 071101 (2004).

\bibitem{moffat}
  J.~W.~Moffat,
  %``Scalar-tensor-vector gravity theory,''
  JCAP {\bf 0603}, 004 (2006); JCAP {\bf 0505}, 003 (2005).



\end{thebibliography}
\end{document}